\documentclass[reprint, prd, superscriptaddress, longbibliography, tightenlines, nofootinbib, eqsecnum, amsfonts, amsmath, amssymb, floatfix, notitlepage,twocolumn]{revtex4-1}
\pdfoutput=1

\usepackage[utf8]{inputenc}
\usepackage{euscript}
\usepackage{epsfig}
\usepackage{graphics}
\usepackage{graphicx}
\usepackage{amsmath}
\usepackage{amssymb}
\usepackage{bm}
\usepackage[usenames,dvipsnames,svgnames,table]{xcolor}
\usepackage{xspace}
\usepackage{times}
\usepackage{appendix}
\usepackage{lipsum}
\usepackage[nolist,nohyperlinks]{acronym}
\usepackage{float}
\usepackage{simplewick}
\usepackage{tabularx}
\usepackage{booktabs}
\usepackage{natbib, ifthen}
\usepackage{hyperref}
\usepackage{comment}

\newcommand{\la}{\langle}
\newcommand{\ra}{\rangle}

\newcommand{\fsky}{f_{\rm sky}}

\providecommand{\planck}{\textit{Planck}}
\providecommand{\Planck}{\planck}

\newcommand{\mksym}[1]{\ifmmode {\rm #1}\else #1\fi}

\newcommand{\taunl}{\ifmmode {\tau_{\rm{NL}}}\else $\tau_{\rm{NL}}$\fi}

\providecommand{\Planck}{\textit{Planck}}
\providecommand{\planck}{\Planck}

\providecommand{\lea}{\la}

\providecommand{\alt}{\lea}

\providecommand{\text}[1]{\rm{#1}}

\newcommand{\half}{{\textstyle \frac{1}{2}}}

\providecommand{\muK}{\mu{\rm K}}
\providecommand{\arcmin}{{\rm arcmin}}

\newcommand{\begm}{\begin{pmatrix}}
\newcommand{\enm}{\end{pmatrix}}

\newcommand\ba{\begin{eqnarray}}
\newcommand\ea{\end{eqnarray}}
\newcommand\bea{\begin{eqnarray}}
\newcommand\eea{\end{eqnarray}}

\newcommand\be{\begin{equation}}
\newcommand\ee{\end{equation}}

\newcommand{\valpha}{{\boldsymbol{\alpha}}}

\newcommand{\ud}{{\rm d}}

\newcommand{\boldvec}[1]{{\mbox{\boldmath{$#1$}}}}

\newcommand{\vl}{\boldvec{l}}

\newcommand{\vn}{\boldvec{n}}

\providecommand{\vr}{\boldvec{r}}
\newcommand{\vs}{\boldvec{s}}

\newcommand{\vx}{\boldvec{x}}
\newcommand{\vy}{\boldvec{y}}

\newcommand{\vnhat}{\hat{\vn}}
\newcommand{\vrhat}{\hat{\vr}}

\graphicspath{{./clfigs/}}

\renewcommand{\boldvec}[1]{\boldsymbol{#1}}
\renewcommand{\vr}{\boldsymbol{r}}

\newcommand{\vDelta}{\boldsymbol{\Delta}}
\newcommand{\meandelta}{\bar{\Delta}}

\newcommand{\websky}{Websky\xspace}

\newcommand{\poissonmean}{\lambda_R}
\newcommand{\poissondelta}{f}
\newcommand{\ximasked}{\tilde \xi_{\rm masked}}

\definecolor{ZurichBlue}{rgb}{.255,.41,.884} 		
\definecolor{ZurichRed}{rgb}{0.9, 0.1, 0} 			
\definecolor{ZurichGreen}{rgb}{.196,.504,.396} 		
\definecolor{ZurichYellow}{rgb}{1,.648,0} 			
\definecolor{dodgerblue}{rgb}{0.12, 0.56, 1.0}
\definecolor{azure}{rgb}{0.0, 0.5, 1.0}
\definecolor{awesome}{rgb}{1.0, 0.13, 0.32}
\definecolor{alizarincrimson}{rgb}{0.82, 0.1, 0.26}
\definecolor{mediumpurple}{rgb}{0.58, 0.44, 0.86}
\definecolor{lasallegreen}{rgb}{0.03, 0.47, 0.19}

\begin{document}

\newcommand{\Sussex}{Department of Physics \& Astronomy, University of Sussex, Brighton BN1 9QH, United Kingdom}
\newcommand{\Ferrara}{Dipartimento di Fisica e Scienze della Terra, Universit\`a degli Studi di Ferrara, via Giuseppe Saragat 1, I-44122 Ferrara, Italy}
\newcommand{\INFN}{Istituto Nazionale di Fisica Nucleare (INFN), Sezione di Ferrara, Via Giuseppe Saragat 1, I-44122 Ferrara, Italy}

\title{Lensed CMB power spectrum biases from masking extragalactic sources}

\author{Giulio Fabbian}\email{G.Fabbian@sussex.ac.uk}
\affiliation{\Sussex}

\author{Julien Carron}
\affiliation{Universit\'e de Gen\`eve, D\'epartement de Physique Th\'eorique et CAP, 24 Quai Ansermet, CH-1211 Gen\`eve 4, Switzerland}
\affiliation{\Sussex}

\author{Antony Lewis}
\affiliation{\Sussex}
\homepage{http://cosmologist.info}

\author{Margherita Lembo}
\affiliation{\Sussex}
\affiliation{\Ferrara}
\affiliation{\INFN}

\begin{abstract}
The cosmic microwave background (CMB) is gravitationally lensed by large-scale structure, which distorts observations of the primordial anisotropies in any given direction. Averaged over the sky, this important effect is routinely modelled with the lensed CMB power spectra. This accounts for the variance of this distortion, where the leading variance effect is quadratic in the lensing deflections. However, we show that if bright extragalactic sources correlated with the large-scale structure are masked in a CMB map, the power spectrum measured over the unmasked area using a standard pseudo-$C_\ell$ estimator has an additional~\emph{linear} lensing effect arising from correlations between the masked area and the lensing. This induces a scale-dependent average demagnification of the unlensed distance between unmasked pairs of observed points and a negative contribution to the CMB correlation function peaking at $\sim 10\,\arcmin$. We give simple analytic models for point sources and a threshold mask constructed on a correlated Gaussian foreground field. We demonstrate the consistency of their predictions for masks removing radio sources and peaks of Sunyaev-Zeldovich and cosmic infrared background emissions using realistic numerical simulations. We discuss simple diagnostics that can be used to test for the effect in the absence of a good model for the masked sources and show that by constructing specific masks the effect can be observed on Planck data. For masks employed in the analysis of Planck and other current data sets, the effect is likely to be negligible, but may become an important subpercent correction for future surveys if substantial populations of resolved sources are masked.
\end{abstract}


\maketitle

\section{Introduction}
\label{sec:intro}

CMB observations are inevitably contaminated at some level by foregrounds, from galactic dust and synchrotron emission to a range of extragalactic signals including the cosmic infrared background (CIB), thermal  Sunyaev-Zeldovich effect (tSZ), and radio point sources. These extragalactic signals are correlated to the matter density at the foreground source redshifts, and point source brightness may also be affected by line-of-sight gravitational lensing. Much of the foreground signal can either be modelled or removed using the distinct frequency dependence. However, bright sources can be problematic and are often masked out.
It is usually tacitly assumed that the CMB power spectra estimated over the unmasked areas are then unbiased estimates that can be used to study cosmology. As long as sources with strong correlation to lensing are not masked, for current data this is likely to be a safe approximation. For future data, where large populations of extragalactic sources will be resolved, corrections may become important. We quantify the likely size of the bias due to mask correlations, as well as proposing empirical consistency tests than can be used in the absence of detailed models or predictions for the source populations.

The CMB is lensed by the large-scale structure along the line of sight, and hence some correlation between extragalactic sources and the CMB lensing convergence is inevitable. The effect of CMB lensing on the full-sky CMB power spectra is well understood and routinely modelled~\cite{Seljak:1995ve,Challinor:2005jy}: the varying magnification and shear of the unlensed acoustic peaks as a function of position on the sky leads to a small smoothing of the peaks in the power spectrum, and the small-scale lenses also increase the power in the CMB damping tail. These are both effects quadratic in the lensing, since over the full sky the convergence and shear average to zero. However, if bright extragalactic sources are masked, due to the correlation of the source density with the lensing this will preferentially be removing peaks of the CMB lensing convergence. If the power spectrum is now estimated only using the unmasked area, there can be additional net effect that is linear in the lensing. The correlation between the deflection angle around convergence peaks is relatively long range, peaking at around $20\,\arcmin$, so every masked peak is associated with a surrounding area of correlated deflection angle that distorts (magnifies) the unlensed CMB.
When these peaks are masked, the corresponding regions of demagnifying deflection angle are no longer fully balanced, and the net effect is a scale-dependent net average demagnification.

The effect of a constant demagnification on the CMB is easily understood: it simply shifts angular scales so that everything looks smaller and the CMB power spectrum is therefore shifted toward higher harmonic multipole $\ell$. At any given observed $\ell$, the CMB power is then the same as at a lower pre-demagnification $\ell$, which on small scales is larger because the CMB power decreases rapidly with $\ell$, leading to an increase in power on small scales (and a corresponding decrease on large scales).
Since the angular acoustic scale is shifted to smaller values, corresponding to the acoustic peaks being shifted to smaller scales, and there is also a strongly oscillatory difference between the power spectra. Due to the steep fall of the CMB spectrum with $\ell$ in the damping tail, a small constant demagnification can lead to non-negligible signatures on the power spectrum. Plausible numbers may be given as follows: removing $2\%$ of the sky on the convergence peaks would give a mean convergence  $\langle\kappa\rangle\approx-0.003$ over the remaining unmasked area. This leads to a significant $1\%$ change in the temperature spectrum at $\ell\sim 2000$, and larger on smaller scales\footnote{The rms of the (assumed Gaussian) convergence field down to $l\sim 2000$ is $\approx 0.06$. The impact of the unmasked large-scale lenses on $l^2 C_l$ may be written to linear order as $\left\langle{\kappa} \right\rangle \frac{d (l^2C_l)}{d \ln l}$ ~\cite[e.g.][]{Lewis:2016tuj}}. This crude estimate is one motivation to the more careful analysis that we give in this paper. For future data, with the CMB power spectrum measured to nearly cosmic variance out to small scales, any small percent-level corrections would have to be carefully accounted for.

In this constant demagnification picture, the effect would be almost degenerate with a change in the angular diameter distance to the CMB (the effect from large-scale lenses would be like a mask-correlated lensing super-sample variance~\cite{Manzotti:2014wca}).
However, this model is not accurate, since the effective net demagnification is both mode-orientation and scale dependent. The degree-scale acoustic features are only slightly affected because the deflection-convergence correlation peaks on smaller scales, about $20\,\arcmin$. The corresponding effect on the power spectrum is therefore distinctive, and important corrections actually arise mostly from relatively smaller-scale lenses.

We start in Sec.~\ref{sec:analytics} by giving a simple leading-order analytic model for the effect in terms of a general mask-deflection correlation function. We give specific analytic forms for the case of masking the most relevant CMB extragalactic foreground emission correlated with CMB lensing: Poisson point sources (an approximate model for radio sources), and peaks above some threshold in a Gaussian isotropic convergence or foreground field (a model for tSZ sources and a component of the infrared sources). We show that this model is sufficient to accurately calculate the effect when these assumptions hold, leaving details of a fully nonperturbative calculation to Appendix~\ref{app:nonpert}.

In Sec.~\ref{sec:results} we test the models and compare analytic predictions with results based on realistic numerical simulations which include non-Gaussian correlated maps of the CMB lensing convergence, tSZ and CIB emission at various frequencies as well as radio sources. In real-world analyses, masks are usually apodized to remove ringing effects when estimating power spectra in harmonic space. Although this case is harder to model fully analytically, we show that semianalytic estimates of the bias based on the mask-lensing correlation measured in the simulated maps describes the bias measured in simulations quite accurately.

In this paper, we focus on the effect of masking on the CMB power spectra. In a companion paper~\cite{Lembo:MaskingPaper} we consider the impact on lensing reconstruction, for which the preliminary investigation of Refs.~\cite{harnois-deraps2016,Liu:2015xfa} suggested a similar effect might be important in particular for cross-correlation between CMB lensing and external matter tracers. Since extragalactic foregrounds are most dominant for the small-scale CMB temperature we focus on that, however some bright extragalactic polarized sources may also have to be masked, so the impact on polarization is also potentially important \cite{Lagache:2019xto}. We include a few numerical and analytic results for polarization for completeness, but leave a more detailed quantitative analysis of the likely impact of masking polarized sources to future work (the effect would be both experiment and spectrum estimator dependent).

\section{Modelling}
\label{sec:analytics}
\newcommand{\w}{ {W}}
\newcommand{\n}{ {\xi_{\rm mask}}}
\newcommand{\hatn}{ {\hat{\xi}_{\rm mask}}}

The effects of masking are largely on small scales, so for simplicity we use the flat-sky approximation in the main text, where the lensed temperature $\tilde{T}(\vx)=T(\vx+\valpha(\vx))$ is related to the unlensed temperature $T$ via the lensing deflection angle $\valpha(\vx)$. In Appendix~\ref{app:curved} we also provide leading-order curved-sky results .

It is convenient to work mostly in position space using a correlation function approach, just as for the usual lensed CMB spectra~\cite{Challinor:2005jy}.
The lensed correlation function is defined by
\begin{equation}
\tilde\xi(r) \equiv \langle \tilde T(\vx) \tilde T(\vx+\vr)\rangle,
\end{equation}
and is independent of $\vx$ and the direction of $\vr$ for a homogeneous statistically isotropic field.
From a statistically isotropic map with a fixed mask an estimator for the lensed CMB correlation function can be built by spatial averaging. In the absence of noise and assuming all distances $r$ can be probed at least once, an estimator is \cite{Szapudi:2000xj,Chon:2003gx}
\begin{equation} \label{eq:hxi}
\hat{\tilde\xi}(r) \equiv \frac{ \left\langle (\w \tilde T)(\vx)  (\w \tilde T)(\vx + \vr)\right\rangle_{\vx, \phi_r}}{{\left\langle \w(\vx) \w(\vx + \vr) \right\rangle}_{\vx, \phi_r}}.
\end{equation}
The normalization in the denominator is required for the estimator to be unbiased in the case where the lensed temperature distribution is independent of the mask. After transforming to the power spectrum, the correlation function estimator is equivalent to a standard ``pseudo-$C_\ell$'' estimator with mask $\w(\vx)$~\cite{Wandelt:2000av}. In the presence of mask-lensing correlations, this estimator is no longer unbiased, since the conditional distribution for the lensed temperature given the fixed mask is no longer statistically isotropic. This is the bias we aim to quantify.

The mask $\w(\vx)$ is a function of position on the sky, which is zero over sources that are masked out.
For an extragalactic source mask, where $\w(\vx)$ is constructed based on the realization of statistically isotropic sources, $\w(\vx)$ can also be viewed as a statistically isotropic random field. The denominator in Eq.~\eqref{eq:hxi} is its empirical two-point correlation function, which we denote $\hatn(r)$. With $f_{\rm sky}$ the average of the mask across the sky, $\hatn(r)$ is a smooth function varying from $f_{\rm sky}^2$ at separations larger than all relevant correlation lengths, to $f_{\rm sky}$ for separations much smaller than the typical mask hole size where both points are almost surely both inside or both outside the mask.

We now turn to the calculation of the expectation values and biases entering the estimator given by Eq.~\eqref{eq:hxi}. We proceed by replacing spatial averages with expectations values over ensembles of $T, W$ at fixed $\vx$ and $\vr$. Since we model the mask as a random field, there is a slight possible ambiguity in this approach. In practice, for simulating CMB data, both CMB and extragalactic foreground skies should be varied at the same time. With the extragalactic part of mask varying with the foregrounds, the CMB correlations must be deconvolved from the mask realization per realization: the estimator mean is the expectation value of the ratio in Eq.~\eqref{eq:hxi}, rather than the ratio of expectation values. However, we show in Appendix~\ref{app:averages} that these are equivalent for binary masks.

The numerator of Eq.~\eqref{eq:hxi} becomes simply the un-normalized pseudocorrelation function of the masked temperature
\begin{align}
\ximasked(r) &\equiv \left\langle (\w \tilde{T})(\vx)(\w\tilde{T})(\vx  +\vr)\right\rangle\nonumber\\
&=\left\langle  T(\vx+\valpha(\vx)) T(\vx'+\valpha(\vx'))\w(\vx)\w(\vx')\right\rangle,
\end{align}
where $\vx' = \vx + \vr$.
Expanding into flat-sky harmonics and
taking the unlensed CMB $T$ to be uncorrelated to anything else, we then have
\begin{equation}
\ximasked(r)= \int \frac{\ud^2 \vl}{(2\pi)^2} C_l e^{i\vl\cdot \vr} \left\langle e^{i\vl\cdot (\valpha(\vx)-\valpha(\vx'))} \w(\vx)\w(\vx')\right\rangle.
\label{lensed_zeta}
\end{equation}
In this equation $C_l$ is the unlensed temperature power spectrum.
The leading correction in $\alpha$ to the masked correlation function from mask correlations is then
\begin{align}
&\Delta \tilde \xi_{\rm masked}(r) \equiv \ximasked(r) -  \tilde\xi(r) \n(r) \\ &\approx \int \frac{\ud^2 \vl}{(2\pi)^2} C_l e^{i\vl\cdot \vr} \left\langle i\vl\cdot (\valpha(\vx)-\valpha(\vx')) \w(\vx)\w(\vx')\right\rangle \nonumber\\
&= \partial_r \xi(r) \left\langle (\alpha_r(\vx)-\alpha_r(\vx')) \w(\vx)\w(\vx')\right\rangle.
\label{eq:delta_exp}
\end{align}
In the last line we introduced $\alpha_r$, the components of the deflection parallel to $\vr$ at $\vx$ and $\vx'$, and the unlensed CMB correlation function $\xi(r)$.
The result for the polarization correlation functions has exactly the same form, with $\xi$ replaced by $\xi_+$ or $\xi_-$ for polarization or $\xi_\times$ for the temperature cross-correlation.
At lowest order, the unlensed correlation function $\xi(r)$ can equally well be replaced by the standard lensed correlation function $\tilde\xi(r)$, which leads to a better approximation as it captures the main nonperturbative standard lensing effects (see Appendix~\ref{app:nonpert} for a more accurate result).
The expectation in Eq.~\eqref{eq:delta_exp} is just the average
difference between the lensed and unlensed distance between any two points (allowing for masking this is positive), and the derivative term then gives how much the correlation function changes due to the mean shift in separation (negative since the correlation falls with distance on relevant scales).

Dividing by $\n(r)$,
the normalized (mask-deconvolved) correction to the correlation function is therefore always of the product form
\begin{equation}
\Delta \tilde{\xi} \approx \partial_r \tilde\xi(r) \meandelta(r),
\label{generaldeconvolved}
\end{equation}
where from Eq.~\eqref{eq:delta_exp} we defined $\meandelta(r)$ as the average over the unmasked area of change in the separation of points due to lensing
\begin{align}
\meandelta(r)\equiv &\frac{\langle [\alpha_r(\vx)-\alpha_r(\vx')]\w(\vx)\w(\vx')\rangle}{\langle \w(\vx)\w(\vx')\rangle}
\\= &2\frac{\langle \alpha_r(\vx) \w(\vx)\w(\vx')\rangle}{\langle \w(\vx)\w(\vx')\rangle},
\label{delta_def}
\end{align}
where in the last equation we used the symmetry properties of $\langle \alpha_r(\vx) \w(\vx)\w(\vx')\rangle$  under the coordinates transformation $\vx\rightarrow\vx'$ (see Sec.~\ref{sec:models}).  The unmasked area can therefore be thought of as having scale-dependent demagnification of the distance between points, with\footnote{As discussed in more detail in Appendix~\ref{app:nonpert} this relation is not exact beyond leading-order, since the lensing of the correlation function is not independent of the local $\meandelta(r)$} $\tilde\xi(r)|_{\textrm{unmasked area}} \sim \tilde\xi(r + \meandelta(r))$.
The product form of Eq.~\eqref{generaldeconvolved} in real space corresponds in harmonic space to a convolution of the CMB temperature-gradient power with the power spectrum corresponding to $\meandelta$.

In Sec.~\ref{sec:empiest} we first give a recipe to estimate the bias in Eq.~\eqref{generaldeconvolved} from simulations. We then proceed with analytic methods in Sec.~\ref{sec:models}. There we start by discussing results for masks built locally from some Gaussian foreground field $f$. We then look in more detail at two mask models: in Sec.~\ref{sec:numask} we discuss thresholding the peaks of $f$, where the effect can be significant, then in Sec.~\ref{sec:poissonmask} we consider masking sources that are modelled as a Poisson sampling of $f$, as a model of masking radio point source (where the effect is typically much smaller). A set of appendices collect details of the calculations related to these two models.

\subsection{Empirical estimation of the bias}\label{sec:empiest}
Equation \eqref{delta_def} can in principle be calculated empirically for any mask construction if the required average can be calculated from simulations that capture the relevant correlations and (potentially non-Gaussian) statistics. The quantity $\langle \alpha_r(\vx)\w(\vx)\w(\vx')\rangle$ appearing in Eq.~\eqref{delta_def} is just the correlation function of the gradient mode of the masked deflection angle with the mask. For any given simulation, where we know $\w$ and $\kappa$ (and hence the deflection $\valpha$), we can estimate $\meandelta(r)$ directly from the cross-spectrum between the masked deflection and the mask measured in that simulation.

More explicitly, if $E(\vl)$  and $B(\vl)$ are the gradient and curl modes of the spin-1 field $\alpha \w$, and $\w(\vl)$ the Fourier coefficients of the spin-0 mask, we may write on the flat-sky
\begin{equation}
	\hat{\vr}\cdot \valpha W =\int \frac{\ud^2 \vl}{2\pi} \left(E(\vl)\hat \vr \cdot \hat \vl  + B(\vl) \hat \vr \star \hat \vl \right) i e^{i\vl\cdot \vx}
	\end{equation}
with $\hat \vr \star \hat \vl = \hat \vr \cdot (-\sin \phi, \cos \phi)$.
Correlating with $W(\vx')$ gives
\begin{equation}
\langle \alpha_r(\vx) \w(\vx)\w(\vx')\rangle =- \int \frac{\ud l}{2\pi} l C_l^{EW} J_1(lr).
\label{eq:EWpower}
\end{equation}
The denominator in Eq.~\eqref{delta_def} can also be calculated directly from the mask power spectrum with a spin-0 (here, $J_0$) transform.

The leading correction to the correlation function can therefore easily be evaluated from corresponding power spectra. For any masking recipe, this therefore provides a straightforward way to calculate the expected bias in the power spectrum measured over the unmasked area. On data, the deflection field is not known, but it may be possible to estimate it, at least crudely, from a correlated field (e.g. the CIB) or lensing reconstruction, providing an internal estimate of the expected bias without having a detailed model for the statistics of the mask.

\subsection{Analytic models}\label{sec:models}
For a first analytic model, we assume that some underlying Gaussian statistically-isotropic foreground field $f(\vx)$ determines the mask probability locally, so that $\w(\vx)$ only depends on some (in general nonlinear) function of $f(\vx)$.
We will consider two specific analytic models for the mask construction, a peak threshold mask (where the effect can be substantial) and Poisson sources (where the effect is generally small). When considering a threshold mask we will consider specifically the case where $f$ is tSZ or CIB fields, or as an extreme limiting case, the CMB lensing convergence $\kappa$ itself. For Poisson sources, $f$ will be the perturbation to the expected number of sources over the area masked out per source, determined by the perturbations to the underlying galaxy populations, which we approximate as Gaussian.
\begin{figure}
\centering
\includegraphics[width=.49\textwidth]{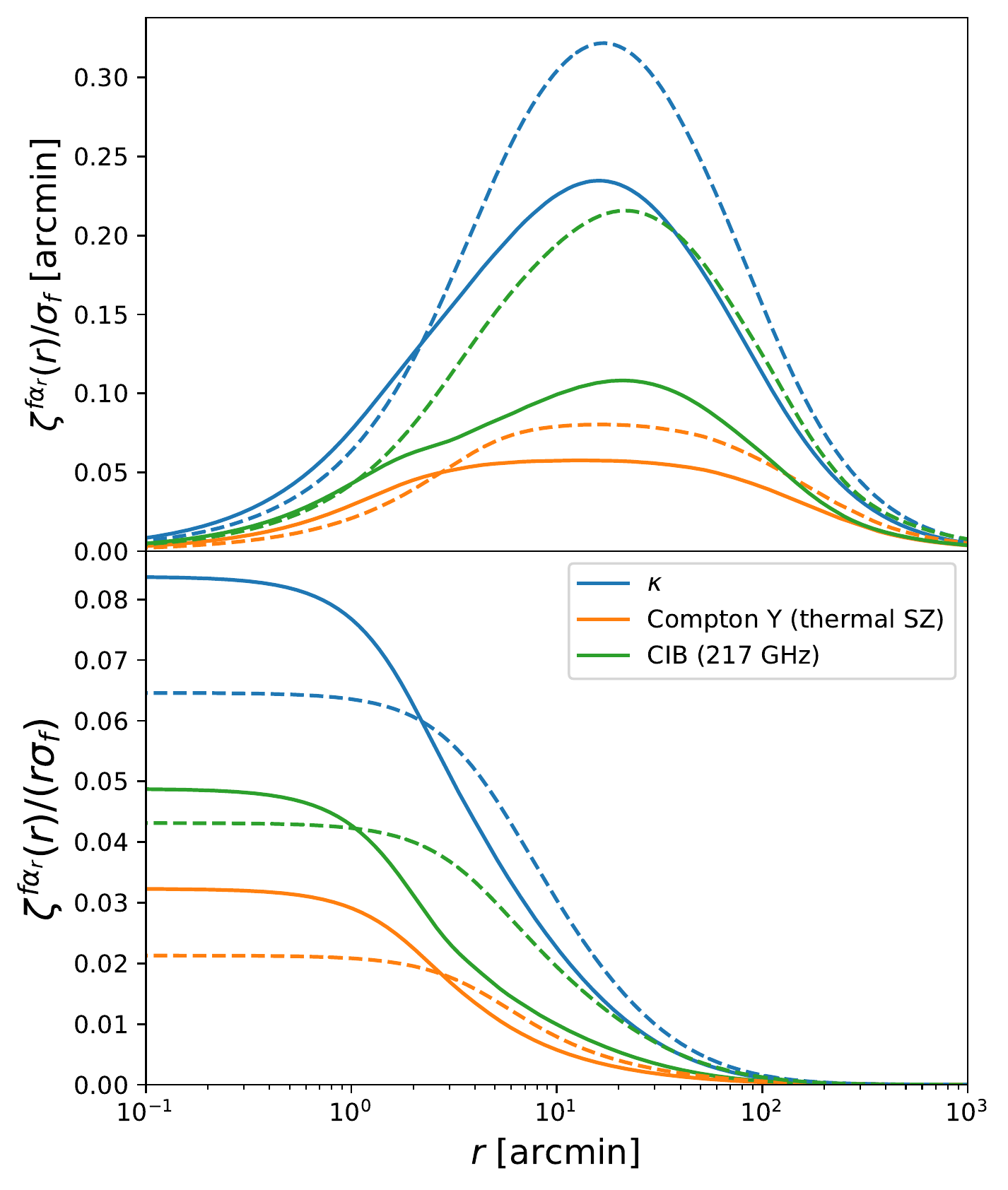}
\caption{The correlation between the size of the (inward pointing) radial lensing deflection about a point and the deviation of the field $f$ at that point from its mean in units of the standard deviation: for a point with a foreground field that is $1\sigma_f$ above the mean, the plots show the mean inward-pointing radial lensing deflection at radius $r$. The top plot shows the result in arcminutes, the bottom shows the fractional change in distance between the points due to the deflection. Different colours correspond to the limiting case of a field that is fully correlated, $f\propto \kappa$ (blue), and the result expected for Compton Y (thermal SZ foreground, orange) or cosmic infrared background foreground at 217 Ghz (green). The latter results are based on a smooth fit to the \websky simulation power spectra~\cite{Stein:2020its}. Solid lines are for the field values after smoothing with a 1.7 FWHM beam, dashed the corresponding result for a 5.1 FWHM beam. The correlation extends to cosmologically important distances, and the fractional change in radius becomes percent level on scales below 10s of arcminutes.}
\label{fig:zetaalpha}
\end{figure}

By symmetry, at a point there is no correlation between the scalar foreground $f$ and the vector deflection angle, $\langle f(\vx)\valpha(\vx)\rangle=0$. However, if $f$ is correlated to large-scale structure it will be correlated to the lensing convergence, and hence have a nonzero correlation $\xi^{f\alpha_i}(\vr) \equiv \langle f(\vx) \alpha_i(\vx')\rangle = -\langle \alpha_i(\vx) f(\vx')\rangle \equiv \xi^{f\alpha_r} \hat{r}_i$, corresponding e.g. to deflection angles around overdensities having an inward-pointing radial direction (positive $\xi^{\kappa\alpha_r}$ for our definition of $\vr\equiv \vx' -\vx $ and $\kappa = -\nabla\cdot\valpha/2$). If $\phi$ is the lensing potential with $\valpha = \nabla \phi$, then its explicit form is
\begin{equation}
\xi^{f\alpha_r}(r) = -\partial_r \xi^{f\phi}(r) =  \int \frac {\ud l}{2\pi} \:l^2 C^{f \phi}_l J_1(lr).
\end{equation}
As shown in Fig.~\ref{fig:zetaalpha}, $\xi^{f\alpha_r}(r)$ peaks somewhere around $r\sim 20\,\arcmin$ depending on the field being considered.

\begin{figure}
\includegraphics[width=\columnwidth]{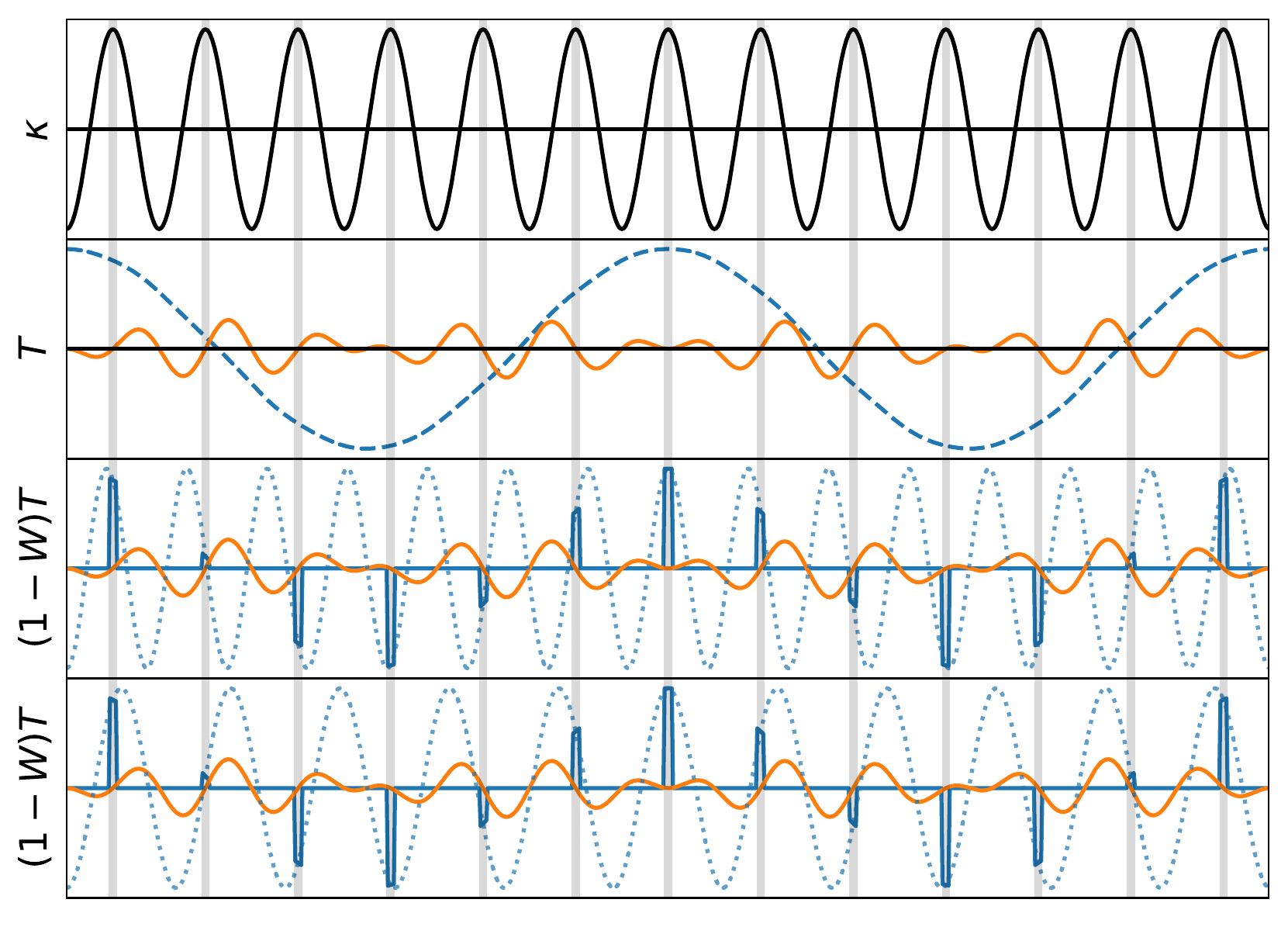}
\caption{An illustration of the lensing effect of a small-wavelength convergence plane wave $\kappa$ (top panel) on an aligned longer-wavelength CMB temperature modes (second panel, dashed line). Remapping points with the corresponding deflection angles gives the lensing correction (solid orange, greatly exaggerated in relative size for illustration).
The vertical bands show the peaks of the convergence, which are preferentially masked if a masked foreground is correlated to the lensing.
The third and bottom panels show the temperature values that are removed by masking the peaks of the convergence to calculate a pseudo power spectrum (solid blue lines). Dotted lines show a Fourier component of these values (at the sum and difference of the lens and CMB frequencies respectively). The higher-frequency component in the third panel is negatively correlated with the lensing signal oscillations shown in orange; the lower-frequency component in the bottom panel is positively correlated with the lensing. Since the temperature is much larger than the lensing signal, the cross-correlation can be similar or larger than the lensing auto spectrum even if only a small area at the peaks is masked. On small scales the temperature and lensing spectrum fall with $l$, so more negative cross-correlation is removed by masking than positive is added, leading to a net positive signal that is linear in the lensing.
This also leads to a positive bias on the power spectrum estimator after deconvolving the pseudospectrum assuming statistical isotropy over the unmasked area.}
\label{fig:lensed_waves}
\end{figure}

Since we are only considering the two-point CMB correlation function, for any choice of coordinates the correlation function is an average over the correlated Gaussian variables $f(\vx), f(\vx'), \valpha(\vx)-\valpha(\vx')$.
The expectations in Eq.~\eqref{eq:delta_exp} can then be evaluated for Gaussian fields to give
\begin{figure}
\includegraphics[width=\columnwidth]{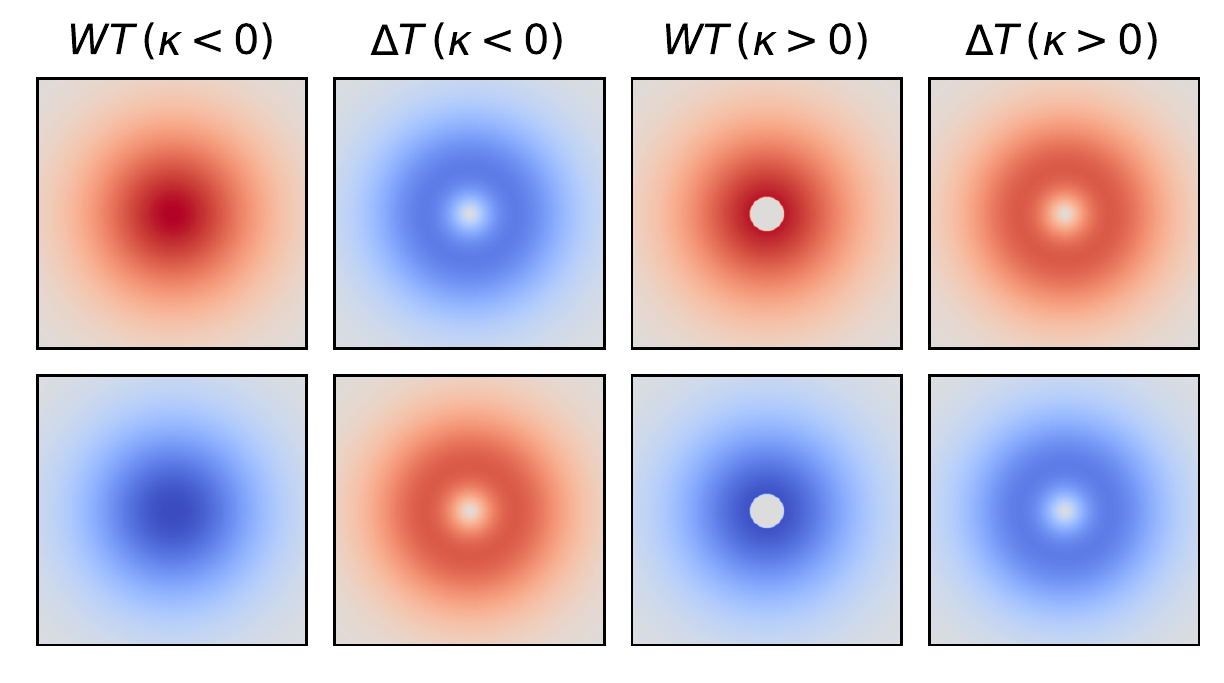}
\caption{An illustration of the lensing effect on a temperature hot spot (top row) or cold spot (bottom row).
The columns show the unlensed temperature and change in temperature due to lensing by the expected radial deflection if there is a convergence minimum (left two columns) or peak (right two columns) in the centre. The unlensed temperature and lensing colour scales are not the same to make the much smaller lensing signal easily visible.
 The temperature $T$ in the centre is unchanged by an aligned lens, and is positively correlated with the ring of lensing signal $\Delta T$ if the lensing is converging ($\kappa>0$), and negatively correlated if the lens is diverging ($\kappa<0)$. For Gaussian fields both signs are equally common, and the correlation averages to zero when there is no masking. If the centre of the lens is preferentially removed by a mask $W$ when $\kappa>0$ due to correlation between foregrounds and the convergence, there will be more points of negative correlation between the centre and the ring, giving a net negative change to the lensed CMB correlation function at radius $r>0$.
In harmonic space, masking of the temperature at a convergence peak leads to net negative correlation between the large-scale temperature and lensing correction, leading to a negative contribution to the large-scale pseudo power spectrum. On small scales (but larger than the hole size), there is a ring-like pattern in both the masked temperature and lensing signal at a convergence peak, hence
there is a positive correlation between them leading to an enhancement of small-scale pseudo power spectrum (which is not removed by a mask deconvolution that does not account for the mask correlation). Since the lensing signal is much smaller than the unlensed temperature, the cross-correlation terms dominate the effect on the power spectrum compared to small changes due to also masking the lensing signal.
}
\label{fig:lensing_pic_2d}
\end{figure}
\begin{equation}
\Delta\tilde \xi(r)
\approx  g(r)\partial_r \tilde{\xi}(r) \frac{ \xi^{f\alpha_r}(r)}{\sigma_f},
\label{eq:generalmeanf}
\end{equation}
where
\begin{equation}
g(r) \equiv - \frac{2\sigma_f\bar{f}(r)}{\sigma_f^2+\xi_f(r)}, \quad \bar{f}(r)\equiv \frac{\left\langle f(\vx) \w(\vx) \w(\vx') \right\rangle}{\left\langle  \w(\vx) \w(\vx')\right\rangle}.
\label{eq:gfbar}
\end{equation}
Here $\bar{f}(r)$ is the mean of the foreground field over the unmasked area weighted by the number of pairs of points each point forms with separation $r$, which is usually negative.
In Eq.~\eqref{eq:generalmeanf}, $g(r)$ is a very smooth prefactor which, in all models we considered, varies by a factor of two across all distances.
For separations large compared to the correlation length ($\xi_f(r)\ll \sigma_f^2$) and hole size, the foreground mean becomes the simple mean over the unmasked area, $\bar f(r)\rightarrow \bar{f}$, hence for large separations we have
\begin{align}
\Delta\tilde \xi(r)
\approx  -2\partial_r \tilde{\xi}(r) \frac{\bar{f}}{\sigma_f} \frac{\xi^{f\alpha_r}(r)}{\sigma_f}.
\label{eq:largescales}
\end{align}
For very small separations, assuming the mask holes have finite size so the two points are almost surely either inside the same hole or both unmasked,
$\w(\vx')\approx \w(\vx)=\w(\vx)^2$ and
 $\xi_f(r)\approx \sigma_f^2$ so that\footnote{
 This is for a binary mask. More generally $\w(\vx')\w(\vx)\approx \w(\vx)^2$, and $\bar{f}$ can then be defined as the $W^2$-weighted mean of $f$.}
 \begin{align}
\Delta\tilde \xi(r)
\approx  -\partial_r \tilde{\xi}(r)  \frac{\bar{f}}{\sigma_f}\frac{\xi^{f\alpha_r}(r)}{\sigma_f}.
\label{eq:smallscales}
\end{align}
The $\frac{\bar{f}}{\sigma_f} \frac{\xi^{f\alpha_r}(r)}{\sigma_f}$ term is simply the mean radial deflection at one of a pair of points separated by $r$ over the unmasked area. For large separations, where the foreground values at the points are uncorrelated, the total relative change in separation of the two points is twice this.
 Equations \eqref{eq:largescales} and~\eqref{eq:smallscales} are of the form of the product of two real space functions. In harmonic space, the result is therefore a convolution, so on large scales compared to the foreground correlation length and hole size the correction to the power spectrum is
\begin{equation}
\Delta \tilde{C}_l \sim -2\frac{\bar{f}}{\sigma_f^2} \int \frac{\ud^2\vl'}{(2\pi)^2} \tilde{C}_{l'} C^{\phi f}_{|\vl-\vl'|} \vl'\cdot (\vl-\vl'),
\label{eq:convolution}
\end{equation}
where $C^{\phi f}_{l'}$ is the cross-spectrum between the lensing potential and the foreground. For a foreground that scales roughly like the convergence, the convolution is with a kernel that goes like the $\alpha\kappa$ spectrum, which has more small-scale power compared to the $\alpha\alpha$ spectrum that enters the convolution for the leading-order standard lensing effect.
 This leads to much broader mixing of scales, giving
a relatively non-peaky result mixing contributions from different acoustic peaks, and efficiently transfers power to small scales where the CMB spectrum is small. If we consider an $l$ in the damping tail (i.e. much higher than the peak of the temperature gradient spectrum at $l\sim 1000$), where the power spectrum is small, most of the integrand comes from $l'\ll l$; in this limit, the leading term is
\begin{equation}
\Delta \left(l^2\tilde{C}_l\right) \sim 2\frac{\bar{f}}{\sigma_f^2}  \frac{\ud C_l^{\kappa f}}{d \ln l} \int \frac{\ud l'}{l'} \frac{{l'}^4 \tilde C_{l'}}{2\pi}.
\end{equation}
Since $\bar{f}$ is negative when masking peaks, the result is positive when $C_l^{\kappa f}$ is decreasing at high $l$ where the limit applies.
It vanishes only when there is no foreground-lensing correlation or a cross-correlation spectrum $C_l^{\kappa f}$ that is constant (white, which corresponds to no spatial correlation between the foreground value and surrounding lensing field).
The integral over the CMB spectrum quantifies the total power from larger scales in the correlation between the CMB temperature and its curvature. The correction spectrum falls much less quickly than the unlensed CMB, and when there are substantial correlations can become a large fractional correction deep in the damping tail; in the high $l$ limit it can become comparable to the standard lensing signal, which is given in this limit by
\begin{equation}
l^2\tilde{C}_l \sim 2  C_l^{\kappa} \int \frac{\ud l'}{l'} \frac{{l'}^4 \tilde C_{l'}}{2\pi}.
\end{equation}
See Fig.~\ref{fig:lensing_pic_2d} for an illustration of the effect on the lensed CMB signal in real space when masking convergence peaks and Fig.~\ref{fig:lensed_waves} for its harmonic domain version.

\subsubsection{Peaks of foreground fields}\label{sec:numask}

For a mask that is constructed by thresholding the foreground to mask out the peaks, i.e. a step function $\w(\vx) = \Theta(\nu\sigma_f-f(\vx))$ where $\nu$ determines the ``sigma" value of the cut, the derivative of $\w(\vx)$ is just a delta function. The remaining Gaussian integral over $f(\vx)$ to calculate the expectation in Eq.~\eqref{eq:gfbar} can then be done to give
\begin{equation} \label{approxresult}
g(r) \n(r) = \frac{e^{-\nu^2/2}}{\sqrt{2\pi}} \left[1 + \textrm{erf}\left(\frac {\nu}{\sqrt{2}} \sqrt{\frac{\sigma_f^2 - \xi_f(r)}{\sigma_f^2 + \xi_f(r)}}\right) \right].
\end{equation}
\begin{figure}
\includegraphics[width=.49\textwidth]{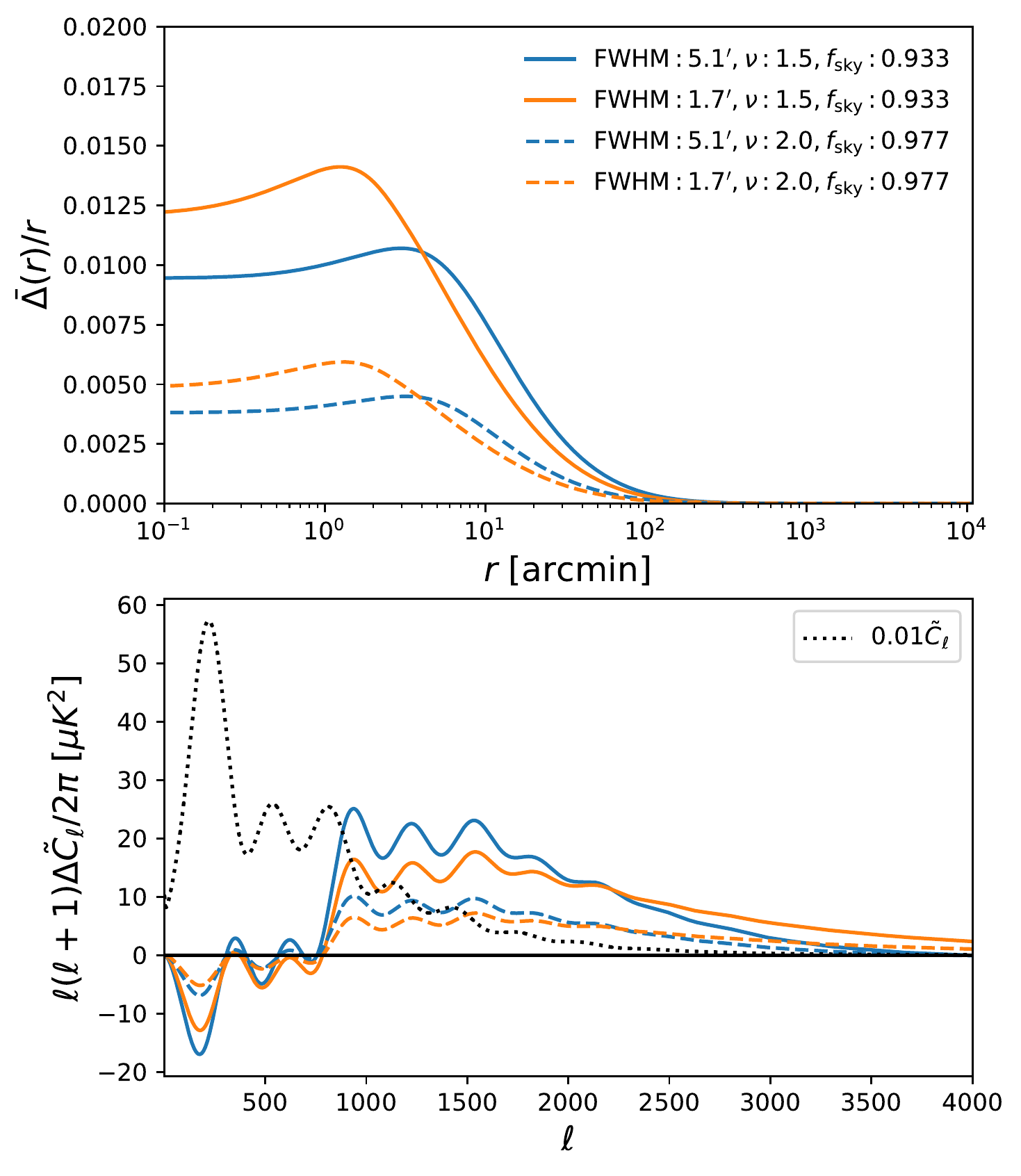}
\caption{Top: the average fractional change in the separation $r$ of pairs of points in the unmasked area as a function of separation for a thresholded foreground field [see Eq.~\eqref{delta_def}]. This is assuming an unapodized mask is constructed by thresholding $\nu\sigma$ of a Gaussian foreground field proportional to the lensing convergence after Gaussian smoothing with the given beam full-width-half-maximum (FWHM). Below the acoustic scale, the effect becomes percent-level, and is relatively more important in the power spectrum due to the rapid fall in power on Silk-damping scales.
Bottom: the corresponding correction to the lensed CMB temperature power spectrum $\tilde C_l$ estimated from the unmasked area (the dotted line shows the one hundredth of the full power spectrum for comparison).
 Results for SZ and CIB-thresholded maps have similar shapes but with lower amplitude proportional to their lower lensing correlation.
}
\label{fig:mean_mag}
\end{figure}

The normalization $\n(r)$ generally needs to be calculated numerically, but
varies between $\sim \fsky^2$ at large $r$ where the foreground fields are nearly uncorrelated, to $\sim\fsky$ for small $r$ where both points are almost surely either both inside or outside the mask. The transition between these values is very smooth and determined by the correlation length of the foreground field. The expected observed sky fraction is
\begin{equation}
\langle \fsky \rangle =\langle \w(\vx)\rangle = \frac 12\left[ 1+ \text{erf}(\nu/\sqrt{2})\right].
\end{equation}
The factor in the square brackets in Eq.~\eqref{approxresult} varies smoothly between $\sim 2\fsky$ when $\xi_f(r)\ll \sigma_f^2$ (for $r$ much larger than the correlation length) to unity on very small scales. The factor $-e^{-\nu^2/2}/\sqrt{2\pi}= \langle f \w\rangle/\sigma_f = \fsky \bar{f}/\sigma_f$ is the mean masked value of $f$ in units of its standard deviation, which is negative, where (as before) $\bar{f}$ as the mean value of $f$ over the unmasked area. It therefore has the general limiting forms given for large $r$ by Eq.~\eqref{eq:largescales} and small $r$ by
Eq.~\eqref{eq:smallscales}.
Note that the result is independent of the scale or normalization of $f$, so the effect is leading order in the perturbations (linear in $\alpha_r$), and will only be negligible when the correlation is very low or $f$ is dominated by very small scales so that $\bar{f}/\sigma_f$ is small.

\begin{figure}
\includegraphics[width=\columnwidth]{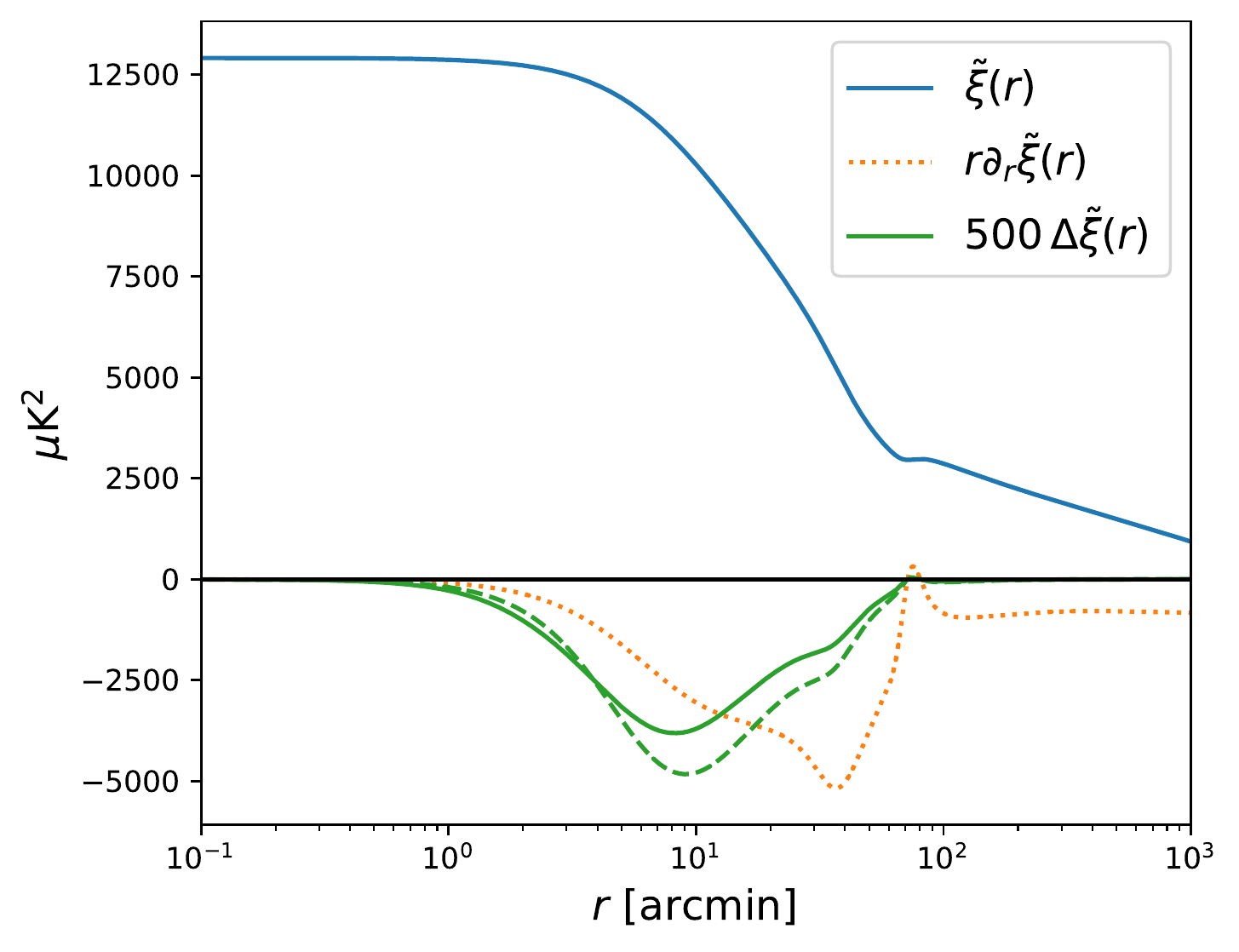}
\caption{The lensed CMB correlation function $\tilde{\xi}(r)$ (blue solid), and its log derivative (dotted orange). The green lines show the correction $\Delta\tilde{\xi}(r)$ due to threshold masking of a fully-correlated foreground with $f\propto \kappa$ smoothed with a $1.7$
(solid) or $5.1$ (dashed) $\arcmin$ FWHM Gaussian beam with a $\nu=2$ threshold mask ($\fsky \approx 0.977$). In the power spectrum the effect is a much larger fractional correction on small scales, since it adds power on scales where the lensed correlation function has little; on large scales the negative sign of the correction corresponds to a small reduction in power at $l\alt 1000$.}
\label{fig:delta_zeta}
\end{figure}

Although the mean deflection at any distance is very small, less than a quarter of an arcminute, the mean \emph{relative} change in distance $\meandelta(r)/r$ is an important percent-level effect at scales of tens of arcminutes and below when masking foreground peaks; see Fig.~\ref{fig:mean_mag}.
A typical full numerical result for the correlation function correction over the unmasked area, $\Delta \tilde\xi(r)$, is shown in Fig.~\ref{fig:delta_zeta} for a smoothed foreground $f\propto \kappa$ that is threshold-masked with $\nu=2$. The signal peaks at around $10\,\arcmin$; on much smaller scales the CMB is very smooth so $\partial_r \tilde{\xi}(r)\rightarrow 0$, and on much larger scales the deflections have little correlation and only a small fractional effect.

If $f$ is band limited or smoothed to a certain scale, so that $C_l^{\phi f}$ starts to fall off sharply with $l$, the signal will also decline at the same scale,  and the approximation limit will no longer be valid.
The full shape of $\meandelta(r)/r$ shown in Fig.~\ref{fig:mean_mag} has a peak in between the small and large-scale limits, determined by the clustering scale of the foreground that determines the size of the mask holes. As scales transition from the large-scale to small-scale limit, this translates into a change in sign of the second derivative, $\partial_r^2 \Delta\tilde{\xi}(r)$.
For the power spectrum, this corresponds to the correction going negative at high $l$ (for high $l$, the integral of $J_0(lr)$ against a smooth function depends on the second derivatives because the fast oscillations average to zero for constant and constant gradient terms).

It may seem quite unintuitive that an effect being sourced from a small sky-fraction mask could be a large fractional effect on the total: doesn't this imply that at each masked point the effect must be very large?
From Fig.~\ref{fig:zetaalpha}, for a 1-sigma convergence peak centred at $r=0$, the radial (inward-pointing) lensing deflection peaks at $\sim 0.3\,\arcmin$ at a radius of $\sim 20\,\arcmin$.
The r.m.s. size of the CMB gradient is $\sim 14 \muK\, \arcmin^{-1}$, so the typical size of the lensing-induced signal at $r \sim 20\,\arcmin$ is therefore $\Delta T \sim \vrhat\cdot\valpha \partial_r T \sim 4\muK$, with a dipole-like pattern about the centre if there is a significant central temperature gradient. However, the correction of interest comes from the fact that the temperature at the centre and the radial temperature gradient at distance $r$ are correlated, with $\langle T(0)\partial_r T(r)\rangle = \partial_r \tilde\xi(r) \sim -190 \muK^2\arcmin^{-1}$ for $r\sim 20\,\arcmin$ (see Fig.~\ref{fig:delta_zeta}). Hence, there is a correlation between the central temperature and size of the surrounding lensing signal: for a $\sim 0.3\,\arcmin$ inward-pointing radial deflection, $\langle T(0) \Delta T(r)\rangle \sim 57 \muK^2$. For example, for convergence peaks located at temperature peaks there is a positive surrounding ring of lensing-induced signal; for lenses located at temperature troughs, there is a negative ring of lensing-induced signal (see Figs.~\ref{fig:lensing_pic_2d} and \ref{fig:lensed_waves}).
This correlation signal is larger than the variance of the deflection signal, which is $\sim 16 \muK^2$ on these scales. Without masking, the signal around lensing overdensities on average cancels with that from underdensities, but when only the peaks of the convergence are masked, there is a net effect that can be significant even if only a small fraction of the sky is masked. For an $n$-sigma peak, the signal is proportionately larger, which also partly offsets the smaller sky area affected for moderate $n$.

In practice, a threshold mask is often enlarged or apodized, which breaks the strict assumption that the mask is a local function of the foreground. The general form of Eq.~\eqref{generaldeconvolved} still holds and can be applied if it can be estimated from simulations, but the specific analytic results do not generalize straightforwardly.

\subsubsection{Poisson point sources}\label{sec:poissonmask}
In CMB frequency bands with $\nu \alt 217{\rm GHz}$ bright extragalactic sources detected in the sky  (and that are later masked) are dominated by radio sources (RS). At higher frequencies dusty star-forming galaxies (DSFGs), which are observed as infrared (IR) sources via their  thermal emission from dust heated by the ultraviolet emission of young
stars, start to dominate~\cite{DeZotti2019,Everett:2020jcf,gralla2020}. Whether a given galaxy contains a bright radio source involves largely stochastic processes determining the generation of an active-galactic nucleus (AGN), the largely random alignment of any radio jet with our line of sight, or the status of star formation processes. They are therefore often modelled as a Poisson process, with a distribution following the distribution of the host galaxies. Since on large scales the universe is homogeneous, to zeroth order this results in an uncorrelated white-noise spectrum of sources.

We make the simple assumption that the probability of an observed bright radio source in a galaxy is independently the same for each galaxy in a population. In redshift interval $\ud z$ the number of sources in solid angle $\ud \Omega$ in direction $\vnhat$ is taken to be $n_g(\vnhat, z)\ud z\ud\Omega$, so for small probability $p_g$ per galaxy, the mean number of sources per solid angle is
\begin{equation}
\lambda(\vnhat) = \int \ud z\: p_g n_g(z)\left[1+\Delta n_g(\vnhat,z) - (2+5s(z))\kappa(\vnhat, z)\right] .
\end{equation}
This neglects small velocity and potential corrections and strong lensing events but accounts for the fact that at first order in perturbations, the number density of galaxies is correlated to the density and hence to CMB lensing. There is therefore a clustered component to the spectrum that will correlate masked sources with the lensing potential. In addition, there are also potentially correlations with CMB lensing induced by  magnification bias (due to the weak lensing convergence $\kappa(z)$ of sources at redshift $z$).
The size of this lensing effect depends on the slope of the source luminosity function $s(z)$ at the flux cut used for the mask~\cite{Tegmark:1996ze,Matsubara:2000pr}. The lensing term should be included for an accurate analysis, but it is usually a small fractional correction. As we shall see the effect of Poisson point source mask is small anyway, so the lensing terms can safely be neglected for our purposes. This is consistent with the numerical simulations that we use, which also do not include the lensing effect on the point source fluxes.

If for each source we mask out an circular area around it of radius $R$, the probability of a given direction being masked ($\w(\vx)=0$), is one minus the probability of the Poisson probability of no point sources over the hole area,
\begin{equation}
P(\w(\vx)=1|\poissonmean(\vx)) = e^{-\poissonmean(\vx)},
\end{equation}
where $\poissonmean(\vx)$ is the hole area mean number field. Here we use the flat sky approximation where
\begin{equation}
\poissonmean(\vx) \equiv  \int \ud\theta_r \int_0^R r\ud r \lambda(\vx+\vr),
\end{equation}
so that in Fourier space $\poissonmean(\vl) = 2\pi R^2 [J_1(lR)/(lR)] \lambda(\vl)$.

If we approximate $\Delta n_g(\vnhat,z)$ and $\kappa(z)$
as Gaussian random fields, or invoke approximate central limit theorem Gaussianization by line of sight averaging, we can take $\poissonmean(\vx) = \poissonmean + \poissondelta(\vx)$ as having the background value $\poissonmean$ plus a perturbation $f$ that is an Gaussian random field with variance $\sigma_f^2$ at any point. The sky fraction after masking is therefore
\begin{align}
\langle \fsky\rangle &= \langle \w(\vx)\rangle
= \langle  e^{-\poissonmean(\vx)} \rangle \nonumber\\
 &= e^{-\poissonmean}e^{\sigma_f^2/2}.
\end{align}
Note that for small perturbations, the masked area is dominated by Poisson sampling of the background source population, with source density $\poissonmean$ per mask area, which has no correlation to the lensing. The $\sigma_f$ term reflects the fact that more clustered matter will have more overlapping mask holes, hence less masked area (higher $\fsky$). For small numbers of sources, $\langle \fsky\rangle \approx 1-\poissonmean$.

Finite-sized point source mask holes in general violate the assumption that $\w(\vx)$ only depends on $f(\vx)$, since if $\w(\vx')$ is masked, $\vx$ may already be inside the same mask hole. However, it does hold for $r$ large enough that the two points are never inside the same mask hole ($r>2R$), so that
\begin{multline}
P(\w(\vx)=1,\w(\vx')=1|\poissonmean(\vx),\poissonmean(\vx')) \\= e^{-\poissonmean(\vx)-\poissonmean(\vx')}.
\end{multline}
For Gaussian $f$ and $r>2R$, the general form of Eq.~\eqref{eq:generalmeanf} holds, with $g(r)$ identically equal to $2$, so that

\begin{equation}
\Delta\tilde\xi(r)
\approx 2\partial_r \tilde{\xi}(r) \xi^{f\alpha_r}(r) .
\label{poissonapprox}
\end{equation}
Although the correlation function $\xi^{f\alpha_r}(r)$ is linear in the deflection angle, it is also linear in the galaxy density perturbations, so the overall correction is small unless the source galaxies are very strongly clustered.
Equation \eqref{poissonapprox} is equivalent to the general limiting form of Eq.~\eqref{eq:largescales} for large separations, since in this case $\bar{f}=-\sigma_f^2$, but here the result is valid for all $r>2R$.

More generally, the result can be calculated on all scales using Eq.~\eqref{generaldeconvolved} where
\begin{align}
\meandelta(r) =
2\frac{\left\langle \alpha_r(\vx) \exp\left(-\int_{A(\vx,\vx')} \lambda(\vy) \ud^2 \vy \right)\right\rangle }{\left\langle\exp\left(-\int_{A(\vx,\vx')} \lambda(\vy) \ud^2 \vy \right)\right\rangle },
\end{align}
where $A(\vx,\vx')$ denotes restricting the integral to the area where a point source would give $W(\vx)=0$ or $W(\vx')=0$. For $r>2R$ the area $A(\vx,\vx')$ is just the two circular regions around each point, and this reduces to Eq.~\eqref{poissonapprox}.
More generally, for Gaussian fields it can be evaluated numerically using
\begin{align}
\Delta\tilde\xi(r) &\approx 2\partial_r \tilde{\xi}(r) \int_{A(\vx,\vx')} \ud^2 \vy \xi^{\lambda \alpha_r}(r_y) \vrhat_y\cdot \vrhat \\
&=  4\partial_r \tilde{\xi}(r)  \int_{\max(R, r-R)}^{R+r} \ud s\:s\: \xi^{\lambda \alpha_r}(s)  \sin(\phi_r(s)),
\label{eq:poissonnumerical}
\end{align}
where $\vr_y \equiv \vy-\vx$ and $\phi_r(s)$ is defined through $\cos \phi_r(s) =( s ^2 + r^2 - R^2)/ 2 s r$. For $r<2R$ the region $A$ is the area inside the two overlapping circles centred at each point.
For $r\ll R$ the limiting form of Eq.~\eqref{eq:smallscales} applies, with $g(r) = 1$, so that on scales much smaller than the holes
\begin{equation}
\Delta\tilde\xi(r)
\approx \partial_r \tilde{\xi}(r) \xi^{f\alpha_r}(r).
\label{eq:poisson_small}
\end{equation}
Equation \eqref{eq:poissonnumerical} smoothly interpolates between the limiting forms of Eq.~\eqref{eq:poisson_small} and Eq.~\eqref{poissonapprox}.

\begin{figure}
\includegraphics[width=.49\textwidth]{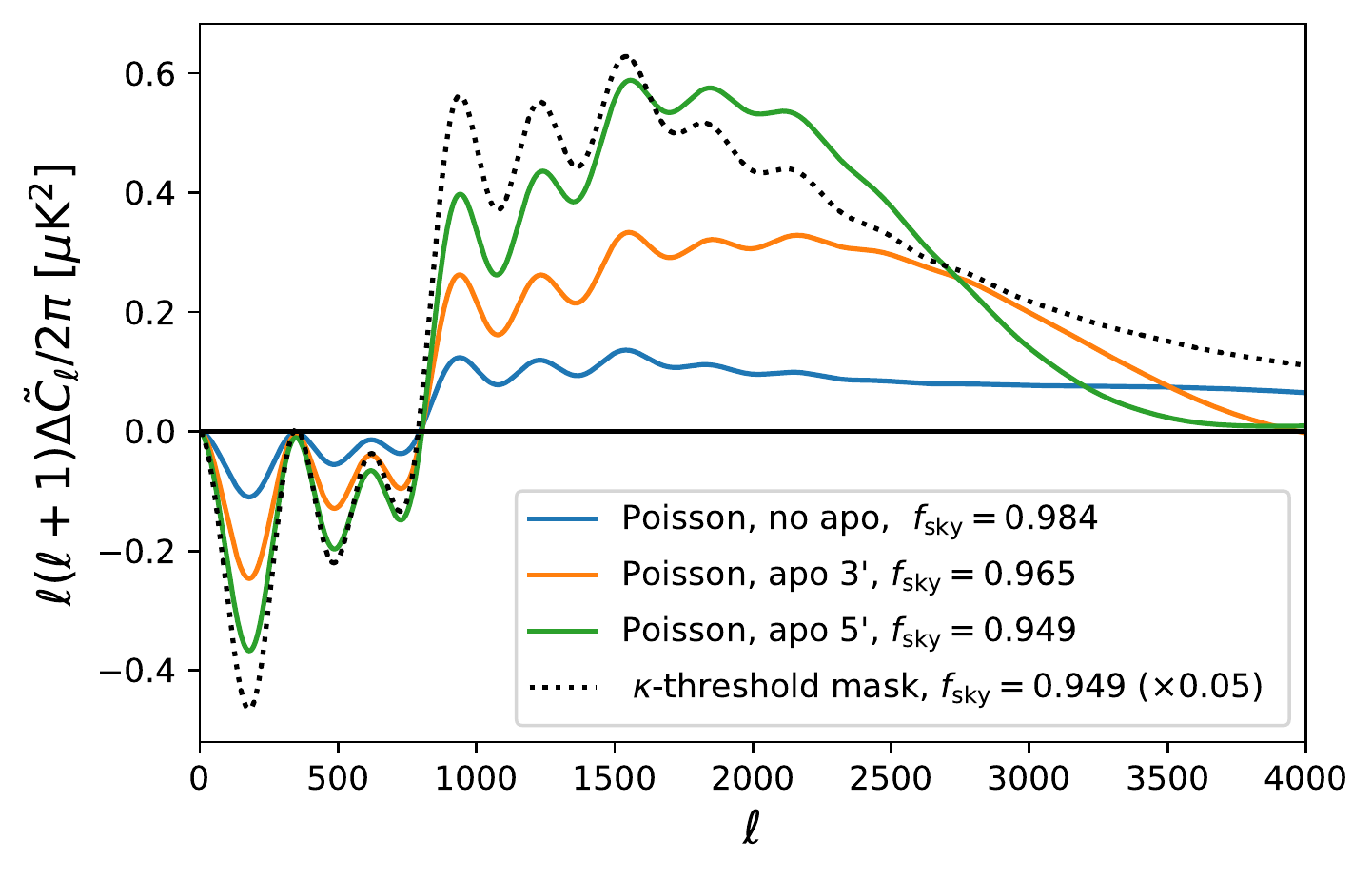}
\caption{
Correction to the lensed CMB temperature power spectrum $\tilde C_l$ for a mask consisting of an ensemble of disks of radius $3\:\arcmin$, centred on sources Poisson-sampling a underlying density field $\delta(\hat n)$, for a total masked (unapodized) sky fraction of $ 1.6\%$. Shown are the cases without apodization (blue) or after apodization as indicated in the legend (orange and green). Apodization increases the masked sky fraction, and introduces a cut-off at the corresponding scale. When sampling a biased matter tracer $b\delta$, the mask traces the peaks better (for $b > 1$) and all the coloured curves scale increase linearly with $b$. For this figure, $\delta$ is taken to be the lensing convergence field $\kappa$ to allow comparison with the $\kappa$-threshold mask results (black dotted, scaled by a factor 0.05; see Fig.~\ref{fig:mean_mag}). For Poisson distributed sources, even highly biased source masks give a much smaller effect than direct thresholding.}
\label{fig:poisson_Dl}
\end{figure}

Figure \ref{fig:poisson_Dl} shows predictions for the power spectrum correction. The blue line shows the prediction of Eq.~\eqref{eq:poissonnumerical}, where disks of 3 $\arcmin$ are drawn for total masked sky fraction of $1.6\%$. The other coloured lines illustrate the impact of apodization of the mask. The apodization procedure is performed as described in Appendix~\ref{app:apo}. The orange and green curves show the case of 3 and 5 $\arcmin$ apodization respectively, and show two main signatures: the increase of the masked sky fraction, boosting the large-scale signal, and the introduction of a cut-off on small scales. For comparison with the threshold mask of the previous section, we have picked for this figure $f(\vx)$ equal to $\kappa(x)$ the lensing convergence field; more realistic point source fields are dealt with in Sec.~\ref{sec:results}. If sources were to form preferentially in peaks of the $\kappa$ field, the relevant Poisson intensity $f$ would be a biased version $b \kappa$, and the coloured curves would scale linearly with $b$.  The black line shows the threshold-mask analytic prediction at the same masked sky fraction than the green curve, reduced by a factor 20. Hence, unless the bias is extremely high, a Poisson-induced signal is typically much smaller.

\subsection{Polarization}

The general result of Eq.~\eqref{eq:generalmeanf} also holds for the polarization or cross-correlation, simply by using the relevant correlation functions in place of the temperature correlation function. However, for B-mode polarization, the choice of estimator is much more important. The mask-normalized pseudo-correlation functions we are analysing here correspond to deconvolved pseudo-$C_\ell$ estimators. It is well known that for polarization, although these estimators are unbiased, they couple cosmic variance of E-modes into B-modes due to E-to-B mixing on the cut sky.
For this reason they are unlikely to be used in practice for analysing future data, where sensitivity to small B-mode signal is a major goal. It is also clear that there are likely to be many fewer polarized sources compared to temperature sources, so a much small masked sky fraction is likely~\cite{Lagache:2019xto}. However, as a baseline for reference and comparison, we do briefly present a few basic results for the pseudo-correlation function estimators. These are likely to remain relevant for many E-mode power spectrum analyses, and we comment in later sections about the impact of using different estimators where the effects on the B-modes may be much smaller.

The form of the correlation function results is basically the same, but the polarization pseudopower spectra are formed from combinations of the two $\xi_\pm$ correlation functions. In harmonic space this still give a convolution-like effect on the power spectra: if we take the unlensed $C_l^B=0$, on large scales compared to the foreground correlation length and hole size, the result corresponding to Eq.~\eqref{eq:convolution} for the temperature is
\begin{equation}
\Delta \tilde{C}^E_l \sim -2\frac{\bar{f}}{\sigma_f^2} \int \frac{\ud^2\vl'}{(2\pi)^2} \tilde{C}^E_{l'} C^{\phi f}_{|\vl-\vl'|} \vl'\cdot (\vl-\vl')\cos^2(\phi_{\vl'}-\phi_{\vl}),
\label{eq:pol_convolution-E}
\end{equation}
\begin{equation}
\Delta \tilde{C}^B_l \sim -2\frac{\bar{f}}{\sigma_f^2} \int \frac{\ud^2\vl'}{(2\pi)^2} \tilde{C}^E_{l'} C^{\phi f}_{|\vl-\vl'|} \vl'\cdot (\vl-\vl')\sin^2(\phi_{\vl'}-\phi_{\vl}).
\label{eq:pol_convolution-B}
\end{equation}
For low $l$, we have the leading order result
\begin{equation}
\Delta \tilde{C}^B_l \sim 2\frac{\bar{f}}{\sigma_f^2} \int \frac{\ud l'}{l'} \frac{{l'}^2\tilde C^E_{l'}}{2\pi} C^{f\kappa}_{l'},
\label{eq:pol_lowl}
\end{equation}
which is white and negative, compared to the standard lensing result
\begin{equation}
\tilde{C}^B_l \sim 2 \int \frac{\ud l'}{l'} \frac{{l'}^2\tilde C^E_{l'}}{2\pi} C^{\kappa\kappa}_{l'}.
\end{equation}
The correction can easily make the total negative on large scales if $f$ is well correlated to $\kappa$ and relatively smooth. Figure \ref{fig:exactnu} in Appendix~\ref{app:nonpert} shows numerical results for a simple test case. On the E-modes and temperature cross spectrum the effect is qualitatively similar to on the temperature spectrum, but the B-mode spectrum picks up a large bias. This large bias is a result of the way that the estimators are combining cut-sky modes, and is entirely driven by the masking effect on E-modes. Using a pure-B estimate of the power spectrum would give a much smaller result.

\section{Numerical results}
\label{sec:results}
\subsection{Simulations and comparison method}
We tested the accuracy of our analytic estimates against numerical simulations that model the relevant effects, in particular the extragalactic foreground emission and their correlation with CMB lensing. For this purpose we used the publicly available \websky simulation suite\footnote{\url{https://mocks.cita.utoronto.ca/index.php/WebSky_Extragalactic_CMB_Mocks}} \cite{Stein:2020its} which includes maps of CMB lensing convergence $\kappa$, radio point sources, CIB, and tSZ produced from the same underlying mass distribution at $z\leq 4.5$. The mass distribution was constructed with the accelerated N-body mass-Peak Patch approach \cite{Stein2019,peak-patch} from a 15.4 Gpc$^3$, 12,2883 particle lightcone
 in a Planck 2018 cosmology. CIB and tSZ emission maps were constructed starting from the same matter distribution and using halo models matched to the latest CMB data from Planck, SPT and ACT as well as Herschel data at frequencies relevant for CMB experiments. We refer the reader to Ref.~\cite{Stein:2020its} for more details of the semianalytic models adopted for these maps.

Since the mask-induced biases are small, and depend on the properties of the underlying matter field which is non-Gaussian, we created two sets Monte Carlo simulations of 100 lensed CMB realizations each. To build the first set, the unlensed CMB realizations were lensed using the same deflection field constructed from the \websky $\kappa$ simulation (NG set). To build the second set, the same unlensed CMB simulations were lensed with different Gaussian random realizations of the deflection field having the same angular power spectrum as the \websky $\kappa$ map (G set). We used the NG set to isolate the bias as it would appear on real data while the G set was used to compute the error bars of our measurements. Hence, the error bars displayed in the figures do not include any non-Gaussian contribution to the covariance. In the following, unless stated otherwise, error bars displayed in figures represent the error on the average measured on the G simulations.

\subsection{Limiting case: 100\% correlated foreground mask}
As a first test we considered the extreme case of a mask constructed from a foreground that is 100\% correlated with CMB lensing, creating a foreground mask $W_{\kappa}$ by simply thresholding the CMB lensing $\kappa$ field. Since the total bias is sensitive to the overall sky fraction removed by the mask, as well to the specific correlation between the mask and the convergence, we tested different configurations. To test the dependency on the sky fraction we thresholded the field masking all the pixels above a specific $\kappa$ value so that a sky fraction $f_{\rm sky}^{\rm mask}$ is removed. This generates masks with large numbers of small holes.
To test the effect of the correlation scale of the deflection field and the shape of the mask, we also created masks by smoothing the $\kappa$ field with Gaussian beams of different full width at half maximum (FWHM, $\theta_{1/2}$) prior to the thresholding step. This results in more regular and connected holes due to the longer correlation length, and also effectively reduces the shot noise of the foreground map (i.e. $\kappa$) due to the finite number of particles in the \websky N-body simulation.

The  bias induced by $W_{\kappa}$ is estimated as the difference between the power spectra obtained using the original (unrotated) mask, and a randomly rotated mask, both using the same NG lensed CMB realizations.
The rotated mask $W_{\kappa}^{\rm rot}$ is derived from a random rotation of the original $W_{\kappa}$ so that it is uncorrelated with $\kappa$, but retains all the other nontrivial mode-coupling effects due to cut sky and hole shapes. The correlated mask bias evaluated in this way is therefore insensitive to numerical effects only due to an incomplete sky coverage\footnote{We neglect the small error from regions near the poles of the rotation axes that are correlated even after random rotation.}. We computed the power spectrum of the masked CMB skies using a pseudo-$C_{\ell}$ method as implemented in the NaMaster package \cite{namaster} and used a $C^2$ function (effectively a cosine) to apodize the mask to control ringing effects in harmonic space. This approach follows common practice in CMB analyses including small angular scales and is described by the analytic modelling presented in the previous sections. As we discuss in Sec.~\ref{sec:mitigation}, alternative estimators capable of effectively recovering the information inside the holes of the mask would give different results and potentially have a reduced effect.\\*

\begin{figure*}[htbp]
\includegraphics[width=\textwidth]{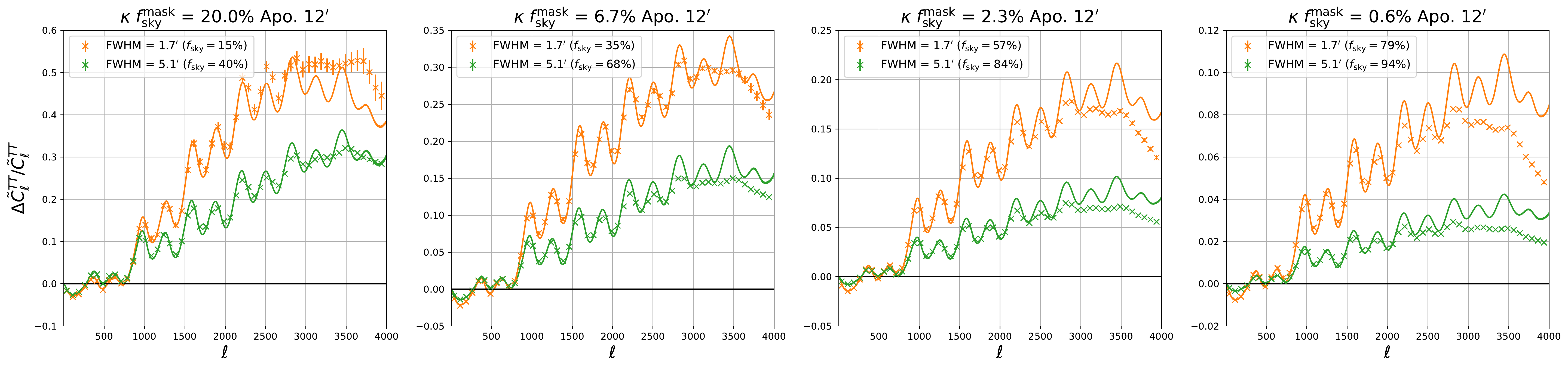}\\
\includegraphics[width=\textwidth]{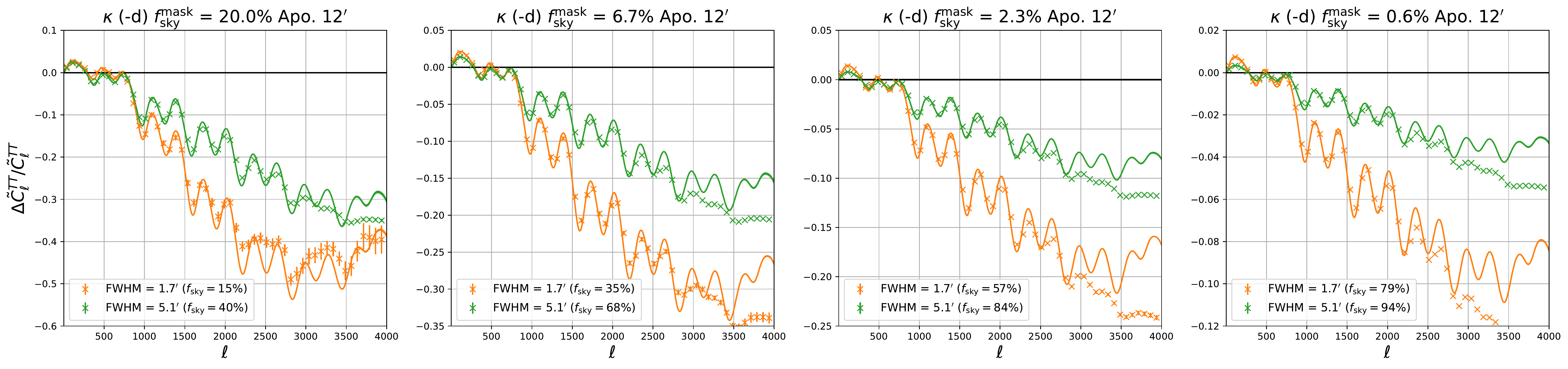}
\caption{Bias induced by masking the lensed CMB temperature with a foreground mask generated by thresholding the \websky CMB lensing $\kappa$ field after smoothing to a scale of $1.7'$ (orange) or $5.1'$ (green). In the top panel data points measured from simulations are compared to perturbative semianalytic predictions in solid. Results obtained by masking different sky areas are shown in different columns. We apodized each mask with a 0.2 deg. apodization length to control ringing effects in the power spectrum estimation step. The masked area prior to apodization is reported in the title and the effective sky area after apodization used to compute $C_{\ell}^{TT}$ is shown in the legend. The bottom panel shows the bias measured on CMB simulations lensed with a deflection field with an inverted sign (NG$^-$ set). Since the leading-order effect of the mask correlation is linear in the lensing, the bias has the opposite sign compared to the case of the top panel on scales where the leading order predictions are accurate. See Sec.~\ref{sec:tsz} for more details.}
\label{fig:frac_diff_kappa}
\end{figure*}

Figure \ref{fig:frac_diff_kappa} shows the measured bias from mask correlations measured in the simulations (shown as data points), compared to the semianalytic perturbative prediction described in Sec.~\ref{sec:empiest}.
To compute the theoretical predictions we measured the required cross-spectra between the mask and the deflection field from simulations, as well as the mask auto spectrum.
The semianalytic model describes the effect on large scales up to $\ell\approx 3000$ remarkably well for all the configurations considered here. This holds also for extreme cases where the relatively blue shape of the \websky $\kappa$ angular power spectrum, the presence of N-body shot-noise and the relatively large apodization length adopted, leads to the mask containing numerous tiny disconnected regions with greatly reduced effective sky area (as low as 15\%, even with no Galactic plane mask).  On smaller scales, the agreement between simulations and predictions gets worse, but a better fit can be achieved using the nonperturbative calculations discussed in Appendix \ref{app:nonpert}.

\subsection{Cosmic infrared background}
The CIB is produced by star-forming galaxies through the absorption of stellar radiation by dust grains which is later reemitted in the infrared. The clustering of halos, and consequently of the galaxies within, then generates the observed CIB intensity fluctuations~\cite{viero2013}. In addition to providing important constraints on the physics of star formation over a wide range of redshifts and halo and galaxy masses, especially for the objects with low luminosity that cannot be studied individually, the CIB acts as an important contaminating emission at microwave frequencies. Due to its spectral energy distribution (SED) similar to thermal dust emission it is difficult to disentangle CIB and galactic dust through component separation and perfectly remove both components, in particular at small angular scales and high observing frequencies. CIB residuals then propagate to data products derived from CMB maps.

\begin{figure*}
\includegraphics[width=\textwidth]{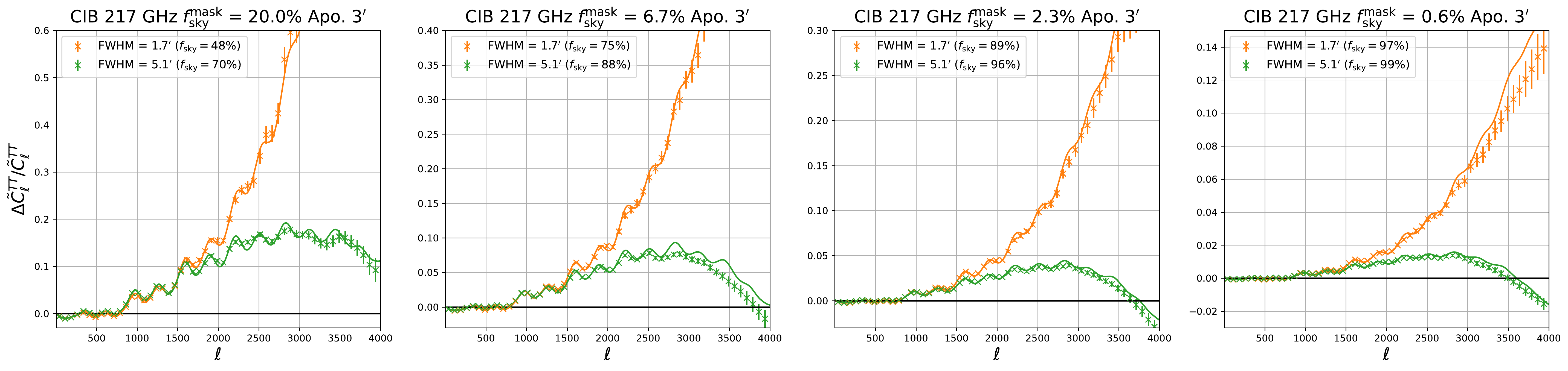}\\
\includegraphics[width=\textwidth]{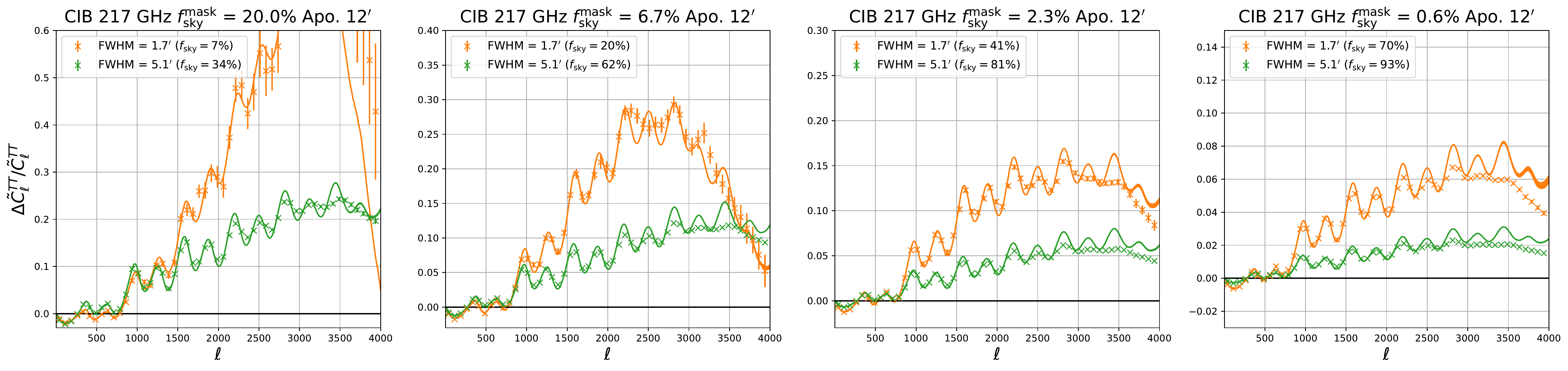}
\caption{Bias induced by masking the lensed CMB temperature with a foreground mask generated by thresholding the CIB map at 217 GHz of the \websky suite after smoothing to a scale of $1.7'$ (orange) or $5.1'$ (green). Data points show the measurements of the bias on simulations while perturbative semianalytic predictions are shown in solid lines. Masks with different sky fractions are shown in different columns.  The top row shows results with $3'$ mask apodization tapering function, the bottom row using a larger $12'$ apodization (giving substantially larger masked areas as shown in the legend). }
\label{fig:frac_diff_cib}
\end{figure*}

For CMB lensing and Compton $y$ maps, CIB residuals are potentially particularly harmful as they are highly correlated with the underlying cosmological signals \cite{Ade:2013aro,planck2015-cibXtsz,planck2015-sz,Song:2002sg}, and hence can bias cosmological analyses. The CIB is therefore an example of a foreground highly correlated with CMB lensing ($\gtrsim 70\%$ for $\ell\lesssim 1000$ where clustering of the emission is important).  We constructed a threshold mask $W_{\rm CIB}$ following the procedure outlined in the previous section starting from the \websky CIB map at 217GHz. This frequency was chosen as it is the highest relevant frequency typically used for CMB power spectrum analysis based on multi-frequency cross-correlation as done for e.g. \Planck. The \websky maps are based on a halo model of CIB previously used to fit Herschel and \Planck~data \cite{shang2012,viero2013,planck2013cib}. The rest-frame SED of CIB in these halos accounts for mass, frequency and redshift evolution as well as frequency decorrelation, and was normalized to reproduce the \Planck~CIB at 545 GHz \cite{planck2013cib,planck2015gnilc}. While improvements to this model have been recently presented in the literature \cite{Maniyar:2020tzw}, it is sufficient to reproduce with good accuracy all the measurements available in the literature from \Planck~and Herschel data between 143 GHz and 857 GHz (see \cite{Stein:2020its} and references therein for more details). Figure \ref{fig:frac_diff_cib} shows the correlated mask bias measured from simulations adopting the same $C^2$ function of the previous section and using two different apodization lengths ($3^\prime$ and $12^\prime$), compared to our semianalytic perturbative predictions.

As for the case of the $W_\kappa$ mask, the theoretical predictions match  the simulation measurements very well up to scales $\ell\lesssim 2500$.
The amplitude of the mask bias at small scales has a peak and then decreases on scales smaller than the characteristic scale imposed by the mask hole size.
Qualitatively this turnover is similar whether the larger hole size is caused by apodization, or by thresholding a smoothed CIB map.
When masking the CIB peaks without applying any smoothing of the CIB maps prior to thresholding (orange lines and points in Fig.~\ref{fig:frac_diff_cib}),
there are many very small holes due to the relatively blue shape of the CIB angular power spectrum.
A larger apodization scale increases the fraction of sky that is masked for fixed underlying hole distribution,
increasing the bias on larger scales (where noise and foreground power is lower, and therefore potentially more important in the analysis of real data).

Although masks on real data are usually not designed to remove peaks of CIB emission per se, the case where we masked the highest peaks so that only the 0.6\% of the sky is removed is of particular interest. Infrared sources that are local dusty galaxies are expected to have a low correlation to CMB lensing due to the short path length. However, chance radial alignments of sources for the CIB, high-redshift protoclusters, and lensed high-redshift galaxies, may make up an important fraction of the point sources detected in CMB maps \cite{vieira2010,vieira2013}, all of which may have a significant correlation to the line of sight CMB lensing \cite{bianchini2015,bianchini2016,pbXhatlas}. The brightest of these objects are usually removed by point sources masks (see later Sec.~\ref{sec:rs}). Despite the reduced masked sky area, the bias in this case is potentially significant and could lead to important detectable effects as we will see in the following sections.

There are however several caveats to our analysis. The \websky CIB simulations do not model specifically the effect of Poisson shot noise for the brightest sources nor include lensing of the infrared galaxies, which potentially make up a significant fraction of the detected objects \cite{negrello2010,gonzalez-nuevo2012}, especially the brightest one. Moreover, objects located at very high redshift above the maximum redshift probed by the LSS included in \websky ($z_{\rm max}=4.5$), despite being very rare,  can still retain a nonzero correlation with CMB lensing as CMB lensing kernel has a non-negligible amplitude in that regime (see e.g. \cite{wilson2019} for a discussion on high-redshift object cross-correlation in the optical band).

\subsection{Thermal SZ}\label{sec:tsz}
Observation of the tSZ effect, the inverse Compton scattering of CMB photons by free electrons, is a well established way to construct roughly mass-limited samples of galaxy clusters that are independent of redshift and thus very powerful cosmological probes \cite{birkinshaw1999,carlstrom-sz,szreview2019}. tSZ clusters mark out large-scale density peaks, and as such have substantial correlation to CMB lensing, at the $30-50\%$ level~\cite{Hill:2013dxa}, and the emission also follows highly non-Gaussian statistics \cite{thiele-tszpdf,coulton2018}. If tSZ clusters are masked out, the CMB lensing-mask correlation can be substantial.

Current CMB surveys from the ground and from space have blindly detected
approximately 3200 tSZ clusters with redshift measurements to date \cite{planck2015-szcatalog,sptpol-szcatalog,advact-szcatalog}. Due to its characteristic spectral signature, tSZ emission can be subtracted from CMB maps using component separation. However, this becomes difficult on small scales where noise becomes important, and foreground-cleaning residuals are less simple to model. The tSZ signal is therefore usually not cleaned for CMB power spectrum analysis, instead its contribution to the observed power spectra is accounted for in the model. Nevertheless, to minimize complex foreground residuals, for various higher-point statistics (including CMB lensing reconstruction) it is often useful and common practice to remove some of this source of highly non-Gaussian signal by masking the SZ clusters (see e.g.~\cite{Osborne:2013nna}). In this case it may also be important to understand what happens to the two-point statistics over the remaining unmasked area. Planck data were shown to be robust to these effects \cite{Aghanim:2018oex}, however future ground-based surveys such as Simons Observatory \cite{Ade:2018sbj} (SO) and CMB-S4 \cite{Abazajian:2019eic} (S4 hereafter) will detect one order of magnitude more clusters and thus cluster masking might potentially soon become a more significant issue.

We followed the same procedure outlined in previous sections and constructed a mask based on the thresholding of the \websky\ tSZ Compton $y$ parameter map $W_{y}$. The \websky\ simulation models the tSZ emission starting from the dark matter halos identified in the simulation, and applies a halo model construction including the effects of non-thermal processes such as radiative cooling, star formation, supernova and AGN feedback in the pressure profile \cite{battaglia2012}. As a result, the $y$ map is highly non-Gaussian with the skewness and kurtosis of its 1-point PDF having values significantly above 1.

\begin{figure*}[htbp]
\includegraphics[width=\textwidth]{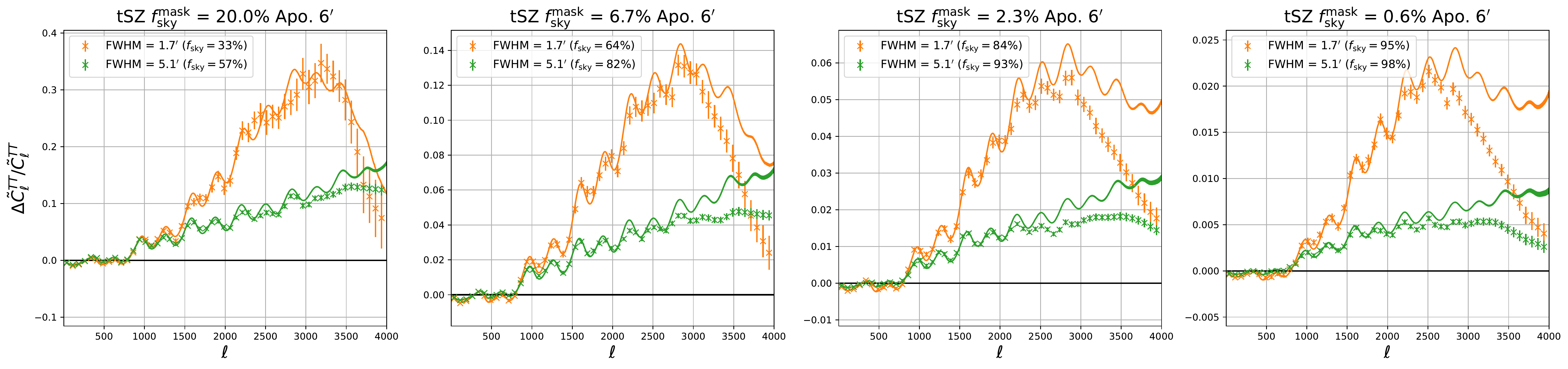}\\
\includegraphics[width=\textwidth]{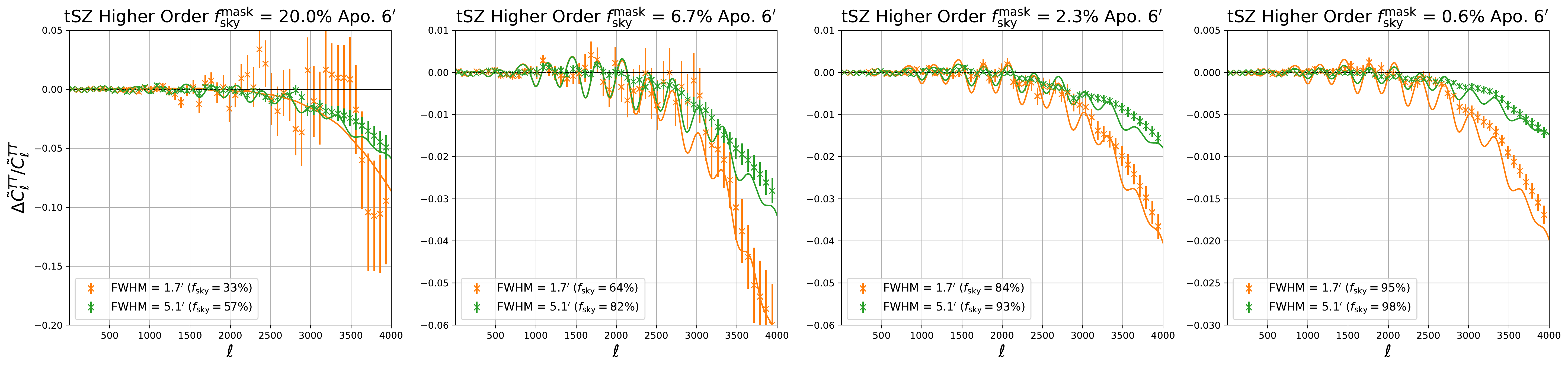}\\
\caption{Top: bias induced by masking the peaks of the tSZ emission (after smoothing to a scale of $1.7'$, orange, or $5.1'$, green) on the lensed CMB temperature as measured on simulations compared to perturbative leading order analytic predictions (solid line). Masks with different sky fractions are shown in different columns. We adopted a $6^\prime$ apodization length for the mask tapering function. Bottom: measurement on simulations of the even higher-order biases responsible for the discrepancies between the leading order predictions and the simulation results shown in the upper panel. Approximate analytic predictions of the second-order terms are shown as solid lines and described in Sec.~\ref{sec:tsz}.}
\label{fig:frac_diff_tsz}
\end{figure*}

In Fig.~\ref{fig:frac_diff_tsz} we show the comparison of our theoretical predictions with the simulation measurements.
Compared to the case of $\kappa$ and CIB thresholding, the agreement between the perturbative model and simulation results is worse, with significant discrepancies observed already at multipoles $\ell\approx 2500$ and reaching a factor between 2 to 4 at $\ell\approx 4000$ in particular when only the highest peaks are masked (right panel). For more aggressive masks where a significant fraction of the peak is masked however the agreement (left panel) between simulations and analytic predictions improve substantially. Since the bulk of the tSZ emission is localized in highly clustered and dense regions at relatively low redshift for a threshold that is sufficiently small, $W_y$ contains holes with a larger angular size around the overdensity corresponding to the galaxy cluster.
 The masked region at each cluster may therefore remove a significant area of high lensing signal associated with the cluster (rather than just a small area at the very peak of the overdensity).
We therefore checked whether higher-order effects beyond the linear term modelled in the previous section could be responsible for the observed discrepancy, for example from the reduction in lensed CMB signal over the cluster mask.

To test higher-order effects we constructed another set of lensed CMB simulations with the same masks as the NG set, but lensed with a deflection field with an inverted sign. We refer to this set of simulations as NG$^-$ in the following. Since the leading-order effect of the mask correlation is linear in the lensing, in these maps it should have opposite sign (see Fig.~\ref{fig:frac_diff_kappa}).
Higher-order effects that are quadratic or involve a higher even power of the lensing can be isolated on simulations using the half sum of the mask biases measured on the NG and NG$^-$ sets using the same threshold mask for both NG and NG$^-$. In the bottom panel of Fig.~\ref{fig:frac_diff_tsz} we show that higher-order effects induce a negative correction to the leading order predictions that explains the discrepancy. When only a reduced fraction of the sky is masked, the higher-order effects become important at $\ell\approx 2000$ and suppress the bias by a factor of 4 compared to the leading order predictions at $\ell\approx 4000$. In the limiting case where we mask a large fraction of the sky, the corrections become relevant at progressively smaller angular scales and their relative importance is reduced.

Corrections that are quadratic in the lensing largely account for a change in the underlying lensed CMB power spectrum due to the masking of areas where the lensing is larger. An approximate analytic estimate of this higher-order bias can be obtained by computing the lensed CMB power spectrum
(approximately a convolution of the CMB lensing and the unlensed CMB power spectra) where the CMB lensing power spectrum is derived from the lensing convergence power spectrum computed over the masked sky using the $W_y$ mask.
Figure \ref{fig:frac_diff_tsz} shows that this simple model describes the effect seen in the simulations quite accurately (a more accurate analytic calculation, including all orders for a Gaussian foreground, is described in Appendix \ref{app:nonpert}).

\subsection{Radio point sources}\label{sec:rs}
The dominant population of bright point sources detected at CMB frequencies are AGN-powered radio sources emitting synchrotron radiation through acceleration of relativistic charged particles \cite{dezotti2010}. The details of the observed emission law of such sources (whose intensity typically decreases as frequency grows) depends on the orientation of the observer relative to the axis of the characteristic jets emerging from the central black hole \cite{padovani2017}.
Because the synchrotron emission is polarized, some of the sources detected in temperature also have a counterpart in CMB polarization maps. So far only a minor fraction of the detected sources in temperature are polarized, but the situation is expected to change in the coming years where hundreds of object will be identified in deep polarization maps \cite{Puglisi:2017lpn, Lagache:2019xto}. These are potentially an important obstacle to the exploitation of small scale E-mode polarization data as well as large scale B-mode polarization if the tensor-to-scalar ratio $r$ is sufficiently low. As such, all these sources are systematically masked in CMB temperature power spectra analyses. Polarization data can be masked separately (using only the detected objects in polarization) or together with temperature data using the same mask \cite{Aghanim:2015xee,Choi:2020ccd,sptpol2020,henning2018}.
Other analyses studying statistically anisotropic effects in CMB maps (e.g. CMB lensing or birefringence reconstructions) adopted different approaches, ranging from keeping the same mask as in power spectrum analysis or using dedicated source-subtracted or inpainted maps \cite{Aghanim:2018oex,bianchini2020,wu2019,Aiola:2020azj,naess2020}.

Halos hosting radio sources, and therefore the radio source distribution (especially the low flux component),
 correlate with large-scale structure  and hence with the tSZ emission, CMB and galaxy lensing and CIB \cite{Holder:2002wb,Shirasaki:2018wdq,Allison:2015fac,dwek2002}. The relatively low amplitude of the clustered component of  $\sim 10{\rm s}-100{\rm s}$  mJy radio sources detected in current generation CMB maps, means that for current masked source densities the mask can be approximated as uncorrelated to the lensing to good accuracy. We used \websky radio sources mock catalogues to test that this is indeed the case, and whether this assumption breaks down for future experiments.

 The radio-source mocks use the halos identified in the simulation box of \websky to implement a halo occupation distribution (HOD) for the Fanaroff-Riley Class I (FR- I) and Class II (FR-II) galaxies described in \cite{wilman2008,sehgal-sim}. The HOD models the occupation numbers of FR-I and FR-II populations as broken power laws and asymmetric Gaussians and a luminosity function given by a broken power law with a luminosity cut-off set to reproduce the luminosity function at 151MHz. The constructed HOD is then resampled to match the observed flux counts $n(S)=\ud N/\ud S$ while keeping the same rank ordering of the original catalogue, mixing in practice HOD and abundance matching techniques (see \cite{websky-rs} for more details\footnote{See also \url{https://github.com/xzackli/XGPaint.jl}.}). The constructed catalogues reproduce with good precision the Planck number counts at frequencies $\nu\leq 143$GHz where the radio galaxies dominate the DSFGs population.

\begin{figure*}[!htbp]
\includegraphics[width=\textwidth]{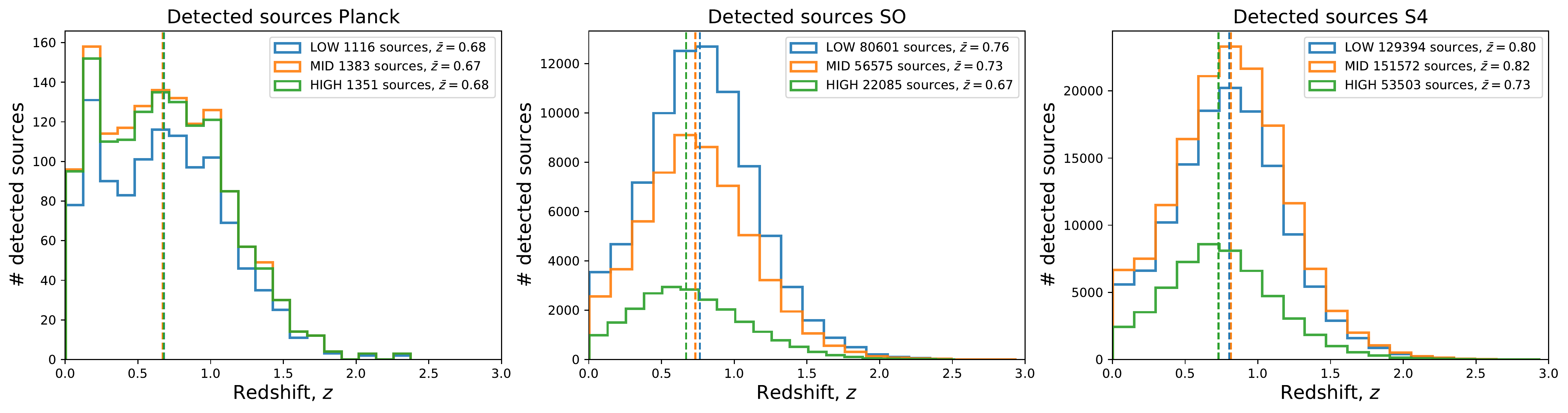}
\caption{Distributions of the \websky simulation radio galaxy population at different frequencies that would be detected using the flux limit for different experiments. Numbers here are over the full sky and are masked in our full-sky analysis. The number of sources detected over the full sky is shown in the label while the median redshift is shown as a dashed vertical line. See Table~\ref{table:specs} for the specification of the LOW, MID, and HIGH frequency channels.}
\label{fig:rs-dndz}
\end{figure*}

To build the RS mask for a given experiment we started from the simulated radio catalogues and selected the sources that have a measured flux above the detection limit of a particular experiment. We focused on \Planck~, SO and S4, and for each of these we selected the sources in the three frequency bands most relevant for small-scale power spectra measurements. We label these LOW, MID, HIGH, with each having a different flux limit and resolution as shown in Table \ref{table:specs}\footnote{The value of the flux limits for SO have been computed using the publicly available noise curves discussed later in the text and the method discussed in Appendix 4 of \cite{s4dsr}, which takes into account uncertainties due to foreground residuals. We note that more accurate estimates including noise inhomogeneity could lead to flux limits that are $\sim 20\%$ lower than those quoted in Table \ref{table:specs} \cite{naess-private}. This would lead to a higher number of detected sources that are then masked. The SO-related results presented in the following can therefore be considered as lower bounds on the amplitude of the mask bias.}. The properties of the selected galaxy samples for each experiment are summarized in Fig.~\ref{fig:rs-dndz}.

\begin{table}
\begin{tabular}{l|c|c|c}
\hline
Channel & $\nu$ (GHz)& Intensity flux cut (mJy)&$\theta_{1/2}$ (arcmin)\\
\hline
Planck LOW &100 &232 & 9.69\\
Planck MID &143 &147 & 7.30\\
Planck HIGH &217 & 127& 5.02\\
SO LOW&93 & 4.37& 2.2\\
SO MID&145& 5.03&1.4\\
SO HIGH&225&9.88&1.0\\
S4 LOW&95 &2.82&2.2\\
S4 MID&143 &1.98&1.4\\
S4 HIGH&220 &4.37&1.0\\
\hline
\end{tabular}
\caption{Point source intensity flux cut values and resolution of the different frequency channels and experiments considered in this work. See \cite{Lagache:2019xto,Aghanim:2015xee,Ade:2018sbj} for more details.}
\label{table:specs}
\end{table}

\begin{figure*}[!htbp]
\includegraphics[width=\textwidth]{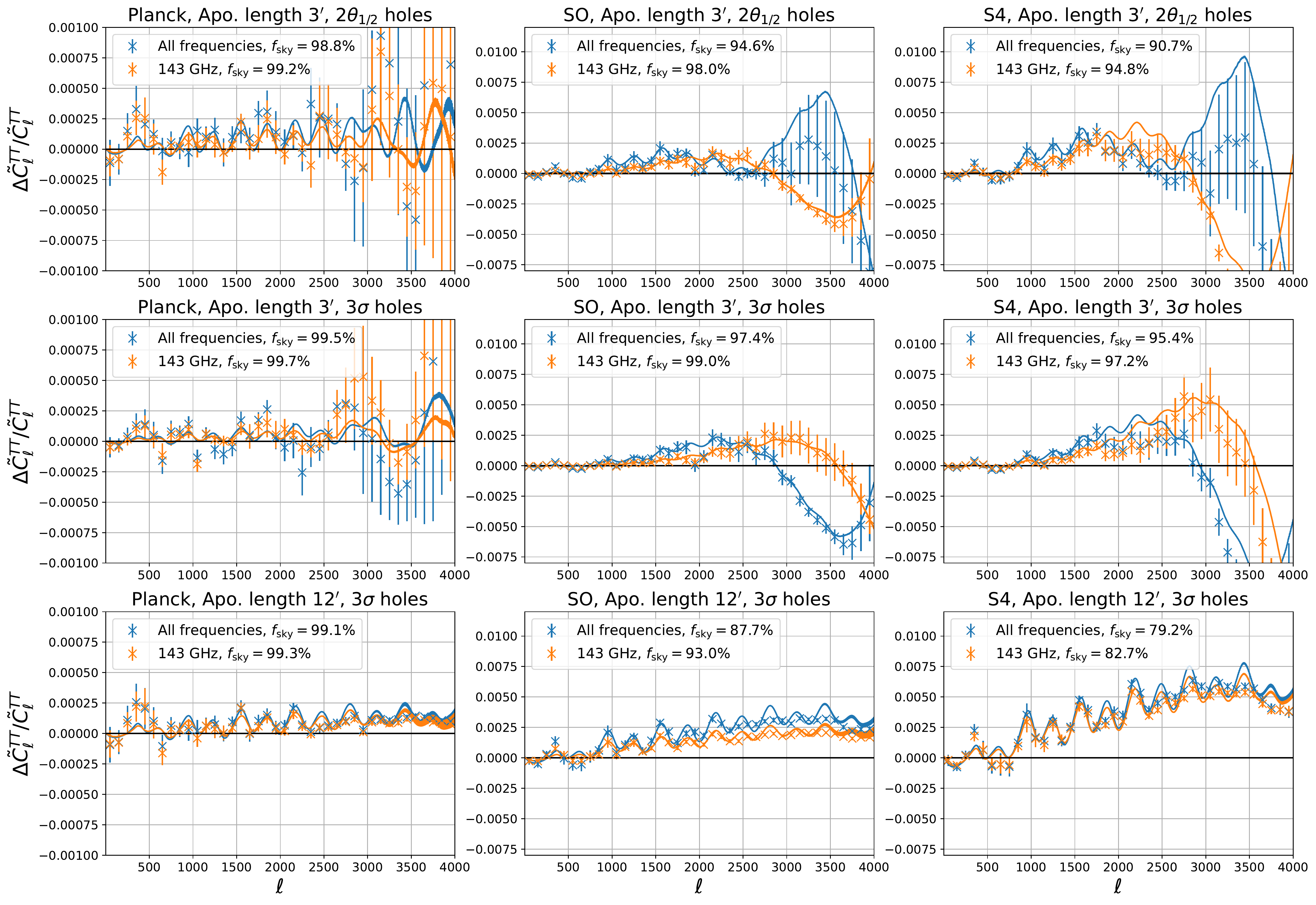}
\caption{Effect of masking radio sources for different experiments considered in this work (left to right).
The MID frequency channel is shown in orange, and the mask derived by taking the product of the masks at each considered frequency channel is shown in blue. Simulation measurements are shown as data points and the semianalytic theory prediction in solid. Different hole sizes are displayed in the top and middle panel. The point source masks were apodized with a $C^2$ tapering having an apodization length of $3^\prime$ (upper and middle panel) and $12^\prime$ for the bottom one.}
\label{fig:rs-summary}
\end{figure*}

In the literature, different experiments adopted different choices for how to mask point sources. \Planck~masked a circle of radius $3\sigma = 3\theta_{1/2}/\sqrt{8\log 2}\approx 1.3\theta_{1/2}$, where $\theta_{1/2}$ is the FWHM of the beam of each frequency channel and a Gaussian tapering function with $30^\prime$ apodization length \cite{Aghanim:2015xee}. Ground-based experiments adopted more conservative choices. ACTpol used holes of a radius of about $\sim 3.5\theta_{1/2}$ radius hole at 98GHz and 150GHz with a sine apodization having a length ranging between $10^\prime$ to $15^\prime$ \cite{Choi:2020ccd}. SPTpol typically masked the sources with a fixed $5^\prime$ radius circle (which is $\sim 3-5 \theta_{1/2}$ at 95-220GHz) and a cosine apodization with $5^\prime$ apodization length \cite{bianchini2020}. For small-scale temperature analysis, they adopted a different masking procedure with larger holes for the brightest sources \cite{reichardt2020}. For wide surveys such as \Planck~ or ACTpol Wide, the fraction of observed sky masked by sources before apodization amounts to $\sim 0.4\%$ while deep surveys like SPTpol and ACT deep removed a few percent of the observed sky. We investigated the impact of different setups in terms of apodization and hole size, and Fig.~\ref{fig:rs-summary} summarizes our findings.

The clustering of the selected galaxies is dominated by the shot noise for all the selected galaxy samples. The mean cross-correlation with CMB lensing is 5\%, below 10\% on all angular scales for the deepest sample of S4, and one order of magnitude lower for \Planck.
The formalism based on Poisson sampling of the density field (see Sec.~\ref{sec:poissonmask}) would thus be appropriate if one had to model the effect from first principles. In Fig.~\ref{fig:rs-summary} we use the empirical model of  Sec.~\ref{sec:empiest} to compute the analytic predictions. A $C^2$ apodization of the mask holes is used to measure the effects on simulations for consistency with the results of the previous sections.

For the case of masks with $3\sigma$ hole radius, and the union mask that removes sources detected at all frequencies (which largely overlap between frequency channels), for \Planck~we find a negligible effect of the order of $\sim 0.01\%$. For SO and S4 however, the effect becomes comparable to the cosmic variance uncertainty and therefore becomes relevant. Increasing the hole radius to $2\theta_{1/2}$ makes the bias shape change significantly, especially at small angular scales where it can grow to  about 1\% and change sign. Increasing the apodization length potentially has a more important effect as all scales are affected by the increased masked area. An apodization such as that of adopted by \Planck~\cite{Aghanim:2015xee} can increase biases by a factor two, however for the specific case of \Planck~shown here, it still keeps the bias below the detection threshold. If instead we mask only the sources detected at a given frequency, we found that the LOW and MID frequency channels are the ones most affected, as they are the ones where the effect is larger and/or have the lowest flux detection threshold.

More conservative approaches to point source masking, as typically adopted in the analysis of ground-based experiments mentioned above, where the hole radius exceeds the $2\theta_{1/2}$ value considered in this work and wider apodization lengths are employed, will lead to a significant increase of the bias and a strong detection if unmodelled. At the SO and S4 level of sensitivity such strategies will need to be reconsidered as it may become necessary to find a compromise between data loss, increase of the mask-induced bias, and foreground contamination. In all cases, however, our analytic model describes the results of simulations well and can be used to estimate or mitigate the bias when required.

\begin{figure}
\includegraphics[width=0.48\textwidth]{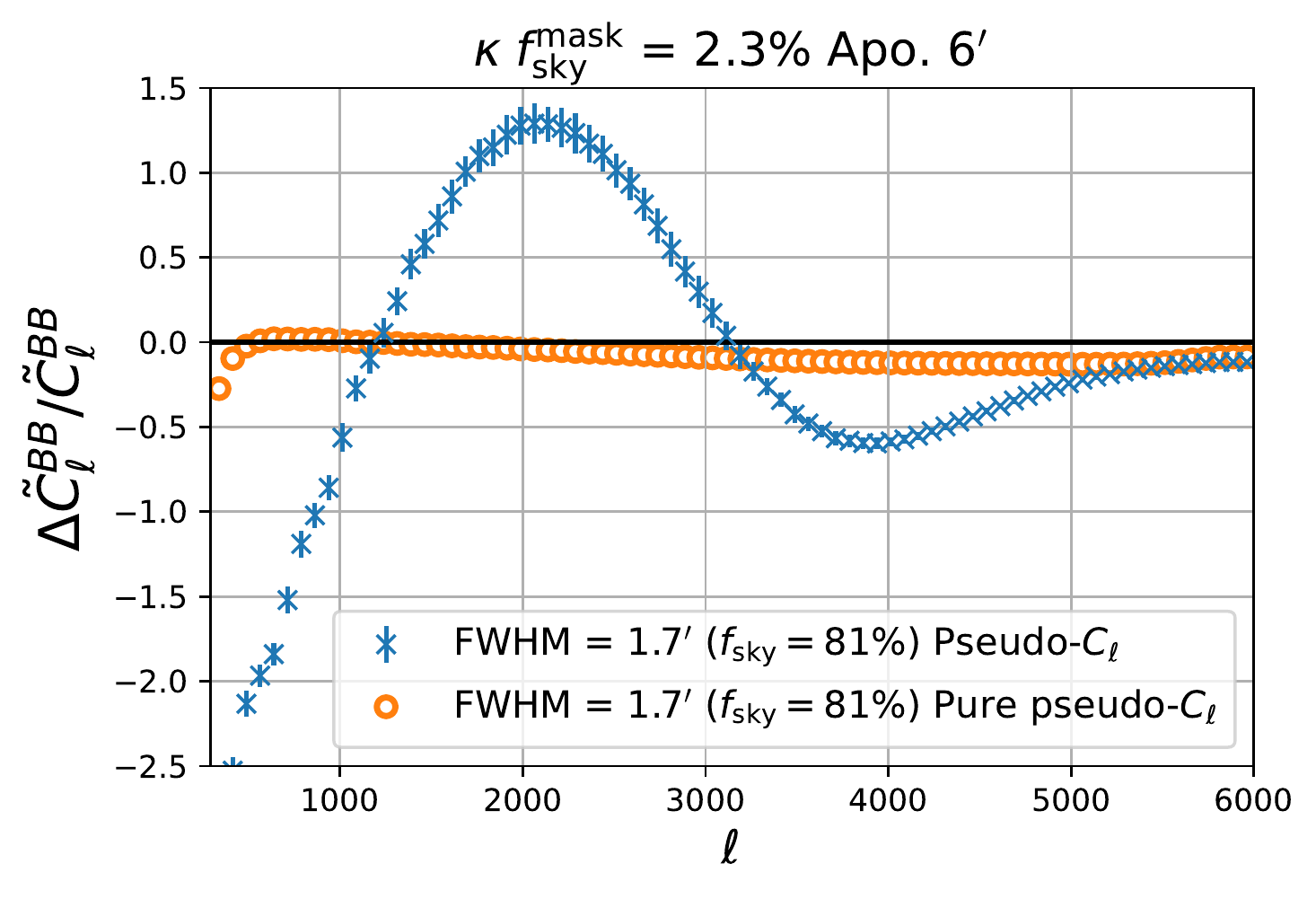}
\caption{Lensing-induced mask bias on the B-mode angular power spectrum for a standard pseudo-$C_{\ell}$ estimator (blue) and for a pure pseudo-$C_{\ell}$ estimator (orange). Because the major source of the bias at leading order is a conversion from E to B-modes, the pure estimator removes almost entirely the bias observed with the standard estimator. The two estimators generate a very similar bias at small scales where the bias is mainly sourced by higher-order terms in the lensing (e.g. the suppression of lensing power due to peak masking).}
\label{fig:pure}
\end{figure}

As shown in Appendix \ref{app:nonpert}, the mask bias observed on E-modes is roughly a factor 2 lower compared to the one observed in the temperature power spectrum. If a common mask between temperature and polarization is adopted, we expect the bias to become relevant for high-sensitivity analysis of small-scale E-mode polarization and be negligible for B-modes on scales $\ell\gtrsim 200$ if a pure-pseudo power spectrum (or more optimal \cite{Tegmark2001}) method is used \cite{Smith:2005gi,grain2009}. In the case of a pure-B estimator the residual observed bias comes mainly from higher-order masking effects suppressing the lensed B-mode power, while the larger bias at linear order involving E-modes converting to B-modes is naturally removed.  An accurate evaluation of the bias on the large angular scale B-mode power from pure pseudo-$C_{\ell}$ methods would depend on the details of the apodization length of the mask. This can be highly nontrivial in presence of masks with complex boundaries, such as those removing radio point sources, and should anyway be optimized given an experimental noise level and a choice of multipole binning to minimize the total B-mode variance \cite{ferte2013,ferte2015}. This is beyond the scope of this paper and we leave this exercise for future work. However, in Fig.~\ref{fig:pure} we show an example of the effect of the B-mode purification on the mask bias for the limiting case of a $W_{\kappa}$ mask. For the more realistic case of a mask that removes radio sources detected at all frequencies, the B-mode power bias for the pure estimator is $\sim 10 - 20\%$ at $\ell\approx 200$ for S4.

\section{Impact on current and future data sets}
\label{sec:mitigation}
\subsection{Detectability, diagnostics and mitigation}
Although the level of the bias can easily be calculated from simulations, in practice it is usually not straightforward to reliably simulate very precisely what is being masked, so some kind of internal measurement or diagnostic would be useful. Fortunately, because the effect is linear in the lensing, it is quite distinctive.

We can expect methods that reconstruct the CMB inside the holes, such as inpainting or CMB Wiener filtering, to be quite effective at reducing the bias (or affecting its shape) if the mask is not too large. This is because
the temperature in a small hole can be predicted accurately by using the large-scale temperature modes that are well measured outside the mask. The correlation between the temperature value in the masked hole and the surrounding lensing (see Fig.~\ref{fig:lensing_pic_2d}) would then be mostly recovered, giving little net bias. The temperature reconstruction may itself bias the result, but in a very different way that allows for consistency checks. The effect can also be isolated in cross-correlation of masked and unmasked (or inpainted) maps, where the bias appears on large-scales with half the amplitude\footnote{Cross-correlating a perfectly inpainted map with the masked map results in a linear order bias proportional to $\left \langle \alpha_r(\vx) \w(\vy) \right\rangle / \left\langle \w(\vy) \right\rangle$, instead of $2\left \langle \alpha_r(\vx) \w(\vx) \w(\vy) \right\rangle /\left\langle \w(\vx)\w(\vy) \right\rangle$ for the autospectrum. For a Gaussian foreground field model, the first is half the second on large scales, but transitions to be equal to it on small scales.}. This has the advantage of not picking up mean white foreground noise from the unmasked foreground peak or some effects of inpainting errors, allowing a direct comparison with the masked auto spectrum.

For example, simulations suggest that the \Planck~ point-source mask of $\fsky \sim 2\%$~\cite{Aghanim:2015xee} could bias cosmological parameters by up to about $1\sigma$ if the mask were highly correlated to the lensing, but assessing exactly the level of correlation from purely theoretical considerations or simulation is difficult. We can instead directly assess the size of the bias by looking at cross-spectra between masked and inpainted maps. Specifically, we consider the difference of power spectra $\rm{SMICA}\times\rm{SMICA} - \rm{SMICA}\times\rm{SMICA}'$, where $\rm{SMICA}$ is the  foreground-cleaned SMICA temperature map~\cite{Akrami:2018mcd} masked by one of the likelihood masks including point source mask, and $\rm{SMICA}'$ is the SMICA map only masked by the galactic mask and inpainted elsewhere. To avoid noise bias in the power spectrum, the first and second map can be taken from different half-mission splits.
For the various frequency masks the smoothed difference is always $\Delta D_\ell< 1\muK^2$ at $1000<\ell<2000$ and $< 4 \muK^2$ on larger scales (with much of the variation expected from cosmic variance over the differing areas), suggesting the level of bias is safely negligible for the default \Planck~ masks.

The effect can also be tested using an estimate of the deflection field over the unmasked area to empirically estimate $\meandelta(r)$ (Eqs.~\eqref{delta_def}, \eqref{eq:EWpower}).  For \Planck~two good tracers of the lensing field are available: the lensing reconstruction~\cite{Aghanim:2018oex} (on large scales) and the cosmic infrared background (which is highly correlated to lensing and well-measured by \Planck~on smaller scales), which can be used to estimate $\meandelta(r)$ and hence the expected impact on the power spectrum. In either case we find these semianalytic predictions consistent with zero.
Each method of assessing the bias has some caveats, but taken together there seems to be good consistency with negligible bias for \Planck~parameters due to mask-lensing correlations. This is consistent with the expectation that the mask is dominated by Poisson radio sources, which have negligible impact given the number densities of sources masked by Planck, and nearby galaxies that are only weakly correlated to the CMB lensing. We reach the same conclusion trying to estimate the $\meandelta(r)$ on ACT DR4~\cite{Aiola:2020azj} D56 and D8 deep regions. For higher-resolution and forthcoming data, where substantially more sources may be resolved, mask bias consistency checks may be much more important.

\begin{figure}[h]
\centering
\includegraphics[width = 0.48 \textwidth]{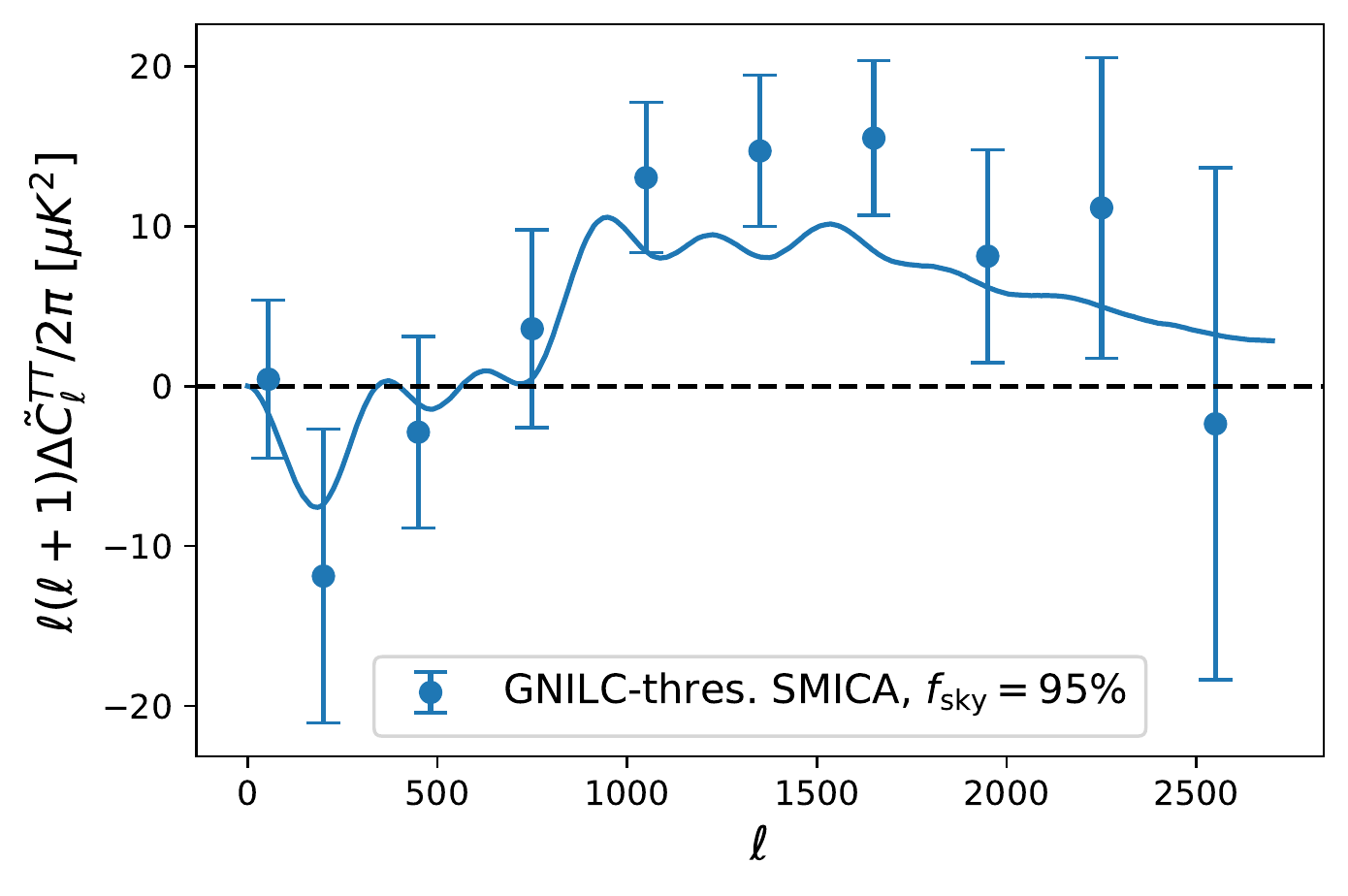}
\caption{Difference between \planck~$\rm{SMICA}$~data power spectra estimated on the \planck~likelihood mask and after removing an additional $5\%$ of the sky, directly thresholding on the GNILC CIB map at $545$GHz. The solid line is the analytic prediction using the Gaussian foreground thresholding model in this paper, where the cross-spectrum of the foreground to the lensing potential is obtained from the cross-spectrum of  GNILC map to the \planck~2018 lensing MV lensing potential quadratic estimate.}
\label{fig:SMICA_GNILCnu}
\end{figure}

On the other hand, we can easily detect biases in Planck's \rm{SMICA} maps when using a modified mask designed for the purpose.  Fig.~\ref{fig:SMICA_GNILCnu} shows results when masking additional $5\%$ of the sky with a foreground threshold mask. The points show the difference in the $\rm{SMICA}$ map power spectrum, as calculated on the union of the \planck~likelihood masks at 143 and 217GHz ($f_{\rm sky} = 43\%$), after masking this additional $5\%$ of the sky by directly thresholding on a foreground map taken here to be the (noisy, and beam-convolved) \planck~CIB observations as captured by the GNILC~\cite{Aghanim:2016pcc} map at 545GHz. The error bars are estimated for each multipole bin from the empirical standard deviation of the spectrum. The blue curve shows the analytic prediction for the bias as obtained with the threshold model of Sec.~\ref{sec:numask}. Along with the foreground autospectrum, the prediction requires its cross-spectrum to the lensing potential. We have used the empirical cross-correlation of the GNILC map to \planck~2018 publicly available lensing map \cite[MV estimate]{Aghanim:2018oex} for this purpose. This could be viewed as a rather nontrivial consistency check of our analysis and several \planck~products.

\subsection{Forecasts for future experiments}
\begin{figure*}[!htbp]
\includegraphics[width=\textwidth]{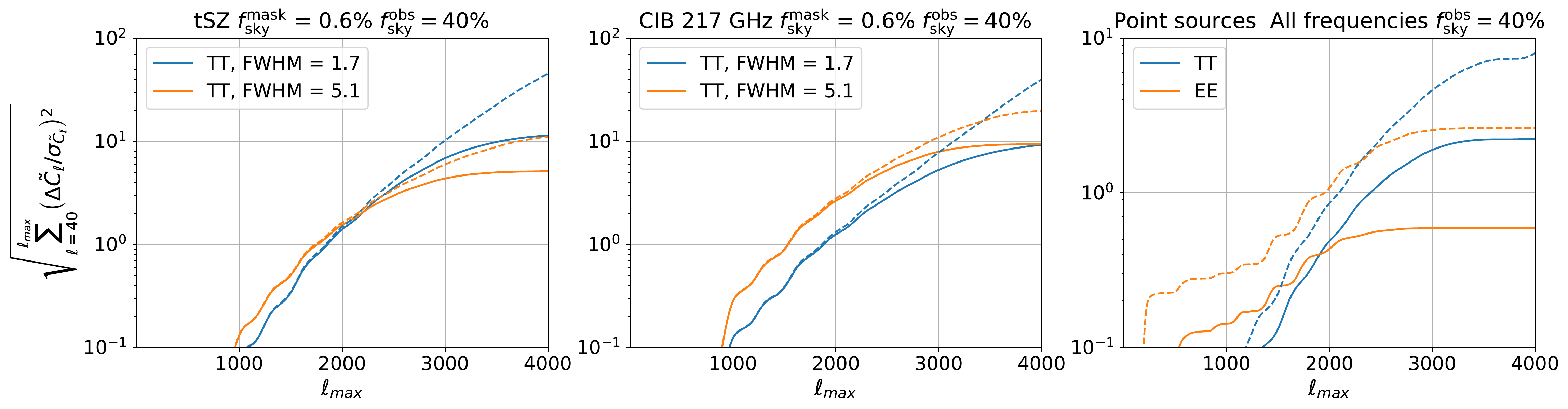}
\caption{Detection significance of the mask bias as a function of maximum multipole $\ell_{\rm max}$ included in the analysis for different sources for future high-resolution ground-based experiments. SO is shown in solid line while S4 is shown in dashed line. For both surveys we assumed a sky coverage of $f_{\rm obs}=40\%$. Unlike Figs.~\ref{fig:frac_diff_cib}, \ref{fig:frac_diff_tsz}, \ref{fig:rs-summary}, no apodization was applied to the mask prior to the computation of the power spectrum bias. This choice is conservative as it allows us to retain the largest observed sky area for a given mask and thus a smaller bias on the majority of angular scales. The  fraction of the sky area removed by masks based on foreground thresholding is shown in the title as $f_{\rm sky}^{\rm mask}$.
The significance reported is assuming the full sky CMB spectra are known perfectly (and an error model only accounting for foreground-cleaned noise).  }
\label{fig:snr}
\end{figure*}

Despite not being detectable on current data sets, in the previous sections we have shown that correlated masks can introduce substantial biases on the power spectrum if not accounted for, and they may be import for forthcoming more sensitive experiments that measure small angular scales. We therefore calculated the detectability of the biases induced by masking of tSZ, CIB and radio sources for SO and S4 assuming a sky coverage of $f_{\rm sky}=40\%$ and the realistic publicly available noise power spectra in temperature and polarization after a component separation procedure based on a standard\footnote{We consider the standard version of the algorithm the one that does not explicitly deproject any extragalactic foreground component.} internal linear combination algorithm\footnote{Details of the noise model for SO can be found at
\\
 \url{https://github.com/simonsobs/so_noise_models}, while the noise specifications for S4 have been taken from \url{https://cmb-s4.org/wiki/index.php/Survey_Performance_Expectations}. For SO we used the so-called baseline noise.}. In Fig.~\ref{fig:snr} we show the detectability of the bias in terms of achievable detection significance as a function of the highest multipole included in the analysis. This approach is simplified and assumes the perfect knowledge of the CMB power spectrum and, if employed, of the nuisance parameters used to describe the foreground residuals. As such, a detectable bias should be interpreted as showing that it is necessary to model the effect to be sure the inference of remaining cosmological (or nuisance) parameters is not biased. We note here that the masks are unapodized.

If tSZ and the brightest regions of CIB emission (which are considered here as proxy of DSFGs and covering only $0.6\%$ of the sky) are masked, biases on $\tilde{C}_{\ell}^{TT}$ will be detected with a statistical significance well above $5\sigma$ for both SO and S4. For RS masking we assumed that a joint mask removing all sources detected at any frequencies with a hole radius of $\sim1\theta_{1/2}$ is applied to both temperature and polarization. Considering only multipoles $\ell\lesssim 3000$ where extragalactic foreground residuals become important, only S4 would detect the effect above $5\sigma$ in $\tilde{C}_{\ell}^{TT}$ while for SO the detection significance is reduced to $\sim 2\sigma$. Measurements of $\tilde{C}_{\ell}^{EE}$ are less affected by the bias and should remain insensitive to it at SO sensitivity, while S4 will need to account for the effect as it should be able to measure it at $\sim 3\sigma$. The size of the bias for $\tilde{C}_{\ell}^{BB}$ is highly dependent on the choice of the estimator. Standard pseudo-$C_\ell$ estimators not accounting for E-to-B leakage due to partial sky coverage in the E-B separation will lead to a very significant detection of the bias also on subdegree scales. However, estimators that remove the E-to-B leakage, such as the pure-pseudo-$C_\ell$, can remove the majority of the bias and leave the residual effect below the detection threshold. Except on scales smaller than the smoothing scale for the mask, the effect on the temperature and E-mode power spectra is mainly an increase in power at small angular scales.
This is likely to partly degenerate with the spectral index $n_{\rm s}$ and other parameters affecting the damping tail, so any analysis neglecting the effect may misestimate these parameters. We stress however that for a given noise level, the quantitative impact of the mask bias in the analysis of real data is ultimately dependent on the details of the final analysis mask and thus on the interplay between the shape of the bias and cosmological, foreground and other nuisance parameters.

\section{Conclusions}\label{sec:conclusions}
We have shown that masks that are correlated to lensing can potentially give large biases in pseudo-$C_\ell$ power spectrum estimators, even if the masked sky area is small. To a good approximation, this results from a scale-dependent demagnification causing an efficient transfer of power from large to small scales. We discussed analytic models which accurately describe the effect of simple masks, and provided a recipe to estimate the bias empirically on simulations or data. We verified on simulations that the predicted change in the CMB power spectra is accurately capturing the main effect of the mask bias, with no significant change to the power spectrum covariances identifiable above the Monte Carlo noise. For current data, where masked source densities are relatively low and CIB and tSZ are usually not masked, the bias appears to be safely negligible. For future data, with much larger populations of resolved sources, care will be required to either include the correlated mask bias in the model, or ensure that mask hole sizes and number densities are sufficiently low that the bias remains negligible.

The bias from masking radio sources is relatively low because the Poisson sampling ensures a mask hole population tracing the background galaxy density, rather than correlating strongly with the density perturbations. However, for the high radio source densities expected in fourth-generation CMB observations, this may also start to become marginally important. If tSZ clusters (or CIB peaks) are included in the mask, the effect could be much larger, producing highly significant biases in the power spectra if left unmodelled. For these contaminants foreground modelling and cleaning is likely to remain the best approach, rather than masking. However, non-Gaussianity and lensing studies that choose to mask these sources may have to also carefully account for the induced change in the power spectrum over the remaining area. We discuss in detail the effect on lensing estimation in our companion paper~\cite{Lembo:MaskingPaper}. A detailed study of the impact on large-scale CMB polarization and delensing is left for future work.

\section*{Acknowledgments}
We thank Anthony Challinor, Giuseppe Puglisi, Kevin Huffenberger, Sigurd Naess, Marina Migliaccio for useful discussions and the \websky team for providing the catalogues of radio sources used in this work. A.L., G.F. and J.C. acknowledge support from the European Research Council under the European Union's Seventh Framework Programme (FP/2007-2013) / ERC Grant Agreement No. [616170], and support by the UK STFC grants No. ST/P000525/1 (A.L.) and No. ST/T000473/1 (A.L. and G.F.). J.C. acknowledges support from a SNSF Eccellenza Professorial Fellowship (No. 186879). Some of the results in this paper have been derived using the healpy/HEALPix package \cite{Zonca2019,healpix}, and NumPy \cite{2020NumPy-Array}, SciPy \cite{2020SciPy-NMeth} and Matplotlib libraries \cite{Hunter:2007}.

\appendix

\section{Correlation function estimators and averages with binary masks} \label{app:averages}
If we define the correlation function on a masked sky by the expectation between unmasked points (assuming a binary mask) we have
\begin{eqnarray}
\tilde\xi(\vx,\vx')
&\equiv& \int \ud \tilde T(\vx) \ud \tilde T(\vx' ) \, \tilde{T}(\vx)\tilde{T}(\vx')\nonumber \\
&&\quad \times  P(\tilde T(\vx),\tilde T(\vx')|\w(\vx)=1,\w(\vx')=1) \nonumber
\\
&=&\int \ud \tilde T(\vx) \ud \tilde T(\vx' )  \left[\tilde{T}(\vx)\w(\vx)\tilde{T}(\vx')\w(\vx')\right]\nonumber\\
&& \times  P(\tilde T(\vx),\tilde T(\vx')|\w(\vx)=1,\w(\vx')=1).\qquad
\end{eqnarray}
For a mask defined on statistically isotropic fields, for $\vx$ and $\vx'$ separated by $r$ we have
\begin{align}
\n(r) &\equiv \langle \w(\vx)\w(\vx')\rangle \nonumber\\
  &= \int \ud W(\vx) \ud W(\vx') W(\vx)W(\vx') P(\w(\vx),\w(\vx)) \nonumber \\
  & = P(\w(\vx)=1,\w(\vx')=1).
\end{align}
The correlation function for unmasked points then becomes
\begin{align}
\tilde\xi(r) &=\int \ud \tilde T(\vx) \ud \tilde T(\vx' ) \left[ \tilde{T}(\vx)\w(\vx)\tilde{T}(\vx')\w(\vx')\right]\nonumber\\
&\qquad \times  \frac{P(\tilde T(\vx),\tilde T(\vx'),\w(\vx)=1,\w(\vx')=1)}{P(\w(\vx)=1,\w(\vx')=1)}
\nonumber\\
&= \frac{1}{\n(r)}\left\langle \tilde{T}(\vx)\w(\vx)\tilde{T}(\vx')\w(\vx')\right\rangle,
\label{meanform}
\end{align}
This is the same as the pseudocorrelation function for the full masked sky normalized by the mask correlation function.

From a single masked sky of data we can estimate the correlation function by an average over the unmasked sky
\begin{align}
\hat{\tilde\xi}(r) & \equiv \left\langle \tilde T(\vx) \tilde T(\vx + \vr)\right\rangle_{\vx, \phi_r, \rm{unmasked}} \nonumber\\
&=\frac{ \left\langle (\w \tilde T)(\vx)  (\w \tilde T)(\vx + \vr)\right\rangle_{\vx, \phi_r,{\rm all}}}{{\left\langle \w(\vx) \w(\vx + \vr) \right\rangle}_{\vx, \phi_r, {\rm all}}},
\label{ratioest}
\end{align}
where angle brackets here denote sums over pairs of points on a fixed sky, mask and area divided by the number of pairs of points in that area.
Since $\la \hat{\tilde\xi}(r) \ra =\tilde\xi(r)$, the expectation of this estimator is also given by Eq.~\eqref{meanform}. So for a binary mask, the expectation of the ratio in Eq.~\eqref{ratioest} is the same as the ratio of the expectations in Eq.~\eqref{meanform}.

\section{Nonperturbative and exact results}
\label{app:nonpert}
We can decompose the difference in the deflection angles $\vDelta\equiv \valpha-\valpha'$ at two points $\vx$ and $\vx'$, into a part correlated with $f(\vx)$, $f(\vx')$ and a part that is not, $\vn$,
\begin{equation}
\alpha_i-\alpha_i' = n_i - \xi^{f\alpha_i}(r)\frac{(f(\vx) + f(\vx'))}{\sigma_f^2 + \xi_f(r)}.
\label{eq:alphasplit}
\end{equation}
From Eq.~\eqref{lensed_zeta} this gives
\begin{multline}
\ximasked(r) = \int \frac{\ud^2 \vl}{(2\pi)^2} C_l e^{i\vl\cdot \vr}\left\langle e^{i\vl\cdot\vn}\right\rangle \\
\left\langle \exp\left(-i \vl\cdot\vrhat \frac{\xi^{f\alpha_r}(f(\vx)+f(\vx'))}{\sigma_f^2+\xi_f(r)} \right)\w(\vx)\w(\vx')\right\rangle,
\label{fullexpect}
\end{multline}
where the second average is now only a 2D integral over the foreground field values. Note that
\begin{equation}
\langle n_i n_j\rangle = \langle \Delta_i\Delta_j\rangle  - 2\frac{\xi^{f\alpha_i}(r)\xi^{f\alpha_j}(r)}{\sigma_f^2+\xi_f(r)},
\end{equation}
so that
\begin{equation}
\left\langle e^{i\vl\cdot\vn}\right\rangle =   e^{-\half\left\langle(\vl\cdot\vn)^2\right\rangle} =
e^{-\half\langle (\vl\cdot \vDelta)^2\rangle} \exp\left(\frac{\left(\vl\cdot \vrhat\xi^{f\alpha_r}(r)\right)^2}{\sigma_f^2+\xi_f(r)}\right).
\end{equation}
The remaining complex exponent on the second line of Eq.~\eqref{fullexpect} is small, since $l\xi^{f\alpha_r}/\sqrt{\sigma_f^2+\xi_f(r)} \ll 1$ for cases of interest at $l\ll 10^{4}$, suggesting a leading-order expansion should be accurate.

\begin{figure*}[t!]
\centering
	\includegraphics[width=\textwidth]{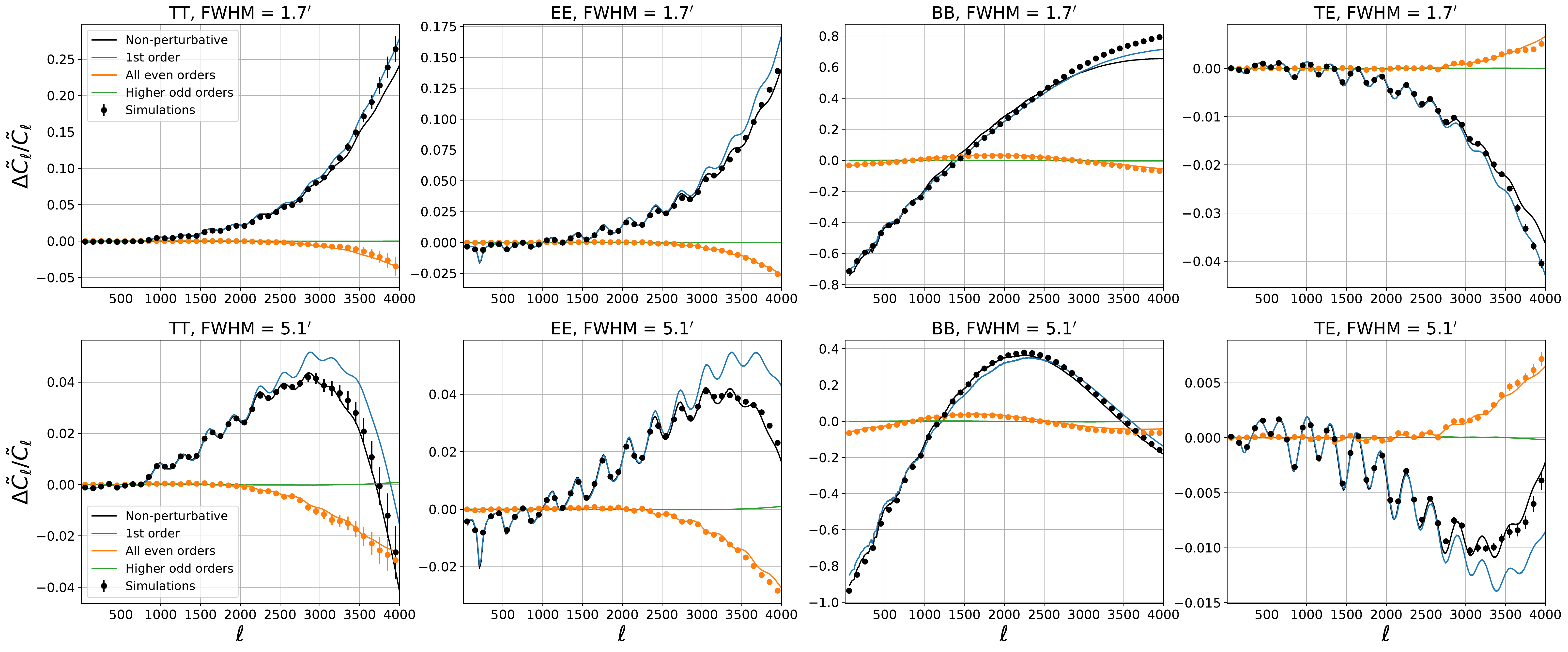}
	\caption{\label{fig:exactnu}Total prediction for the fractional bias on the CMB power spectra (black) together with perturbative contributions of order 1 to 3 in $C_\ell^{f \alpha}$ (blue, orange and green), for an $f_{\rm sky} = 97.7\%$ mask built thresholding $f$, the convergence map smoothed with a beam of $1.7\,\arcmin$ (top, SO-like) or $5.0\,\arcmin$ (bottom, Planck-like).
	The predictions are for deconvolved pseudo power spectra, where the E-modes dominates both polarization spectra biases. E/B-separated or more optimal polarization estimators would produce a much reduced BB bias. In the case of TE, instead of showing the fractional deviation we plot $\Delta \tilde{C}_\ell^{TE} / \sqrt{\tilde{C}_\ell^{TT}\tilde{C}_\ell^{EE}}$, which is the relevant ratio for CMB likelihoods. Note that on small scales $\tilde C_\ell^{TE}$ is mainly negative, so the oppositely signed bias term has qualitatively the same effect on the relative size of the signal as for the autospectra.
In all curves the Gaussian lensing effects are exactly accounted for. The biases measured in simulations are shown with a dot marker. The orange points that isolate even higher-order effects have been computed using the method outlined in Sec.~\ref{sec:tsz}.}
\end{figure*}

Expanding perturbatively to lowest order in $\xi^{f\alpha_r}$ and using
\begin{multline}
\left\langle (f(\vx)+f(\vx'))\w(\vx)\w(\vx')\right\rangle = \\
  \left(\sigma_f^2 +\xi_f(r)\right) \left\langle \frac{\partial \w(\vx)} {\partial f(\vx)} \w(\vx') + \frac{\partial \w(\vx')} {\partial f(\vx')} \w(\vx)\right\rangle
\end{multline}
gives
\begin{align}
\Delta\ximasked(r)
\approx -2\frac{\chi(r)}{r} \xi^{f\alpha_r} \left\langle \frac{\partial \w(\vx)} {\partial f(\vx)} \w(\vx')\right\rangle
\label{eq:exactlinear}
\end{align}
where
\begin{equation}
\chi(r) = \int \frac{\ud^2 \vl}{(2\pi)^2} C_l e^{i\vl\cdot \vr}(i\vr \cdot\vl)e^{-\half\langle (\vl\cdot \vDelta)^2\rangle}.
\end{equation}
This can be evaluated as for standard lensed correlation functions, where $\chi(r)$ is as defined in Eq. C1 of Ref.~\cite{Lewis:2011fk}, related to  $\tilde C_l^{T\nabla T}$ by
\begin{align}
\frac{\chi(r)}{r} &= \vrhat \cdot \langle \widetilde{\nabla T(\vx)}\widetilde{T}(\vx')\rangle \\
&= -\int \ud l \frac{l^2 \tilde C^{T\nabla T}_i}{2\pi}  J_1(lr).
\end{align}
Equation \eqref{eq:exactlinear} is a version of Eq.~\eqref{eq:generalmeanf} that is exact to linear order in $\xi^{f\alpha_r}$.
In the limit of no lensing, $\chi(r) \rightarrow r\partial_r \xi(r)$.
The gradient spectrum $C^{T\nabla T}_i$ is close to the standard lensed CMB power spectrum except on the smallest scales; since the mask correction on the most relevant scales mainly transfers larger-scale power to smaller scales, it is a also a good approximation to just use the lensed correlation function, taking $\chi(r) \approx r\partial_r \tilde\xi(r)$ as in the main text.

In the case of a simple threshold mask, we can further simplify Eq.~\eqref{fullexpect} and put it in a form suitable for numerical evaluation. The expectation in the second line only depends on the sum of the two foregrounds, while their difference is unconstrained by the mask definition. This motivates transforming to the Gaussian independent variables $f_\pm \equiv \frac{(f(\vx)\pm f(\vx'))}{\sigma_f \sqrt{2} }$, with full sky variances $\sigma^2_{f_\pm} = 1 \pm \xi_f / \sigma^2_f$. After masking, the constraints $f(\vx) < \nu \sigma_f$ and  $f(\vx') < \nu \sigma_f$ leave $f_{-}$ unconstrained but $f_+ < \sqrt{2} \nu - |f_-|$. The $f_+$ integral results in a complex error function, giving
\begin{align}
	&\ximasked(r)  = \int \frac{\ud^2\vl}{(2\pi)^2} C_{\vl} \:e^{i \vl\cdot \vr -\half\langle (\vl\cdot \vDelta)^2\rangle} \\
	\cdot &\left[1-\int_0^\infty \frac{\ud t}{\pi} e^{-t^2} \textrm{erfc}\left(\frac{\nu + i (\vl\cdot\hat \vr)\xi^{f\alpha_r}(r) / \sigma_f - t \sigma_{f_-}(r)}{\sigma_{f_+}(r)}\right)\right] \nonumber
\end{align}
The integrand is very smooth, and the derivatives of the complementary error function are exceedingly simple. Hence, this equation can be used to get the exact result for the masked lensed correlation function, or look at the contributions order by order in $\xi^{f\alpha_r}(r)$. This is shown on Fig.~\ref{fig:exactnu}, with the conclusion that the linear approximation of the main text is accurate except at the highest multipoles.

For Poisson sources, the expectations in Eq.~\eqref{fullexpect} are also easily evaluated, but the mask bias is generally very small anyway and the linear term basically exact for all practical purposes. Similar results could be derived for more general cases, for example constructing masks based on multiple different foreground fields, or forming cross-spectra between maps with different masks.

\subsection{Curved-sky expressions}\label{app:curved}
Finally, we give the curved-sky formulation of the biases, in the approximation leading to Eq.~\ref{eq:generalmeanf}. These expressions also provide for convenient implementations since they are very fast to evaluate and free of any flat-to-curved sky remapping ambiguities. To obtain the corresponding result, it is convenient to work in the spin-weight formalism, where the CMB response to lensing~\cite{Challinor:2002cd} can be written in terms of the spin-1 deflection field $_1\alpha$ to leading order as
\begin{equation}
\tilde T(\hat n) \approx T(\hat n) - \frac 12 \left( _{1} \alpha (\hat n)\bar \eth T(\hat n ) + \:_{-1}\alpha(\hat n)  \eth T(\hat n) \right),
\end{equation}
where $\eth$ and $\bar \eth$ (or $\eth^{+}$ and $\eth^{-}$ in what follows) are the spin-raising and spin-lowering operators. Expanding, using
\begin{equation}
	\eth^{\pm} \:_{s}Y_{\ell m} = \pm \sqrt{(\ell \mp s)(\ell \pm s + 1)}\:_{s \pm 1}Y_{\ell m},
\end{equation}
and replacing the unlensed CMB spectrum by the lensed spectrum and the flat-sky distance $r$ by the angular distance $\beta$,
one gets
\begin{align}
\Delta\tilde \xi(\beta)&
\approx  -g(\beta)\eth \tilde{\xi}(\beta) \xi^{\bar\eth \phi f}(\beta) \\ \label{eq:xi2cl_cs}
\Delta \tilde C_\ell &= 2\pi \int_{-1}^1 \ud \cos \beta \:\Delta\tilde \xi(\beta)\: d^\ell_{00}(\beta),
\end{align}
with
\begin{align} \label{eq:dxi_cs}
	\eth\tilde\xi(\beta) &\equiv \sum_{\ell} \left( \frac{2\ell + 1}{4\pi} \right)\sqrt{\ell(\ell + 1)}\: \tilde C_\ell \:d^\ell_{1 0}(\beta)
	\\
	\xi_f(\beta) &\equiv \sum_{\ell} \left( \frac{2\ell + 1}{4\pi} \right)\: C^f_\ell d^\ell_{0 0}(\beta)
	\\
	\xi^{\bar \eth \phi f}(\beta) &\equiv  -\sum_{\ell} \left( \frac{2\ell + 1}{4\pi} \right) \sqrt{\ell(\ell + 1)}\: C^{f\phi}_\ell \:d^\ell_{-1 0}(\beta).
\end{align}
For polarization, Eq.~\eqref{eq:dxi_cs} must be changed to the corresponding derivative of $\xi_+, \xi_-$ or $\xi_\times$, and the spins in the transform Eq.~\eqref{eq:xi2cl_cs} must be adapted accordingly.
\section{Apodization}\label{app:apo}
In practice, a sharply defined mask will be apodized to reduce harmonic-space mixing. We can attempt to include this in our analytic model by considering a mask built by the convolution of a binary mask with an apodization function with a well-defined scale. For example, for a desired apodization length $a$, one may build an apodized mask as follows: first, the mask is extended by $a/2$ and second this extended mask is convolved with an apodization function with support extending to $a/2$. This ensures that all masked points remain masked after convolution, and that the new mask transitions smoothly beyond the edges. This differs somewhat from the most common ways of apodizing a mask in CMB analysis, where a smooth function of the distance to the nearest pixel is applied to the unmasked pixels. However, in the case of disks masks centred on sources, the apodization function can be tuned to match the resulting mask profile. Slight differences might remain in regions close to two disks, but empirically our approximate analytic procedure is working well. For other masks, such as the threshold masks, it is difficult to treat analytically the mask expansion and this prescription remains very crude.

This procedure leads to some minimal changes in the pseudo-$C_\ell$ prediction that we describe now. Let $\w_s(\vx)$ be the sharp binary mask, with corresponding deflection-mask correlators $\left\langle{\alpha_r(\vx)\w_s(\vx)\w_s(\vx')}\right\rangle$. Convolving with an apodization function $a_p(\vx)$, we must evaluate this now as a function of three arguments, say $\left\langle{\alpha_r(\vx)\w_s(\vy)\w_s(\vy')}\right\rangle$, where $\vy$ is close to $\vx$. In simple cases (as for the threshold or other masks defined through the local value of a Gaussian foreground field), we may always write the exact relation $\left\langle{\alpha_r(\vx)\w_s(\vy)\w_s(\vy')}\right\rangle = g(r)\left(\xi^{\alpha_r f}(\vx - \vy) + \xi^{\alpha_r f}(\vx - \vy') \right)$, for some function $g(r = |\vy'-\vy|)$. Let the $\star$ symbol denote a convolution (a multiplication in harmonic space), and the $\cdot $ symbol a pointwise product in real space. Then the separation change becomes
\begin{align} \label{apeq:aww}
	& \left\langle{(\alpha_r(\vx)-\alpha_r(\vx')) (a_p\star \w_s)(\vx)(a_p\star\w_s)(\vx')}\right\rangle \\&=2\left(\left( \xi^{\alpha_r f} \cdot \left(g \star a_p\right)\right) \star a_p \right)(r) \nonumber \\&+2\left(\left( \xi^{\alpha_r f} \cdot a_p\right) \star \left(g \star a_p \right)\right)(r). \nonumber
\end{align}
For no apodization ($a_p(\vx) = \delta^D(\vx)$), the second term vanishes (since $ \xi^{\alpha_r f}(0) = 0)$ and we recover the result of the main text. To get the bias, we also need the apodized mask correlation function $\left\langle \w(\vx) \w(\vy) \right \rangle = \n(r)$. This is simply
\begin{equation}\label{apeq:ww}
	\left\langle (a_p \star \w_s)(\vx) (a_p \star \w_s)(\vy)\right\rangle  = \left(a_p \star \n_{s} \star a_p\right)(r).
\end{equation}
\newcommand{\D}[0]{{\mathcal{D}_{R}}}
Equation \eqref{apeq:aww} is exact for locally defined masks. In the Poisson case, it holds only for $r > 2R$, where $R$ is the sharp point source mask radius, in which case the disks do not overlap.
For $r < 2R $ we can proceed as follows. If $\D(r)$ is the indicator function of the disk of radius $R$ (being unity inside and zero outside), the indicator function of the area drawn by two disks at $\vy$ and $\vy'$ may be defined by $\D(\vx - \vy)  + \D(\vx - \vy') - \D(\vx - \vy) \D(\vx - \vy')$.  The first two terms (that is, the case neglecting the overlap) can be treated exactly with Eq.~\eqref{apeq:aww}. The exact form of the deflection-mask correlator $\left\langle{\alpha_r(\vx)\w(\vx)\w(\vx')}\right\rangle$ of the overlap is
\begin{align}
	&a_p(\vx -\vy)a_p(\vx' -\vy')g(\vy - \vy')\\ \cdot &\xi^{\alpha_r  \lambda}(\vx-\vs) \D(\vs - \vy)  \D(\vs - \vy'),
\end{align}
where there is implicit integration over all positions except $\vx$ and $\vx'$.
The function $g$ is very smooth and varies very little (for the Poisson case, $g(r)$ is simply $\n_s(r)$, which slowly transitions from $f_{\rm{sky}}$ to $f^2_{\rm{sky}}$), and $\vy, \vy'$ are at most an apodization length away from $\vx$ and $\vy$ respectively. This motivates expanding $g(\vy - \vy')$
    around $\vr = \vx - \vx'$. This expansion produces terms only involving real-space and convolution products, and hence can easily be evaluated numerically. The leading term is simply
    \begin{align}
    &\left\langle{\alpha_r(\vx)\w_{\rm ap}(\vx)\w_{\rm ap}(\vx')}\right\rangle \textrm{   (Poisson, overlap term)}\\&\simeq -
    	g(r) \cdot \left[ \left(\xi^{\alpha_r \lambda} \cdot (\D \star a_p) \right) \star (\D \star a_p) \right](r).
    \end{align}
Here $\D \star a_p$ is the profile of the apodized disk mask. As expected we found the first correction to this, proportional to $\partial_r g(r)$, to be negligible for realistic sky fractions close to unity, even for apodization length comparable to or greater than the disk size.

\bibliography{texbase/antony,texbase/cosmomc,julbib,maskbias}

\providecommand{\aj}{Astron. J. }\providecommand{\apj}{ApJ
  }\providecommand{\apjl}{ApJ
  }\providecommand{\mnras}{MNRAS}\providecommand{\prl}{PRL}\providecommand{\prd}{PRD}\providecommand{\jcap}{JCAP}\providecommand{\aap}{A\&A}
\begin{thebibliography}{86}%
\makeatletter
\providecommand \@ifxundefined [1]{%
 \@ifx{#1\undefined}
}%
\providecommand \@ifnum [1]{%
 \ifnum #1\expandafter \@firstoftwo
 \else \expandafter \@secondoftwo
 \fi
}%
\providecommand \@ifx [1]{%
 \ifx #1\expandafter \@firstoftwo
 \else \expandafter \@secondoftwo
 \fi
}%
\providecommand \natexlab [1]{#1}%
\providecommand \enquote  [1]{``#1''}%
\providecommand \bibnamefont  [1]{#1}%
\providecommand \bibfnamefont [1]{#1}%
\providecommand \citenamefont [1]{#1}%
\providecommand \href@noop [0]{\@secondoftwo}%
\providecommand \href [0]{\begingroup \@sanitize@url \@href}%
\providecommand \@href[1]{\@@startlink{#1}\@@href}%
\providecommand \@@href[1]{\endgroup#1\@@endlink}%
\providecommand \@sanitize@url [0]{\catcode `\\12\catcode `\$12\catcode
  `\&12\catcode `\#12\catcode `\^12\catcode `\_12\catcode `\%12\relax}%
\providecommand \@@startlink[1]{}%
\providecommand \@@endlink[0]{}%
\providecommand \url  [0]{\begingroup\@sanitize@url \@url }%
\providecommand \@url [1]{\endgroup\@href {#1}{\urlprefix }}%
\providecommand \urlprefix  [0]{URL }%
\providecommand \Eprint [0]{\href }%
\providecommand \doibase [0]{http://dx.doi.org/}%
\providecommand \selectlanguage [0]{\@gobble}%
\providecommand \bibinfo  [0]{\@secondoftwo}%
\providecommand \bibfield  [0]{\@secondoftwo}%
\providecommand \translation [1]{[#1]}%
\providecommand \BibitemOpen [0]{}%
\providecommand \bibitemStop [0]{}%
\providecommand \bibitemNoStop [0]{.\EOS\space}%
\providecommand \EOS [0]{\spacefactor3000\relax}%
\providecommand \BibitemShut  [1]{\csname bibitem#1\endcsname}%
\let\auto@bib@innerbib\@empty
\bibitem [{\citenamefont {Seljak}(1996)}]{Seljak:1995ve}%
  \BibitemOpen
  \bibfield  {author} {\bibinfo {author} {\bibfnamefont {Uros}\ \bibnamefont
  {Seljak}},\ }\bibfield  {title} {\enquote {\bibinfo {title} {{Gravitational
  lensing effect on cosmic microwave background anisotropies: A Power spectrum
  approach}},}\ }\href {\doibase 10.1086/177218} {\bibfield  {journal}
  {\bibinfo  {journal} {\apj}\ }\textbf {\bibinfo {volume} {463}},\ \bibinfo
  {pages} {1} (\bibinfo {year} {1996})},\ \Eprint
  {http://arxiv.org/abs/astro-ph/9505109} {arXiv:astro-ph/9505109 [astro-ph]}
  \BibitemShut {NoStop}%
\bibitem [{\citenamefont {Challinor}\ and\ \citenamefont
  {Lewis}(2005)}]{Challinor:2005jy}%
  \BibitemOpen
  \bibfield  {author} {\bibinfo {author} {\bibfnamefont {Anthony}\ \bibnamefont
  {Challinor}}\ and\ \bibinfo {author} {\bibfnamefont {Antony}\ \bibnamefont
  {Lewis}},\ }\bibfield  {title} {\enquote {\bibinfo {title} {{Lensed CMB power
  spectra from all-sky correlation functions}},}\ }\href {\doibase
  10.1103/PhysRevD.71.103010} {\bibfield  {journal} {\bibinfo  {journal}
  {\prd}\ }\textbf {\bibinfo {volume} {71}},\ \bibinfo {pages} {103010}
  (\bibinfo {year} {2005})},\ \Eprint {http://arxiv.org/abs/astro-ph/0502425}
  {arXiv:astro-ph/0502425 [astro-ph]} \BibitemShut {NoStop}%
\bibitem [{\citenamefont {Lewis}\ and\ \citenamefont
  {Pratten}(2016)}]{Lewis:2016tuj}%
  \BibitemOpen
  \bibfield  {author} {\bibinfo {author} {\bibfnamefont {Antony}\ \bibnamefont
  {Lewis}}\ and\ \bibinfo {author} {\bibfnamefont {Geraint}\ \bibnamefont
  {Pratten}},\ }\bibfield  {title} {\enquote {\bibinfo {title} {{Effect of
  lensing non-Gaussianity on the CMB power spectra}},}\ }\href {\doibase
  10.1088/1475-7516/2016/12/003} {\bibfield  {journal} {\bibinfo  {journal}
  {\jcap}\ }\textbf {\bibinfo {volume} {1612}},\ \bibinfo {pages} {003}
  (\bibinfo {year} {2016})},\ \Eprint {http://arxiv.org/abs/1608.01263}
  {arXiv:1608.01263 [astro-ph.CO]} \BibitemShut {NoStop}%
\bibitem [{\citenamefont {Manzotti}\ \emph {et~al.}(2014)\citenamefont
  {Manzotti}, \citenamefont {Hu},\ and\ \citenamefont
  {Benoit-L\'evy}}]{Manzotti:2014wca}%
  \BibitemOpen
  \bibfield  {author} {\bibinfo {author} {\bibfnamefont {Alessandro}\
  \bibnamefont {Manzotti}}, \bibinfo {author} {\bibfnamefont {Wayne}\
  \bibnamefont {Hu}}, \ and\ \bibinfo {author} {\bibfnamefont {Aur\'elien}\
  \bibnamefont {Benoit-L\'evy}},\ }\bibfield  {title} {\enquote {\bibinfo
  {title} {{Super-Sample CMB Lensing}},}\ }\href {\doibase
  10.1103/PhysRevD.90.023003} {\bibfield  {journal} {\bibinfo  {journal} {Phys.
  Rev. D}\ }\textbf {\bibinfo {volume} {90}},\ \bibinfo {pages} {023003}
  (\bibinfo {year} {2014})},\ \Eprint {http://arxiv.org/abs/1401.7992}
  {arXiv:1401.7992 [astro-ph.CO]} \BibitemShut {NoStop}%
\bibitem [{\citenamefont {Lembo}\ \emph {et~al.}()\citenamefont {Lembo},
  \citenamefont {Fabbian}, \citenamefont {Carron},\ and\ \citenamefont
  {Lewis}}]{Lembo:MaskingPaper}%
  \BibitemOpen
  \bibfield  {author} {\bibinfo {author} {\bibfnamefont {Margherita}\
  \bibnamefont {Lembo}}, \bibinfo {author} {\bibfnamefont {Giulio}\
  \bibnamefont {Fabbian}}, \bibinfo {author} {\bibfnamefont {Julien}\
  \bibnamefont {Carron}}, \ and\ \bibinfo {author} {\bibfnamefont {Antony}\
  \bibnamefont {Lewis}},\ }\href@noop {} {\enquote {\bibinfo {title} {{CMB}
  lensing reconstruction biases from masking extragalactic sources},}\
  }\bibinfo {note} {In prep.}\BibitemShut {Stop}%
\bibitem [{\citenamefont {Harnois-D\'eraps}\ \emph {et~al.}(2016)\citenamefont
  {Harnois-D\'eraps} \emph {et~al.}}]{harnois-deraps2016}%
  \BibitemOpen
  \bibfield  {author} {\bibinfo {author} {\bibfnamefont {Joachim}\ \bibnamefont
  {Harnois-D\'eraps}} \emph {et~al.},\ }\bibfield  {title} {\enquote {\bibinfo
  {title} {{CFHTLenS and RCSLenS Cross-Correlation with Planck Lensing Detected
  in Fourier and Configuration Space}},}\ }\href {\doibase
  10.1093/mnras/stw947} {\bibfield  {journal} {\bibinfo  {journal} {Mon. Not.
  Roy. Astron. Soc.}\ }\textbf {\bibinfo {volume} {460}},\ \bibinfo {pages}
  {434--457} (\bibinfo {year} {2016})},\ \Eprint
  {http://arxiv.org/abs/1603.07723} {arXiv:1603.07723 [astro-ph.CO]}
  \BibitemShut {NoStop}%
\bibitem [{\citenamefont {Liu}\ and\ \citenamefont {Hill}(2015)}]{Liu:2015xfa}%
  \BibitemOpen
  \bibfield  {author} {\bibinfo {author} {\bibfnamefont {Jia}\ \bibnamefont
  {Liu}}\ and\ \bibinfo {author} {\bibfnamefont {J.~Colin}\ \bibnamefont
  {Hill}},\ }\bibfield  {title} {\enquote {\bibinfo {title} {{Cross-correlation
  of Planck CMB Lensing and CFHTLenS Galaxy Weak Lensing Maps}},}\ }\href
  {\doibase 10.1103/PhysRevD.92.063517} {\bibfield  {journal} {\bibinfo
  {journal} {\prd}\ }\textbf {\bibinfo {volume} {92}},\ \bibinfo {pages}
  {063517} (\bibinfo {year} {2015})},\ \Eprint
  {http://arxiv.org/abs/1504.05598} {arXiv:1504.05598 [astro-ph.CO]}
  \BibitemShut {NoStop}%
\bibitem [{\citenamefont {Lagache}\ \emph {et~al.}(2020)\citenamefont
  {Lagache}, \citenamefont {B\'ethermin}, \citenamefont {Montier},
  \citenamefont {Serra},\ and\ \citenamefont {Tucci}}]{Lagache:2019xto}%
  \BibitemOpen
  \bibfield  {author} {\bibinfo {author} {\bibfnamefont {G.}~\bibnamefont
  {Lagache}}, \bibinfo {author} {\bibfnamefont {M.}~\bibnamefont
  {B\'ethermin}}, \bibinfo {author} {\bibfnamefont {L.}~\bibnamefont
  {Montier}}, \bibinfo {author} {\bibfnamefont {P.}~\bibnamefont {Serra}}, \
  and\ \bibinfo {author} {\bibfnamefont {M.}~\bibnamefont {Tucci}},\ }\bibfield
   {title} {\enquote {\bibinfo {title} {{Impact of polarised extragalactic
  sources on the measurement of CMB B-mode anisotropies}},}\ }\href {\doibase
  10.1051/0004-6361/201937147} {\bibfield  {journal} {\bibinfo  {journal}
  {Astron. Astrophys.}\ }\textbf {\bibinfo {volume} {642}},\ \bibinfo {pages}
  {A232} (\bibinfo {year} {2020})},\ \Eprint {http://arxiv.org/abs/1911.09466}
  {arXiv:1911.09466 [astro-ph.CO]} \BibitemShut {NoStop}%
\bibitem [{\citenamefont {Szapudi}\ \emph {et~al.}(2001)\citenamefont
  {Szapudi}, \citenamefont {Prunet}, \citenamefont {Pogosyan}, \citenamefont
  {Szalay},\ and\ \citenamefont {Bond}}]{Szapudi:2000xj}%
  \BibitemOpen
  \bibfield  {author} {\bibinfo {author} {\bibfnamefont {Istvan}\ \bibnamefont
  {Szapudi}}, \bibinfo {author} {\bibfnamefont {Simon}\ \bibnamefont {Prunet}},
  \bibinfo {author} {\bibfnamefont {Dmitry}\ \bibnamefont {Pogosyan}}, \bibinfo
  {author} {\bibfnamefont {Alexander~S.}\ \bibnamefont {Szalay}}, \ and\
  \bibinfo {author} {\bibfnamefont {J.~Richard}\ \bibnamefont {Bond}},\
  }\bibfield  {title} {\enquote {\bibinfo {title} {Fast cmb analyses via
  correlation functions},}\ }\href@noop {} {\bibfield  {journal} {\bibinfo
  {journal} {\apj Lett.}\ }\textbf {\bibinfo {volume} {548}},\ \bibinfo {pages}
  {115--118} (\bibinfo {year} {2001})},\ \Eprint
  {http://arxiv.org/abs/astro-ph/0010256} {astro-ph/0010256} \BibitemShut
  {NoStop}%
\bibitem [{\citenamefont {Chon}\ \emph {et~al.}(2004)\citenamefont {Chon},
  \citenamefont {Challinor}, \citenamefont {Prunet}, \citenamefont {Hivon},\
  and\ \citenamefont {Szapudi}}]{Chon:2003gx}%
  \BibitemOpen
  \bibfield  {author} {\bibinfo {author} {\bibfnamefont {Gayoung}\ \bibnamefont
  {Chon}}, \bibinfo {author} {\bibfnamefont {Anthony}\ \bibnamefont
  {Challinor}}, \bibinfo {author} {\bibfnamefont {Simon}\ \bibnamefont
  {Prunet}}, \bibinfo {author} {\bibfnamefont {Eric}\ \bibnamefont {Hivon}}, \
  and\ \bibinfo {author} {\bibfnamefont {Istvan}\ \bibnamefont {Szapudi}},\
  }\bibfield  {title} {\enquote {\bibinfo {title} {Fast estimation of
  polarization power spectra using correlation functions},}\ }\href@noop {}
  {\bibfield  {journal} {\bibinfo  {journal} {\mnras}\ }\textbf {\bibinfo
  {volume} {350}},\ \bibinfo {pages} {914} (\bibinfo {year} {2004})},\ \Eprint
  {http://arxiv.org/abs/astro-ph/0303414} {astro-ph/0303414} \BibitemShut
  {NoStop}%
\bibitem [{\citenamefont {Wandelt}\ \emph {et~al.}(2001)\citenamefont
  {Wandelt}, \citenamefont {Hivon},\ and\ \citenamefont
  {Gorski}}]{Wandelt:2000av}%
  \BibitemOpen
  \bibfield  {author} {\bibinfo {author} {\bibfnamefont {Benjamin~D.}\
  \bibnamefont {Wandelt}}, \bibinfo {author} {\bibfnamefont {Eric}\
  \bibnamefont {Hivon}}, \ and\ \bibinfo {author} {\bibfnamefont
  {Krzysztof~M.}\ \bibnamefont {Gorski}},\ }\bibfield  {title} {\enquote
  {\bibinfo {title} {The {Pseudo-$C_l$} method: Cosmic microwave background
  anisotropy power spectrum statistics for high precision cosmology},}\
  }\href@noop {} {\bibfield  {journal} {\bibinfo  {journal} {\prd}\ }\textbf
  {\bibinfo {volume} {64}},\ \bibinfo {pages} {083003} (\bibinfo {year}
  {2001})},\ \Eprint {http://arxiv.org/abs/astro-ph/0008111} {astro-ph/0008111}
  \BibitemShut {NoStop}%
\bibitem [{\citenamefont {Stein}\ \emph {et~al.}(2020)\citenamefont {Stein},
  \citenamefont {Alvarez}, \citenamefont {Bond}, \citenamefont {van Engelen},\
  and\ \citenamefont {Battaglia}}]{Stein:2020its}%
  \BibitemOpen
  \bibfield  {author} {\bibinfo {author} {\bibfnamefont {George}\ \bibnamefont
  {Stein}}, \bibinfo {author} {\bibfnamefont {Marcelo~A.}\ \bibnamefont
  {Alvarez}}, \bibinfo {author} {\bibfnamefont {J.~Richard}\ \bibnamefont
  {Bond}}, \bibinfo {author} {\bibfnamefont {Alexander}\ \bibnamefont {van
  Engelen}}, \ and\ \bibinfo {author} {\bibfnamefont {Nicholas}\ \bibnamefont
  {Battaglia}},\ }\bibfield  {title} {\enquote {\bibinfo {title} {{The Websky
  Extragalactic CMB Simulations}},}\ }\href {\doibase
  10.1088/1475-7516/2020/10/012} {\bibfield  {journal} {\bibinfo  {journal}
  {JCAP}\ }\textbf {\bibinfo {volume} {10}},\ \bibinfo {pages} {012} (\bibinfo
  {year} {2020})},\ \Eprint {http://arxiv.org/abs/2001.08787} {arXiv:2001.08787
  [astro-ph.CO]} \BibitemShut {NoStop}%
\bibitem [{\citenamefont {{De Zotti}}\ \emph {et~al.}(2019)\citenamefont {{De
  Zotti}}, \citenamefont {{Bonato}}, \citenamefont {{Negrello}}, \citenamefont
  {{Trombetti}}, \citenamefont {{Burigana}}, \citenamefont {{Herranz}},
  \citenamefont {{L{\'o}pez-Caniego}}, \citenamefont {{Cai}}, \citenamefont
  {{Bonavera}},\ and\ \citenamefont {{Gonz{\'a}lez-Nuevo}}}]{DeZotti2019}%
  \BibitemOpen
  \bibfield  {author} {\bibinfo {author} {\bibfnamefont {Gianfranco}\
  \bibnamefont {{De Zotti}}}, \bibinfo {author} {\bibfnamefont {Matteo}\
  \bibnamefont {{Bonato}}}, \bibinfo {author} {\bibfnamefont {Mattia}\
  \bibnamefont {{Negrello}}}, \bibinfo {author} {\bibfnamefont {Tiziana}\
  \bibnamefont {{Trombetti}}}, \bibinfo {author} {\bibfnamefont {Carlo}\
  \bibnamefont {{Burigana}}}, \bibinfo {author} {\bibfnamefont {Diego}\
  \bibnamefont {{Herranz}}}, \bibinfo {author} {\bibfnamefont {Marcos}\
  \bibnamefont {{L{\'o}pez-Caniego}}}, \bibinfo {author} {\bibfnamefont
  {Zhen-Yi}\ \bibnamefont {{Cai}}}, \bibinfo {author} {\bibfnamefont {Laura}\
  \bibnamefont {{Bonavera}}}, \ and\ \bibinfo {author} {\bibfnamefont
  {Joaquin}\ \bibnamefont {{Gonz{\'a}lez-Nuevo}}},\ }\bibfield  {title}
  {\enquote {\bibinfo {title} {{Extragalactic astrophysics with next-generation
  CMB experiments}},}\ }\href {\doibase 10.3389/fspas.2019.00053} {\bibfield
  {journal} {\bibinfo  {journal} {Frontiers in Astronomy and Space Sciences}\
  }\textbf {\bibinfo {volume} {6}},\ \bibinfo {eid} {53} (\bibinfo {year}
  {2019})},\ \Eprint {http://arxiv.org/abs/1907.05323} {arXiv:1907.05323
  [astro-ph.GA]} \BibitemShut {NoStop}%
\bibitem [{\citenamefont {Everett}\ \emph {et~al.}(2020)\citenamefont {Everett}
  \emph {et~al.}}]{Everett:2020jcf}%
  \BibitemOpen
  \bibfield  {author} {\bibinfo {author} {\bibfnamefont {W.B.}\ \bibnamefont
  {Everett}} \emph {et~al.} (\bibinfo {collaboration} {SPT}),\ }\bibfield
  {title} {\enquote {\bibinfo {title} {{Millimeter-wave Point Sources from the
  2500-square-degree SPT-SZ Survey: Catalog and Population Statistics}},}\
  }\href {\doibase 10.3847/1538-4357/ab9df7} {\bibfield  {journal} {\bibinfo
  {journal} {Astrophys. J.}\ }\textbf {\bibinfo {volume} {900}},\ \bibinfo
  {pages} {55} (\bibinfo {year} {2020})},\ \Eprint
  {http://arxiv.org/abs/2003.03431} {arXiv:2003.03431 [astro-ph.IM]}
  \BibitemShut {NoStop}%
\bibitem [{\citenamefont {Gralla}\ \emph {et~al.}(2019)\citenamefont {Gralla}
  \emph {et~al.}}]{gralla2020}%
  \BibitemOpen
  \bibfield  {author} {\bibinfo {author} {\bibfnamefont {Megan~B.}\
  \bibnamefont {Gralla}} \emph {et~al.},\ }\bibfield  {title} {\enquote
  {\bibinfo {title} {{Atacama Cosmology Telescope: Dusty star-forming galaxies
  and active galactic nuclei in the equatorial survey}},}\ }\href {\doibase
  10.3847/1538-4357/ab7915} {\  (\bibinfo {year} {2019}),\
  10.3847/1538-4357/ab7915},\ \Eprint {http://arxiv.org/abs/1905.04592}
  {arXiv:1905.04592 [astro-ph.GA]} \BibitemShut {NoStop}%
\bibitem [{\citenamefont {Tegmark}\ and\ \citenamefont
  {Villumsen}(1997)}]{Tegmark:1996ze}%
  \BibitemOpen
  \bibfield  {author} {\bibinfo {author} {\bibfnamefont {Max}\ \bibnamefont
  {Tegmark}}\ and\ \bibinfo {author} {\bibfnamefont {Jens~V.}\ \bibnamefont
  {Villumsen}},\ }\bibfield  {title} {\enquote {\bibinfo {title} {{Is lensing
  of point sources a problem for future CMB experiments?}}}\ }\href {\doibase
  10.1093/mnras/289.1.169} {\bibfield  {journal} {\bibinfo  {journal} {Mon.
  Not. Roy. Astron. Soc.}\ }\textbf {\bibinfo {volume} {289}},\ \bibinfo
  {pages} {169--174} (\bibinfo {year} {1997})},\ \Eprint
  {http://arxiv.org/abs/astro-ph/9608173} {arXiv:astro-ph/9608173} \BibitemShut
  {NoStop}%
\bibitem [{\citenamefont {Matsubara}(2000)}]{Matsubara:2000pr}%
  \BibitemOpen
  \bibfield  {author} {\bibinfo {author} {\bibfnamefont {Takahiko}\
  \bibnamefont {Matsubara}},\ }\bibfield  {title} {\enquote {\bibinfo {title}
  {{The Gravitational Lensing in Redshift-space Correlation Functions of
  Galaxies and Quasars}},}\ }\href@noop {} {\bibfield  {journal} {\bibinfo
  {journal} {\apj Lett.}\ }\textbf {\bibinfo {volume} {537}},\ \bibinfo {pages}
  {77--80} (\bibinfo {year} {2000})},\ \Eprint
  {http://arxiv.org/abs/astro-ph/0004392} {arXiv:astro-ph/0004392} \BibitemShut
  {NoStop}%
\bibitem [{\citenamefont {{Stein}}\ \emph {et~al.}(2019)\citenamefont
  {{Stein}}, \citenamefont {{Alvarez}},\ and\ \citenamefont
  {{Bond}}}]{Stein2019}%
  \BibitemOpen
  \bibfield  {author} {\bibinfo {author} {\bibfnamefont {George}\ \bibnamefont
  {{Stein}}}, \bibinfo {author} {\bibfnamefont {Marcelo~A.}\ \bibnamefont
  {{Alvarez}}}, \ and\ \bibinfo {author} {\bibfnamefont {J.~Richard}\
  \bibnamefont {{Bond}}},\ }\bibfield  {title} {\enquote {\bibinfo {title}
  {{The mass-Peak Patch algorithm for fast generation of deep all-sky dark
  matter halo catalogues and its N-body validation}},}\ }\href {\doibase
  10.1093/mnras/sty3226} {\bibfield  {journal} {\bibinfo  {journal} {\mnras}\
  }\textbf {\bibinfo {volume} {483}},\ \bibinfo {pages} {2236--2250} (\bibinfo
  {year} {2019})},\ \Eprint {http://arxiv.org/abs/1810.07727} {arXiv:1810.07727
  [astro-ph.CO]} \BibitemShut {NoStop}%
\bibitem [{\citenamefont {{Bond}}\ and\ \citenamefont
  {{Myers}}(1996)}]{peak-patch}%
  \BibitemOpen
  \bibfield  {author} {\bibinfo {author} {\bibfnamefont {J.~R.}\ \bibnamefont
  {{Bond}}}\ and\ \bibinfo {author} {\bibfnamefont {S.~T.}\ \bibnamefont
  {{Myers}}},\ }\bibfield  {title} {\enquote {\bibinfo {title} {{The Peak-Patch
  Picture of Cosmic Catalogs. I. Algorithms}},}\ }\href {\doibase
  10.1086/192267} {\bibfield  {journal} {\bibinfo  {journal} {\apjsup}\
  }\textbf {\bibinfo {volume} {103}},\ \bibinfo {pages} {1} (\bibinfo {year}
  {1996})}\BibitemShut {NoStop}%
\bibitem [{\citenamefont {{Alonso}}\ \emph {et~al.}(2019)\citenamefont
  {{Alonso}}, \citenamefont {{Sanchez}}, \citenamefont {{Slosar}},\ and\
  \citenamefont {{LSST Dark Energy Science Collaboration}}}]{namaster}%
  \BibitemOpen
  \bibfield  {author} {\bibinfo {author} {\bibfnamefont {David}\ \bibnamefont
  {{Alonso}}}, \bibinfo {author} {\bibfnamefont {Javier}\ \bibnamefont
  {{Sanchez}}}, \bibinfo {author} {\bibfnamefont {An{\v{z}}e}\ \bibnamefont
  {{Slosar}}}, \ and\ \bibinfo {author} {\bibnamefont {{LSST Dark Energy
  Science Collaboration}}},\ }\bibfield  {title} {\enquote {\bibinfo {title}
  {{A unified pseudo-C$_{{\ensuremath{\ell}}}$ framework}},}\ }\href {\doibase
  10.1093/mnras/stz093} {\bibfield  {journal} {\bibinfo  {journal} {\mnras}\
  }\textbf {\bibinfo {volume} {484}},\ \bibinfo {pages} {4127--4151} (\bibinfo
  {year} {2019})},\ \Eprint {http://arxiv.org/abs/1809.09603} {arXiv:1809.09603
  [astro-ph.CO]} \BibitemShut {NoStop}%
\bibitem [{\citenamefont {Viero}\ \emph {et~al.}(2013)\citenamefont {Viero}
  \emph {et~al.}}]{viero2013}%
  \BibitemOpen
  \bibfield  {author} {\bibinfo {author} {\bibfnamefont {M.P.}\ \bibnamefont
  {Viero}} \emph {et~al.},\ }\bibfield  {title} {\enquote {\bibinfo {title}
  {{HerMES: Cosmic Infrared Background Anisotropies and the Clustering of Dusty
  Star-Forming Galaxies}},}\ }\href {\doibase 10.1088/0004-637X/772/1/77}
  {\bibfield  {journal} {\bibinfo  {journal} {Astrophys. J.}\ }\textbf
  {\bibinfo {volume} {772}},\ \bibinfo {pages} {77} (\bibinfo {year} {2013})},\
  \Eprint {http://arxiv.org/abs/1208.5049} {arXiv:1208.5049 [astro-ph.CO]}
  \BibitemShut {NoStop}%
\bibitem [{\citenamefont {Ade}\ \emph {et~al.}(2014{\natexlab{a}})\citenamefont
  {Ade} \emph {et~al.}}]{Ade:2013aro}%
  \BibitemOpen
  \bibfield  {author} {\bibinfo {author} {\bibfnamefont {P.A.R.}\ \bibnamefont
  {Ade}} \emph {et~al.} (\bibinfo {collaboration} {Planck}),\ }\bibfield
  {title} {\enquote {\bibinfo {title} {{Planck 2013 results. XVIII. The
  gravitational lensing-infrared background correlation}},}\ }\href {\doibase
  10.1051/0004-6361/201321540} {\bibfield  {journal} {\bibinfo  {journal}
  {Astron. Astrophys.}\ }\textbf {\bibinfo {volume} {571}},\ \bibinfo {pages}
  {A18} (\bibinfo {year} {2014}{\natexlab{a}})},\ \Eprint
  {http://arxiv.org/abs/1303.5078} {arXiv:1303.5078 [astro-ph.CO]} \BibitemShut
  {NoStop}%
\bibitem [{\citenamefont {Ade}\ \emph {et~al.}(2016{\natexlab{a}})\citenamefont
  {Ade} \emph {et~al.}}]{planck2015-cibXtsz}%
  \BibitemOpen
  \bibfield  {author} {\bibinfo {author} {\bibfnamefont {P.A.R.}\ \bibnamefont
  {Ade}} \emph {et~al.} (\bibinfo {collaboration} {Planck}),\ }\bibfield
  {title} {\enquote {\bibinfo {title} {{Planck 2015 results. XXIII. The thermal
  Sunyaev-Zeldovich effect--cosmic infrared background correlation}},}\ }\href
  {\doibase 10.1051/0004-6361/201527418} {\bibfield  {journal} {\bibinfo
  {journal} {Astron. Astrophys.}\ }\textbf {\bibinfo {volume} {594}},\ \bibinfo
  {pages} {A23} (\bibinfo {year} {2016}{\natexlab{a}})},\ \Eprint
  {http://arxiv.org/abs/1509.06555} {arXiv:1509.06555 [astro-ph.CO]}
  \BibitemShut {NoStop}%
\bibitem [{\citenamefont {Aghanim}\ \emph
  {et~al.}(2016{\natexlab{a}})\citenamefont {Aghanim} \emph
  {et~al.}}]{planck2015-sz}%
  \BibitemOpen
  \bibfield  {author} {\bibinfo {author} {\bibfnamefont {N.}~\bibnamefont
  {Aghanim}} \emph {et~al.} (\bibinfo {collaboration} {Planck}),\ }\bibfield
  {title} {\enquote {\bibinfo {title} {{Planck 2015 results. XXII. A map of the
  thermal Sunyaev-Zeldovich effect}},}\ }\href {\doibase
  10.1051/0004-6361/201525826} {\bibfield  {journal} {\bibinfo  {journal}
  {Astron. Astrophys.}\ }\textbf {\bibinfo {volume} {594}},\ \bibinfo {pages}
  {A22} (\bibinfo {year} {2016}{\natexlab{a}})},\ \Eprint
  {http://arxiv.org/abs/1502.01596} {arXiv:1502.01596 [astro-ph.CO]}
  \BibitemShut {NoStop}%
\bibitem [{\citenamefont {Song}\ \emph {et~al.}(2003)\citenamefont {Song},
  \citenamefont {Cooray}, \citenamefont {Knox},\ and\ \citenamefont
  {Zaldarriaga}}]{Song:2002sg}%
  \BibitemOpen
  \bibfield  {author} {\bibinfo {author} {\bibfnamefont {Yong-Seon}\
  \bibnamefont {Song}}, \bibinfo {author} {\bibfnamefont {Asantha}\
  \bibnamefont {Cooray}}, \bibinfo {author} {\bibfnamefont {Lloyd}\
  \bibnamefont {Knox}}, \ and\ \bibinfo {author} {\bibfnamefont {Matias}\
  \bibnamefont {Zaldarriaga}},\ }\bibfield  {title} {\enquote {\bibinfo {title}
  {The far-infrared background correlation with cmb lensing},}\ }\href@noop {}
  {\bibfield  {journal} {\bibinfo  {journal} {\apj}\ }\textbf {\bibinfo
  {volume} {590}},\ \bibinfo {pages} {664--672} (\bibinfo {year} {2003})},\
  \Eprint {http://arxiv.org/abs/astro-ph/0209001} {astro-ph/0209001}
  \BibitemShut {NoStop}%
\bibitem [{\citenamefont {{Shang}}\ \emph {et~al.}(2012)\citenamefont
  {{Shang}}, \citenamefont {{Haiman}}, \citenamefont {{Knox}},\ and\
  \citenamefont {{Oh}}}]{shang2012}%
  \BibitemOpen
  \bibfield  {author} {\bibinfo {author} {\bibfnamefont {Cien}\ \bibnamefont
  {{Shang}}}, \bibinfo {author} {\bibfnamefont {Zolt{\'a}n.}\ \bibnamefont
  {{Haiman}}}, \bibinfo {author} {\bibfnamefont {Lloyd}\ \bibnamefont
  {{Knox}}}, \ and\ \bibinfo {author} {\bibfnamefont {S.~Peng}\ \bibnamefont
  {{Oh}}},\ }\bibfield  {title} {\enquote {\bibinfo {title} {{Improved models
  for cosmic infrared background anisotropies: new constraints on the infrared
  galaxy population}},}\ }\href {\doibase 10.1111/j.1365-2966.2012.20510.x}
  {\bibfield  {journal} {\bibinfo  {journal} {\mnras}\ }\textbf {\bibinfo
  {volume} {421}},\ \bibinfo {pages} {2832--2845} (\bibinfo {year} {2012})},\
  \Eprint {http://arxiv.org/abs/1109.1522} {arXiv:1109.1522 [astro-ph.CO]}
  \BibitemShut {NoStop}%
\bibitem [{\citenamefont {Ade}\ \emph {et~al.}(2014{\natexlab{b}})\citenamefont
  {Ade} \emph {et~al.}}]{planck2013cib}%
  \BibitemOpen
  \bibfield  {author} {\bibinfo {author} {\bibfnamefont {P.A.R.}\ \bibnamefont
  {Ade}} \emph {et~al.} (\bibinfo {collaboration} {Planck}),\ }\bibfield
  {title} {\enquote {\bibinfo {title} {{Planck 2013 results. XXX. Cosmic
  infrared background measurements and implications for star formation}},}\
  }\href {\doibase 10.1051/0004-6361/201322093} {\bibfield  {journal} {\bibinfo
   {journal} {Astron. Astrophys.}\ }\textbf {\bibinfo {volume} {571}},\
  \bibinfo {pages} {A30} (\bibinfo {year} {2014}{\natexlab{b}})},\ \Eprint
  {http://arxiv.org/abs/1309.0382} {arXiv:1309.0382 [astro-ph.CO]} \BibitemShut
  {NoStop}%
\bibitem [{\citenamefont {Aghanim}\ \emph
  {et~al.}(2016{\natexlab{b}})\citenamefont {Aghanim} \emph
  {et~al.}}]{planck2015gnilc}%
  \BibitemOpen
  \bibfield  {author} {\bibinfo {author} {\bibfnamefont {N.}~\bibnamefont
  {Aghanim}} \emph {et~al.} (\bibinfo {collaboration} {Planck}),\ }\bibfield
  {title} {\enquote {\bibinfo {title} {{Planck intermediate results. XLVIII.
  Disentangling Galactic dust emission and cosmic infrared background
  anisotropies}},}\ }\href {\doibase 10.1051/0004-6361/201629022} {\bibfield
  {journal} {\bibinfo  {journal} {Astron. Astrophys.}\ }\textbf {\bibinfo
  {volume} {596}},\ \bibinfo {pages} {A109} (\bibinfo {year}
  {2016}{\natexlab{b}})},\ \Eprint {http://arxiv.org/abs/1605.09387}
  {arXiv:1605.09387 [astro-ph.CO]} \BibitemShut {NoStop}%
\bibitem [{\citenamefont {Maniyar}\ \emph {et~al.}(2021)\citenamefont
  {Maniyar}, \citenamefont {B\'ethermin},\ and\ \citenamefont
  {Lagache}}]{Maniyar:2020tzw}%
  \BibitemOpen
  \bibfield  {author} {\bibinfo {author} {\bibfnamefont {A.}~\bibnamefont
  {Maniyar}}, \bibinfo {author} {\bibfnamefont {M.}~\bibnamefont
  {B\'ethermin}}, \ and\ \bibinfo {author} {\bibfnamefont {G.}~\bibnamefont
  {Lagache}},\ }\bibfield  {title} {\enquote {\bibinfo {title} {{Simple halo
  model formalism for the cosmic infrared background and its correlation with
  the thermal Sunyaev Zel'dovich effect}},}\ }\href {\doibase
  10.1051/0004-6361/202038790} {\bibfield  {journal} {\bibinfo  {journal}
  {\aap}\ }\textbf {\bibinfo {volume} {645}},\ \bibinfo {pages} {A40} (\bibinfo
  {year} {2021})},\ \Eprint {http://arxiv.org/abs/2006.16329} {arXiv:2006.16329
  [astro-ph.CO]} \BibitemShut {NoStop}%
\bibitem [{\citenamefont {Vieira}\ \emph {et~al.}(2010)\citenamefont {Vieira}
  \emph {et~al.}}]{vieira2010}%
  \BibitemOpen
  \bibfield  {author} {\bibinfo {author} {\bibfnamefont {J.D.}\ \bibnamefont
  {Vieira}} \emph {et~al.},\ }\bibfield  {title} {\enquote {\bibinfo {title}
  {{Extragalactic millimeter-wave sources in South Pole Telescope survey data:
  source counts, catalog, and statistics for an 87 square-degree field}},}\
  }\href {\doibase 10.1088/0004-637X/719/1/763} {\bibfield  {journal} {\bibinfo
   {journal} {Astrophys. J.}\ }\textbf {\bibinfo {volume} {719}},\ \bibinfo
  {pages} {763--783} (\bibinfo {year} {2010})},\ \Eprint
  {http://arxiv.org/abs/0912.2338} {arXiv:0912.2338 [astro-ph.CO]} \BibitemShut
  {NoStop}%
\bibitem [{\citenamefont {Vieira}\ \emph {et~al.}(2013)\citenamefont {Vieira}
  \emph {et~al.}}]{vieira2013}%
  \BibitemOpen
  \bibfield  {author} {\bibinfo {author} {\bibfnamefont {J.D.}\ \bibnamefont
  {Vieira}} \emph {et~al.},\ }\bibfield  {title} {\enquote {\bibinfo {title}
  {{Dusty starburst galaxies in the early Universe as revealed by gravitational
  lensing}},}\ }\href {\doibase 10.1038/nature12001} {\bibfield  {journal}
  {\bibinfo  {journal} {Nature}\ }\textbf {\bibinfo {volume} {495}},\ \bibinfo
  {pages} {344} (\bibinfo {year} {2013})},\ \Eprint
  {http://arxiv.org/abs/1303.2723} {arXiv:1303.2723 [astro-ph.CO]} \BibitemShut
  {NoStop}%
\bibitem [{\citenamefont {Bianchini}\ \emph {et~al.}(2015)\citenamefont
  {Bianchini} \emph {et~al.}}]{bianchini2015}%
  \BibitemOpen
  \bibfield  {author} {\bibinfo {author} {\bibfnamefont {F.}~\bibnamefont
  {Bianchini}} \emph {et~al.} (\bibinfo {collaboration} {Herschel ATLAS}),\
  }\bibfield  {title} {\enquote {\bibinfo {title} {{Cross-correlation between
  the CMB lensing potential measured by Planck and high-z sub-mm galaxies
  detected by the Herschel-ATLAS survey}},}\ }\href {\doibase
  10.1088/0004-637X/802/1/64} {\bibfield  {journal} {\bibinfo  {journal}
  {Astrophys. J.}\ }\textbf {\bibinfo {volume} {802}},\ \bibinfo {pages} {64}
  (\bibinfo {year} {2015})},\ \Eprint {http://arxiv.org/abs/1410.4502}
  {arXiv:1410.4502 [astro-ph.CO]} \BibitemShut {NoStop}%
\bibitem [{\citenamefont {Bianchini}\ \emph {et~al.}(2016)\citenamefont
  {Bianchini} \emph {et~al.}}]{bianchini2016}%
  \BibitemOpen
  \bibfield  {author} {\bibinfo {author} {\bibfnamefont {Federico}\
  \bibnamefont {Bianchini}} \emph {et~al.},\ }\bibfield  {title} {\enquote
  {\bibinfo {title} {{Toward a tomographic analysis of the cross-correlation
  between Planck CMB lensing and H-ATLAS galaxies}},}\ }\href {\doibase
  10.3847/0004-637X/825/1/24} {\bibfield  {journal} {\bibinfo  {journal}
  {Astrophys. J.}\ }\textbf {\bibinfo {volume} {825}},\ \bibinfo {pages} {24}
  (\bibinfo {year} {2016})},\ \Eprint {http://arxiv.org/abs/1511.05116}
  {arXiv:1511.05116 [astro-ph.CO]} \BibitemShut {NoStop}%
\bibitem [{\citenamefont {Aguilar~Fa\'undez}\ \emph {et~al.}(2019)\citenamefont
  {Aguilar~Fa\'undez} \emph {et~al.}}]{pbXhatlas}%
  \BibitemOpen
  \bibfield  {author} {\bibinfo {author} {\bibfnamefont {M.}~\bibnamefont
  {Aguilar~Fa\'undez}} \emph {et~al.} (\bibinfo {collaboration} {Polarbear}),\
  }\bibfield  {title} {\enquote {\bibinfo {title} {{Cross-correlation of
  POLARBEAR CMB Polarization Lensing with High-$z$ Sub-mm Herschel-ATLAS
  galaxies}},}\ }\href {\doibase 10.3847/1538-4357/ab4a78} {\bibfield
  {journal} {\bibinfo  {journal} {Astrophys. J.}\ }\textbf {\bibinfo {volume}
  {886}},\ \bibinfo {eid} {38} (\bibinfo {year} {2019})},\ \Eprint
  {http://arxiv.org/abs/1903.07046} {arXiv:1903.07046 [astro-ph.CO]}
  \BibitemShut {NoStop}%
\bibitem [{\citenamefont {Negrello}\ \emph {et~al.}(2010)\citenamefont
  {Negrello} \emph {et~al.}}]{negrello2010}%
  \BibitemOpen
  \bibfield  {author} {\bibinfo {author} {\bibfnamefont {Mattia}\ \bibnamefont
  {Negrello}} \emph {et~al.},\ }\bibfield  {title} {\enquote {\bibinfo {title}
  {{The Detection of a Population of Submillimeter-Bright, Strongly-Lensed
  Galaxies}},}\ }\href {\doibase 10.1126/science.1193420} {\bibfield  {journal}
  {\bibinfo  {journal} {Science}\ }\textbf {\bibinfo {volume} {330}},\ \bibinfo
  {pages} {800} (\bibinfo {year} {2010})},\ \Eprint
  {http://arxiv.org/abs/1011.1255} {arXiv:1011.1255 [astro-ph.CO]} \BibitemShut
  {NoStop}%
\bibitem [{\citenamefont {Gonzalez-Nuevo}\ \emph {et~al.}(2012)\citenamefont
  {Gonzalez-Nuevo} \emph {et~al.}}]{gonzalez-nuevo2012}%
  \BibitemOpen
  \bibfield  {author} {\bibinfo {author} {\bibfnamefont {J.}~\bibnamefont
  {Gonzalez-Nuevo}} \emph {et~al.} (\bibinfo {collaboration} {Herschel
  ATLAS}),\ }\bibfield  {title} {\enquote {\bibinfo {title} {{Herschel-ATLAS:
  towards a sample of \textasciitilde{}1000 strongly-lensed galaxies}},}\
  }\href {\doibase 10.1088/0004-637X/749/1/65} {\bibfield  {journal} {\bibinfo
  {journal} {Astrophys. J.}\ }\textbf {\bibinfo {volume} {749}},\ \bibinfo
  {pages} {65} (\bibinfo {year} {2012})},\ \Eprint
  {http://arxiv.org/abs/1202.0402} {arXiv:1202.0402 [astro-ph.CO]} \BibitemShut
  {NoStop}%
\bibitem [{\citenamefont {{Wilson}}\ and\ \citenamefont
  {{White}}(2019)}]{wilson2019}%
  \BibitemOpen
  \bibfield  {author} {\bibinfo {author} {\bibfnamefont {M.~J.}\ \bibnamefont
  {{Wilson}}}\ and\ \bibinfo {author} {\bibfnamefont {Martin}\ \bibnamefont
  {{White}}},\ }\bibfield  {title} {\enquote {\bibinfo {title} {{Cosmology with
  dropout selection: straw-man surveys \& CMB lensing}},}\ }\href {\doibase
  10.1088/1475-7516/2019/10/015} {\bibfield  {journal} {\bibinfo  {journal}
  {\jcap}\ }\textbf {\bibinfo {volume} {2019}},\ \bibinfo {eid} {015} (\bibinfo
  {year} {2019})},\ \Eprint {http://arxiv.org/abs/1904.13378} {arXiv:1904.13378
  [astro-ph.CO]} \BibitemShut {NoStop}%
\bibitem [{\citenamefont {{Birkinshaw}}(1999)}]{birkinshaw1999}%
  \BibitemOpen
  \bibfield  {author} {\bibinfo {author} {\bibfnamefont {M.}~\bibnamefont
  {{Birkinshaw}}},\ }\bibfield  {title} {\enquote {\bibinfo {title} {{The
  Sunyaev-Zel'dovich effect}},}\ }\href {\doibase
  10.1016/S0370-1573(98)00080-5} {\bibfield  {journal} {\bibinfo  {journal}
  {\physrep}\ }\textbf {\bibinfo {volume} {310}},\ \bibinfo {pages} {97--195}
  (\bibinfo {year} {1999})},\ \Eprint {http://arxiv.org/abs/astro-ph/9808050}
  {arXiv:astro-ph/9808050 [astro-ph]} \BibitemShut {NoStop}%
\bibitem [{\citenamefont {{Carlstrom}}\ \emph {et~al.}(2002)\citenamefont
  {{Carlstrom}}, \citenamefont {{Holder}},\ and\ \citenamefont
  {{Reese}}}]{carlstrom-sz}%
  \BibitemOpen
  \bibfield  {author} {\bibinfo {author} {\bibfnamefont {John~E.}\ \bibnamefont
  {{Carlstrom}}}, \bibinfo {author} {\bibfnamefont {Gilbert~P.}\ \bibnamefont
  {{Holder}}}, \ and\ \bibinfo {author} {\bibfnamefont {Erik~D.}\ \bibnamefont
  {{Reese}}},\ }\bibfield  {title} {\enquote {\bibinfo {title} {{Cosmology with
  the Sunyaev-Zel'dovich Effect}},}\ }\href {\doibase
  10.1146/annurev.astro.40.060401.093803} {\bibfield  {journal} {\bibinfo
  {journal} {\araa}\ }\textbf {\bibinfo {volume} {40}},\ \bibinfo {pages}
  {643--680} (\bibinfo {year} {2002})},\ \Eprint
  {http://arxiv.org/abs/astro-ph/0208192} {arXiv:astro-ph/0208192 [astro-ph]}
  \BibitemShut {NoStop}%
\bibitem [{\citenamefont {Mroczkowski}\ \emph {et~al.}(2019)\citenamefont
  {Mroczkowski} \emph {et~al.}}]{szreview2019}%
  \BibitemOpen
  \bibfield  {author} {\bibinfo {author} {\bibfnamefont {Tony}\ \bibnamefont
  {Mroczkowski}} \emph {et~al.},\ }\bibfield  {title} {\enquote {\bibinfo
  {title} {{Astrophysics with the Spatially and Spectrally Resolved
  Sunyaev-Zeldovich Effects}: {A Millimetre/Submillimetre Probe of the Warm and
  Hot Universe}},}\ }\href {\doibase 10.1007/s11214-019-0581-2} {\bibfield
  {journal} {\bibinfo  {journal} {Space Sci. Rev.}\ }\textbf {\bibinfo {volume}
  {215}},\ \bibinfo {pages} {17} (\bibinfo {year} {2019})},\ \Eprint
  {http://arxiv.org/abs/1811.02310} {arXiv:1811.02310 [astro-ph.CO]}
  \BibitemShut {NoStop}%
\bibitem [{\citenamefont {Hill}\ and\ \citenamefont
  {Spergel}(2014)}]{Hill:2013dxa}%
  \BibitemOpen
  \bibfield  {author} {\bibinfo {author} {\bibfnamefont {J.Colin}\ \bibnamefont
  {Hill}}\ and\ \bibinfo {author} {\bibfnamefont {David~N.}\ \bibnamefont
  {Spergel}},\ }\bibfield  {title} {\enquote {\bibinfo {title} {{Detection of
  thermal SZ-CMB lensing cross-correlation in Planck nominal mission data}},}\
  }\href {\doibase 10.1088/1475-7516/2014/02/030} {\bibfield  {journal}
  {\bibinfo  {journal} {JCAP}\ }\textbf {\bibinfo {volume} {02}},\ \bibinfo
  {pages} {030} (\bibinfo {year} {2014})},\ \Eprint
  {http://arxiv.org/abs/1312.4525} {arXiv:1312.4525 [astro-ph.CO]} \BibitemShut
  {NoStop}%
\bibitem [{\citenamefont {Thiele}\ \emph {et~al.}(2019)\citenamefont {Thiele},
  \citenamefont {Hill},\ and\ \citenamefont {Smith}}]{thiele-tszpdf}%
  \BibitemOpen
  \bibfield  {author} {\bibinfo {author} {\bibfnamefont {Leander}\ \bibnamefont
  {Thiele}}, \bibinfo {author} {\bibfnamefont {J.~Colin}\ \bibnamefont {Hill}},
  \ and\ \bibinfo {author} {\bibfnamefont {Kendrick~M.}\ \bibnamefont
  {Smith}},\ }\bibfield  {title} {\enquote {\bibinfo {title} {{Accurate
  analytic model for the thermal Sunyaev-Zel\textquoteright{}dovich one-point
  probability distribution function}},}\ }\href {\doibase
  10.1103/PhysRevD.99.103511} {\bibfield  {journal} {\bibinfo  {journal} {Phys.
  Rev. D}\ }\textbf {\bibinfo {volume} {99}},\ \bibinfo {pages} {103511}
  (\bibinfo {year} {2019})},\ \Eprint {http://arxiv.org/abs/1812.05584}
  {arXiv:1812.05584 [astro-ph.CO]} \BibitemShut {NoStop}%
\bibitem [{\citenamefont {Coulton}\ \emph {et~al.}(2018)\citenamefont {Coulton}
  \emph {et~al.}}]{coulton2018}%
  \BibitemOpen
  \bibfield  {author} {\bibinfo {author} {\bibfnamefont {William~R.}\
  \bibnamefont {Coulton}} \emph {et~al.},\ }\bibfield  {title} {\enquote
  {\bibinfo {title} {{Non-Gaussianity of secondary anisotropies from ACTPol and
  Planck}},}\ }\href {\doibase 10.1088/1475-7516/2018/09/022} {\bibfield
  {journal} {\bibinfo  {journal} {JCAP}\ }\textbf {\bibinfo {volume} {09}},\
  \bibinfo {pages} {022} (\bibinfo {year} {2018})},\ \Eprint
  {http://arxiv.org/abs/1711.07879} {arXiv:1711.07879 [astro-ph.CO]}
  \BibitemShut {NoStop}%
\bibitem [{\citenamefont {Ade}\ \emph {et~al.}(2016{\natexlab{b}})\citenamefont
  {Ade} \emph {et~al.}}]{planck2015-szcatalog}%
  \BibitemOpen
  \bibfield  {author} {\bibinfo {author} {\bibfnamefont {P.A.R.}\ \bibnamefont
  {Ade}} \emph {et~al.} (\bibinfo {collaboration} {Planck}),\ }\bibfield
  {title} {\enquote {\bibinfo {title} {{Planck 2015 results. XXVII. The Second
  Planck Catalogue of Sunyaev-Zeldovich Sources}},}\ }\href {\doibase
  10.1051/0004-6361/201525823} {\bibfield  {journal} {\bibinfo  {journal}
  {Astron. Astrophys.}\ }\textbf {\bibinfo {volume} {594}},\ \bibinfo {pages}
  {A27} (\bibinfo {year} {2016}{\natexlab{b}})},\ \Eprint
  {http://arxiv.org/abs/1502.01598} {arXiv:1502.01598 [astro-ph.CO]}
  \BibitemShut {NoStop}%
\bibitem [{\citenamefont {Bleem}\ \emph {et~al.}(2020)\citenamefont {Bleem}
  \emph {et~al.}}]{sptpol-szcatalog}%
  \BibitemOpen
  \bibfield  {author} {\bibinfo {author} {\bibfnamefont {L.E.}\ \bibnamefont
  {Bleem}} \emph {et~al.} (\bibinfo {collaboration} {SPT, DES}),\ }\bibfield
  {title} {\enquote {\bibinfo {title} {{The SPTpol Extended Cluster Survey}},}\
  }\href {\doibase 10.3847/1538-4365/ab6993} {\bibfield  {journal} {\bibinfo
  {journal} {Astrophys. J. Suppl.}\ }\textbf {\bibinfo {volume} {247}},\
  \bibinfo {pages} {25} (\bibinfo {year} {2020})},\ \Eprint
  {http://arxiv.org/abs/1910.04121} {arXiv:1910.04121 [astro-ph.CO]}
  \BibitemShut {NoStop}%
\bibitem [{\citenamefont {Hilton}\ \emph {et~al.}(2020)\citenamefont {Hilton}
  \emph {et~al.}}]{advact-szcatalog}%
  \BibitemOpen
  \bibfield  {author} {\bibinfo {author} {\bibfnamefont {M.}~\bibnamefont
  {Hilton}} \emph {et~al.} (\bibinfo {collaboration} {ACT, DES}),\ }\bibfield
  {title} {\enquote {\bibinfo {title} {{The Atacama Cosmology Telescope: A
  Catalog of \ensuremath{>} 4000 Sunyaev-Zel'dovich Galaxy Clusters}},}\
  }\href@noop {} {\  (\bibinfo {year} {2020})},\ \Eprint
  {http://arxiv.org/abs/2009.11043} {arXiv:2009.11043 [astro-ph.CO]}
  \BibitemShut {NoStop}%
\bibitem [{\citenamefont {Osborne}\ \emph {et~al.}(2014)\citenamefont
  {Osborne}, \citenamefont {Hanson},\ and\ \citenamefont
  {Doré}}]{Osborne:2013nna}%
  \BibitemOpen
  \bibfield  {author} {\bibinfo {author} {\bibfnamefont {Stephen~J.}\
  \bibnamefont {Osborne}}, \bibinfo {author} {\bibfnamefont {Duncan}\
  \bibnamefont {Hanson}}, \ and\ \bibinfo {author} {\bibfnamefont {Olivier}\
  \bibnamefont {Doré}},\ }\bibfield  {title} {\enquote {\bibinfo {title}
  {{Extragalactic Foreground Contamination in Temperature-based CMB Lens
  Reconstruction}},}\ }\href {\doibase 10.1088/1475-7516/2014/03/024}
  {\bibfield  {journal} {\bibinfo  {journal} {\jcap}\ }\textbf {\bibinfo
  {volume} {1403}},\ \bibinfo {pages} {024} (\bibinfo {year} {2014})},\ \Eprint
  {http://arxiv.org/abs/1310.7547} {arXiv:1310.7547 [astro-ph.CO]} \BibitemShut
  {NoStop}%
\bibitem [{\citenamefont {Aghanim}\ \emph {et~al.}(2020)\citenamefont {Aghanim}
  \emph {et~al.}}]{Aghanim:2018oex}%
  \BibitemOpen
  \bibfield  {author} {\bibinfo {author} {\bibfnamefont {N.}~\bibnamefont
  {Aghanim}} \emph {et~al.} (\bibinfo {collaboration} {Planck}),\ }\bibfield
  {title} {\enquote {\bibinfo {title} {{Planck 2018 results. VIII.
  Gravitational lensing}},}\ }\href {\doibase 10.1051/0004-6361/201833886}
  {\bibfield  {journal} {\bibinfo  {journal} {Astron. Astrophys.}\ }\textbf
  {\bibinfo {volume} {641}},\ \bibinfo {pages} {A8} (\bibinfo {year} {2020})},\
  \Eprint {http://arxiv.org/abs/1807.06210} {arXiv:1807.06210 [astro-ph.CO]}
  \BibitemShut {NoStop}%
\bibitem [{\citenamefont {Aguirre}\ \emph {et~al.}(2019)\citenamefont {Aguirre}
  \emph {et~al.}}]{Ade:2018sbj}%
  \BibitemOpen
  \bibfield  {author} {\bibinfo {author} {\bibfnamefont {James}\ \bibnamefont
  {Aguirre}} \emph {et~al.} (\bibinfo {collaboration} {Simons Observatory}),\
  }\bibfield  {title} {\enquote {\bibinfo {title} {{The Simons Observatory:
  Science goals and forecasts}},}\ }\href {\doibase
  10.1088/1475-7516/2019/02/056} {\bibfield  {journal} {\bibinfo  {journal}
  {JCAP}\ }\textbf {\bibinfo {volume} {1902}},\ \bibinfo {pages} {056}
  (\bibinfo {year} {2019})},\ \Eprint {http://arxiv.org/abs/1808.07445}
  {arXiv:1808.07445 [astro-ph.CO]} \BibitemShut {NoStop}%
\bibitem [{\citenamefont {Abazajian}\ \emph
  {et~al.}(2019{\natexlab{a}})\citenamefont {Abazajian} \emph
  {et~al.}}]{Abazajian:2019eic}%
  \BibitemOpen
  \bibfield  {author} {\bibinfo {author} {\bibfnamefont {Kevork}\ \bibnamefont
  {Abazajian}} \emph {et~al.},\ }\bibfield  {title} {\enquote {\bibinfo {title}
  {{CMB-S4 Science Case, Reference Design, and Project Plan}},}\ }\href@noop {}
  {\  (\bibinfo {year} {2019}{\natexlab{a}})},\ \Eprint
  {http://arxiv.org/abs/1907.04473} {arXiv:1907.04473 [astro-ph.IM]}
  \BibitemShut {NoStop}%
\bibitem [{\citenamefont {{Battaglia}}\ \emph {et~al.}(2012)\citenamefont
  {{Battaglia}}, \citenamefont {{Bond}}, \citenamefont {{Pfrommer}},\ and\
  \citenamefont {{Sievers}}}]{battaglia2012}%
  \BibitemOpen
  \bibfield  {author} {\bibinfo {author} {\bibfnamefont {N.}~\bibnamefont
  {{Battaglia}}}, \bibinfo {author} {\bibfnamefont {J.~R.}\ \bibnamefont
  {{Bond}}}, \bibinfo {author} {\bibfnamefont {C.}~\bibnamefont {{Pfrommer}}},
  \ and\ \bibinfo {author} {\bibfnamefont {J.~L.}\ \bibnamefont {{Sievers}}},\
  }\bibfield  {title} {\enquote {\bibinfo {title} {{On the Cluster Physics of
  Sunyaev-Zel'dovich and X-Ray Surveys. II. Deconstructing the Thermal SZ Power
  Spectrum}},}\ }\href {\doibase 10.1088/0004-637X/758/2/75} {\bibfield
  {journal} {\bibinfo  {journal} {\apj}\ }\textbf {\bibinfo {volume} {758}},\
  \bibinfo {eid} {75} (\bibinfo {year} {2012})},\ \Eprint
  {http://arxiv.org/abs/1109.3711} {arXiv:1109.3711 [astro-ph.CO]} \BibitemShut
  {NoStop}%
\bibitem [{\citenamefont {{de Zotti}}\ \emph {et~al.}(2010)\citenamefont {{de
  Zotti}}, \citenamefont {{Massardi}}, \citenamefont {{Negrello}},\ and\
  \citenamefont {{Wall}}}]{dezotti2010}%
  \BibitemOpen
  \bibfield  {author} {\bibinfo {author} {\bibfnamefont {Gianfranco}\
  \bibnamefont {{de Zotti}}}, \bibinfo {author} {\bibfnamefont {Marcella}\
  \bibnamefont {{Massardi}}}, \bibinfo {author} {\bibfnamefont {Mattia}\
  \bibnamefont {{Negrello}}}, \ and\ \bibinfo {author} {\bibfnamefont {Jasper}\
  \bibnamefont {{Wall}}},\ }\bibfield  {title} {\enquote {\bibinfo {title}
  {{Radio and millimeter continuum surveys and their astrophysical
  implications}},}\ }\href {\doibase 10.1007/s00159-009-0026-0} {\bibfield
  {journal} {\bibinfo  {journal} {\aapr}\ }\textbf {\bibinfo {volume} {18}},\
  \bibinfo {pages} {1--65} (\bibinfo {year} {2010})},\ \Eprint
  {http://arxiv.org/abs/0908.1896} {arXiv:0908.1896 [astro-ph.CO]} \BibitemShut
  {NoStop}%
\bibitem [{\citenamefont {{Padovani}}\ \emph {et~al.}(2017)\citenamefont
  {{Padovani}}, \citenamefont {{Alexander}}, \citenamefont {{Assef}},
  \citenamefont {{De Marco}}, \citenamefont {{Giommi}}, \citenamefont
  {{Hickox}}, \citenamefont {{Richards}}, \citenamefont {{Smol{\v{c}}i{\'c}}},
  \citenamefont {{Hatziminaoglou}}, \citenamefont {{Mainieri}},\ and\
  \citenamefont {{Salvato}}}]{padovani2017}%
  \BibitemOpen
  \bibfield  {author} {\bibinfo {author} {\bibfnamefont {P.}~\bibnamefont
  {{Padovani}}}, \bibinfo {author} {\bibfnamefont {D.~M.}\ \bibnamefont
  {{Alexander}}}, \bibinfo {author} {\bibfnamefont {R.~J.}\ \bibnamefont
  {{Assef}}}, \bibinfo {author} {\bibfnamefont {B.}~\bibnamefont {{De Marco}}},
  \bibinfo {author} {\bibfnamefont {P.}~\bibnamefont {{Giommi}}}, \bibinfo
  {author} {\bibfnamefont {R.~C.}\ \bibnamefont {{Hickox}}}, \bibinfo {author}
  {\bibfnamefont {G.~T.}\ \bibnamefont {{Richards}}}, \bibinfo {author}
  {\bibfnamefont {V.}~\bibnamefont {{Smol{\v{c}}i{\'c}}}}, \bibinfo {author}
  {\bibfnamefont {E.}~\bibnamefont {{Hatziminaoglou}}}, \bibinfo {author}
  {\bibfnamefont {V.}~\bibnamefont {{Mainieri}}}, \ and\ \bibinfo {author}
  {\bibfnamefont {M.}~\bibnamefont {{Salvato}}},\ }\bibfield  {title} {\enquote
  {\bibinfo {title} {{Active galactic nuclei: what's in a name?}}}\ }\href
  {\doibase 10.1007/s00159-017-0102-9} {\bibfield  {journal} {\bibinfo
  {journal} {\aapr}\ }\textbf {\bibinfo {volume} {25}},\ \bibinfo {eid} {2}
  (\bibinfo {year} {2017})},\ \Eprint {http://arxiv.org/abs/1707.07134}
  {arXiv:1707.07134 [astro-ph.GA]} \BibitemShut {NoStop}%
\bibitem [{\citenamefont {Puglisi}\ \emph {et~al.}(2018)\citenamefont
  {Puglisi}, \citenamefont {Galluzzi}, \citenamefont {Bonavera}, \citenamefont
  {Gonzalez-Nuevo}, \citenamefont {Lapi}, \citenamefont {Massardi},
  \citenamefont {Perrotta}, \citenamefont {Baccigalupi}, \citenamefont
  {Celotti},\ and\ \citenamefont {Danese}}]{Puglisi:2017lpn}%
  \BibitemOpen
  \bibfield  {author} {\bibinfo {author} {\bibfnamefont {G.}~\bibnamefont
  {Puglisi}}, \bibinfo {author} {\bibfnamefont {V.}~\bibnamefont {Galluzzi}},
  \bibinfo {author} {\bibfnamefont {L.}~\bibnamefont {Bonavera}}, \bibinfo
  {author} {\bibfnamefont {J.}~\bibnamefont {Gonzalez-Nuevo}}, \bibinfo
  {author} {\bibfnamefont {A.}~\bibnamefont {Lapi}}, \bibinfo {author}
  {\bibfnamefont {M.}~\bibnamefont {Massardi}}, \bibinfo {author}
  {\bibfnamefont {F.}~\bibnamefont {Perrotta}}, \bibinfo {author}
  {\bibfnamefont {C.}~\bibnamefont {Baccigalupi}}, \bibinfo {author}
  {\bibfnamefont {A.}~\bibnamefont {Celotti}}, \ and\ \bibinfo {author}
  {\bibfnamefont {L.}~\bibnamefont {Danese}},\ }\bibfield  {title} {\enquote
  {\bibinfo {title} {{Forecasting the Contribution of Polarized Extragalactic
  Radio Sources in CMB Observations}},}\ }\href {\doibase
  10.3847/1538-4357/aab3c7} {\bibfield  {journal} {\bibinfo  {journal}
  {Astrophys. J.}\ }\textbf {\bibinfo {volume} {858}},\ \bibinfo {pages} {85}
  (\bibinfo {year} {2018})},\ \Eprint {http://arxiv.org/abs/1712.09639}
  {arXiv:1712.09639 [astro-ph.CO]} \BibitemShut {NoStop}%
\bibitem [{\citenamefont {{Planck Collaboration XI}}(2015)}]{Aghanim:2015xee}%
  \BibitemOpen
  \bibfield  {author} {\bibinfo {author} {\bibnamefont {{Planck Collaboration
  XI}}} (\bibinfo {collaboration} {Planck}),\ }\bibfield  {title} {\enquote
  {\bibinfo {title} {{Planck 2015 results. XI. CMB power spectra, likelihoods,
  and robustness of parameters}},}\ }\href {\doibase
  10.1051/0004-6361/201526926} {\bibfield  {journal} {\bibinfo  {journal}
  {\aap}\ } (\bibinfo {year} {2015}),\ 10.1051/0004-6361/201526926},\ \Eprint
  {http://arxiv.org/abs/1507.02704} {arXiv:1507.02704 [astro-ph.CO]}
  \BibitemShut {NoStop}%
\bibitem [{\citenamefont {Choi}\ \emph {et~al.}(2020)\citenamefont {Choi} \emph
  {et~al.}}]{Choi:2020ccd}%
  \BibitemOpen
  \bibfield  {author} {\bibinfo {author} {\bibfnamefont {Steve~K.}\
  \bibnamefont {Choi}} \emph {et~al.} (\bibinfo {collaboration} {ACT}),\
  }\bibfield  {title} {\enquote {\bibinfo {title} {{The Atacama Cosmology
  Telescope: A Measurement of the Cosmic Microwave Background Power Spectra at
  98 and 150 GHz}},}\ }\href {\doibase 10.1088/1475-7516/2020/12/045}
  {\bibfield  {journal} {\bibinfo  {journal} {JCAP}\ }\textbf {\bibinfo
  {volume} {12}},\ \bibinfo {pages} {045} (\bibinfo {year} {2020})},\ \Eprint
  {http://arxiv.org/abs/2007.07289} {arXiv:2007.07289 [astro-ph.CO]}
  \BibitemShut {NoStop}%
\bibitem [{\citenamefont {Sayre}\ \emph {et~al.}(2020)\citenamefont {Sayre}
  \emph {et~al.}}]{sptpol2020}%
  \BibitemOpen
  \bibfield  {author} {\bibinfo {author} {\bibfnamefont {J.T.}\ \bibnamefont
  {Sayre}} \emph {et~al.} (\bibinfo {collaboration} {SPT}),\ }\bibfield
  {title} {\enquote {\bibinfo {title} {{Measurements of B-mode Polarization of
  the Cosmic Microwave Background from 500 Square Degrees of SPTpol Data}},}\
  }\href {\doibase 10.1103/PhysRevD.101.122003} {\bibfield  {journal} {\bibinfo
   {journal} {Phys. Rev. D}\ }\textbf {\bibinfo {volume} {101}},\ \bibinfo
  {pages} {122003} (\bibinfo {year} {2020})},\ \Eprint
  {http://arxiv.org/abs/1910.05748} {arXiv:1910.05748 [astro-ph.CO]}
  \BibitemShut {NoStop}%
\bibitem [{\citenamefont {Henning}\ \emph {et~al.}(2018)\citenamefont {Henning}
  \emph {et~al.}}]{henning2018}%
  \BibitemOpen
  \bibfield  {author} {\bibinfo {author} {\bibfnamefont {J.W.}\ \bibnamefont
  {Henning}} \emph {et~al.} (\bibinfo {collaboration} {SPT}),\ }\bibfield
  {title} {\enquote {\bibinfo {title} {{Measurements of the Temperature and
  E-Mode Polarization of the CMB from 500 Square Degrees of SPTpol Data}},}\
  }\href {\doibase 10.3847/1538-4357/aa9ff4} {\bibfield  {journal} {\bibinfo
  {journal} {Astrophys. J.}\ }\textbf {\bibinfo {volume} {852}},\ \bibinfo
  {pages} {97} (\bibinfo {year} {2018})},\ \Eprint
  {http://arxiv.org/abs/1707.09353} {arXiv:1707.09353 [astro-ph.CO]}
  \BibitemShut {NoStop}%
\bibitem [{\citenamefont {Bianchini}\ \emph {et~al.}(2020)\citenamefont
  {Bianchini} \emph {et~al.}}]{bianchini2020}%
  \BibitemOpen
  \bibfield  {author} {\bibinfo {author} {\bibfnamefont {F.}~\bibnamefont
  {Bianchini}} \emph {et~al.} (\bibinfo {collaboration} {SPT}),\ }\bibfield
  {title} {\enquote {\bibinfo {title} {{Searching for Anisotropic Cosmic
  Birefringence with Polarization Data from SPTpol}},}\ }\href {\doibase
  10.1103/PhysRevD.102.083504} {\bibfield  {journal} {\bibinfo  {journal}
  {Phys. Rev. D}\ }\textbf {\bibinfo {volume} {102}},\ \bibinfo {pages}
  {083504} (\bibinfo {year} {2020})},\ \Eprint
  {http://arxiv.org/abs/2006.08061} {arXiv:2006.08061 [astro-ph.CO]}
  \BibitemShut {NoStop}%
\bibitem [{\citenamefont {Wu}\ \emph {et~al.}(2019)\citenamefont {Wu} \emph
  {et~al.}}]{wu2019}%
  \BibitemOpen
  \bibfield  {author} {\bibinfo {author} {\bibfnamefont {W.L.K.}\ \bibnamefont
  {Wu}} \emph {et~al.},\ }\bibfield  {title} {\enquote {\bibinfo {title} {{A
  Measurement of the Cosmic Microwave Background Lensing Potential and Power
  Spectrum from 500 deg$^2$ of SPTpol Temperature and Polarization Data}},}\
  }\href {\doibase 10.3847/1538-4357/ab4186} {\bibfield  {journal} {\bibinfo
  {journal} {Astrophys. J.}\ }\textbf {\bibinfo {volume} {884}},\ \bibinfo
  {pages} {70} (\bibinfo {year} {2019})},\ \Eprint
  {http://arxiv.org/abs/1905.05777} {arXiv:1905.05777 [astro-ph.CO]}
  \BibitemShut {NoStop}%
\bibitem [{\citenamefont {Aiola}\ \emph {et~al.}(2020)\citenamefont {Aiola}
  \emph {et~al.}}]{Aiola:2020azj}%
  \BibitemOpen
  \bibfield  {author} {\bibinfo {author} {\bibfnamefont {Simone}\ \bibnamefont
  {Aiola}} \emph {et~al.} (\bibinfo {collaboration} {ACT}),\ }\bibfield
  {title} {\enquote {\bibinfo {title} {{The Atacama Cosmology Telescope: DR4
  Maps and Cosmological Parameters}},}\ }\href {\doibase
  10.1088/1475-7516/2020/12/047} {\bibfield  {journal} {\bibinfo  {journal}
  {JCAP}\ }\textbf {\bibinfo {volume} {12}},\ \bibinfo {pages} {047} (\bibinfo
  {year} {2020})},\ \Eprint {http://arxiv.org/abs/2007.07288} {arXiv:2007.07288
  [astro-ph.CO]} \BibitemShut {NoStop}%
\bibitem [{\citenamefont {Naess}\ \emph {et~al.}(2020)\citenamefont {Naess}
  \emph {et~al.}}]{naess2020}%
  \BibitemOpen
  \bibfield  {author} {\bibinfo {author} {\bibfnamefont {Sigurd}\ \bibnamefont
  {Naess}} \emph {et~al.},\ }\bibfield  {title} {\enquote {\bibinfo {title}
  {{The Atacama Cosmology Telescope: arcminute-resolution maps of 18,000 square
  degrees of the microwave sky from ACT 2008-2018 data combined with
  Planck}},}\ }\href {\doibase 10.1088/1475-7516/2020/12/046} {\bibfield
  {journal} {\bibinfo  {journal} {JCAP}\ }\textbf {\bibinfo {volume} {12}},\
  \bibinfo {pages} {046} (\bibinfo {year} {2020})},\ \Eprint
  {http://arxiv.org/abs/2007.07290} {arXiv:2007.07290 [astro-ph.IM]}
  \BibitemShut {NoStop}%
\bibitem [{\citenamefont {Holder}(2002)}]{Holder:2002wb}%
  \BibitemOpen
  \bibfield  {author} {\bibinfo {author} {\bibfnamefont {Gilbert~P.}\
  \bibnamefont {Holder}},\ }\bibfield  {title} {\enquote {\bibinfo {title}
  {{Radio point sources and the thermal sz power spectrum}},}\ }\href {\doibase
  10.1086/343094} {\bibfield  {journal} {\bibinfo  {journal} {Astrophys. J.}\
  }\textbf {\bibinfo {volume} {580}},\ \bibinfo {pages} {36--41} (\bibinfo
  {year} {2002})},\ \Eprint {http://arxiv.org/abs/astro-ph/0205467}
  {arXiv:astro-ph/0205467} \BibitemShut {NoStop}%
\bibitem [{\citenamefont {Shirasaki}(2019)}]{Shirasaki:2018wdq}%
  \BibitemOpen
  \bibfield  {author} {\bibinfo {author} {\bibfnamefont {Masato}\ \bibnamefont
  {Shirasaki}},\ }\bibfield  {title} {\enquote {\bibinfo {title} {{Impact of
  radio sources and cosmic infrared background on thermal Sunyaev--Zel'dovich
  -- gravitational lensing cross-correlation}},}\ }\href {\doibase
  10.1093/mnras/sty3162} {\bibfield  {journal} {\bibinfo  {journal} {Mon. Not.
  Roy. Astron. Soc.}\ }\textbf {\bibinfo {volume} {483}},\ \bibinfo {pages}
  {342--351} (\bibinfo {year} {2019})},\ \Eprint
  {http://arxiv.org/abs/1807.09412} {arXiv:1807.09412 [astro-ph.CO]}
  \BibitemShut {NoStop}%
\bibitem [{\citenamefont {Allison}\ \emph {et~al.}(2015)\citenamefont {Allison}
  \emph {et~al.}}]{Allison:2015fac}%
  \BibitemOpen
  \bibfield  {author} {\bibinfo {author} {\bibfnamefont {Rupert}\ \bibnamefont
  {Allison}} \emph {et~al.} (\bibinfo {collaboration} {ACT}),\ }\bibfield
  {title} {\enquote {\bibinfo {title} {{The Atacama Cosmology Telescope:
  measuring radio galaxy bias through cross-correlation with lensing}},}\
  }\href {\doibase 10.1093/mnras/stv991} {\bibfield  {journal} {\bibinfo
  {journal} {Mon. Not. Roy. Astron. Soc.}\ }\textbf {\bibinfo {volume} {451}},\
  \bibinfo {pages} {849--858} (\bibinfo {year} {2015})},\ \Eprint
  {http://arxiv.org/abs/1502.06456} {arXiv:1502.06456 [astro-ph.CO]}
  \BibitemShut {NoStop}%
\bibitem [{\citenamefont {Dwek}\ and\ \citenamefont {Barker}(2002)}]{dwek2002}%
  \BibitemOpen
  \bibfield  {author} {\bibinfo {author} {\bibfnamefont {Eli}\ \bibnamefont
  {Dwek}}\ and\ \bibinfo {author} {\bibfnamefont {Michael~K.}\ \bibnamefont
  {Barker}},\ }\bibfield  {title} {\enquote {\bibinfo {title} {The cosmic radio
  and infrared backgrounds connection},}\ }\href {\doibase 10.1086/341143}
  {\bibfield  {journal} {\bibinfo  {journal} {\apj}\ }\textbf {\bibinfo
  {volume} {575}},\ \bibinfo {pages} {7--11} (\bibinfo {year}
  {2002})}\BibitemShut {NoStop}%
\bibitem [{\citenamefont {{Wilman}}\ \emph {et~al.}(2008)\citenamefont
  {{Wilman}}, \citenamefont {{Miller}}, \citenamefont {{Jarvis}}, \citenamefont
  {{Mauch}}, \citenamefont {{Levrier}}, \citenamefont {{Abdalla}},
  \citenamefont {{Rawlings}}, \citenamefont {{Kl{\"o}ckner}}, \citenamefont
  {{Obreschkow}}, \citenamefont {{Olteanu}},\ and\ \citenamefont
  {{Young}}}]{wilman2008}%
  \BibitemOpen
  \bibfield  {author} {\bibinfo {author} {\bibfnamefont {R.~J.}\ \bibnamefont
  {{Wilman}}}, \bibinfo {author} {\bibfnamefont {L.}~\bibnamefont {{Miller}}},
  \bibinfo {author} {\bibfnamefont {M.~J.}\ \bibnamefont {{Jarvis}}}, \bibinfo
  {author} {\bibfnamefont {T.}~\bibnamefont {{Mauch}}}, \bibinfo {author}
  {\bibfnamefont {F.}~\bibnamefont {{Levrier}}}, \bibinfo {author}
  {\bibfnamefont {F.~B.}\ \bibnamefont {{Abdalla}}}, \bibinfo {author}
  {\bibfnamefont {S.}~\bibnamefont {{Rawlings}}}, \bibinfo {author}
  {\bibfnamefont {H.~R.}\ \bibnamefont {{Kl{\"o}ckner}}}, \bibinfo {author}
  {\bibfnamefont {D.}~\bibnamefont {{Obreschkow}}}, \bibinfo {author}
  {\bibfnamefont {D.}~\bibnamefont {{Olteanu}}}, \ and\ \bibinfo {author}
  {\bibfnamefont {S.}~\bibnamefont {{Young}}},\ }\bibfield  {title} {\enquote
  {\bibinfo {title} {{A semi-empirical simulation of the extragalactic radio
  continuum sky for next generation radio telescopes}},}\ }\href {\doibase
  10.1111/j.1365-2966.2008.13486.x} {\bibfield  {journal} {\bibinfo  {journal}
  {\mnras}\ }\textbf {\bibinfo {volume} {388}},\ \bibinfo {pages} {1335--1348}
  (\bibinfo {year} {2008})},\ \Eprint {http://arxiv.org/abs/0805.3413}
  {arXiv:0805.3413 [astro-ph]} \BibitemShut {NoStop}%
\bibitem [{\citenamefont {{Sehgal}}\ \emph {et~al.}(2010)\citenamefont
  {{Sehgal}}, \citenamefont {{Bode}}, \citenamefont {{Das}}, \citenamefont
  {{Hernand ez-Monteagudo}}, \citenamefont {{Huffenberger}}, \citenamefont
  {{Lin}}, \citenamefont {{Ostriker}},\ and\ \citenamefont
  {{Trac}}}]{sehgal-sim}%
  \BibitemOpen
  \bibfield  {author} {\bibinfo {author} {\bibfnamefont {Neelima}\ \bibnamefont
  {{Sehgal}}}, \bibinfo {author} {\bibfnamefont {Paul}\ \bibnamefont {{Bode}}},
  \bibinfo {author} {\bibfnamefont {Sudeep}\ \bibnamefont {{Das}}}, \bibinfo
  {author} {\bibfnamefont {Carlos}\ \bibnamefont {{Hernand ez-Monteagudo}}},
  \bibinfo {author} {\bibfnamefont {Kevin}\ \bibnamefont {{Huffenberger}}},
  \bibinfo {author} {\bibfnamefont {Yen-Ting}\ \bibnamefont {{Lin}}}, \bibinfo
  {author} {\bibfnamefont {Jeremiah~P.}\ \bibnamefont {{Ostriker}}}, \ and\
  \bibinfo {author} {\bibfnamefont {Hy}~\bibnamefont {{Trac}}},\ }\bibfield
  {title} {\enquote {\bibinfo {title} {{Simulations of the Microwave Sky}},}\
  }\href {\doibase 10.1088/0004-637X/709/2/920} {\bibfield  {journal} {\bibinfo
   {journal} {\apj}\ }\textbf {\bibinfo {volume} {709}},\ \bibinfo {pages}
  {920--936} (\bibinfo {year} {2010})},\ \Eprint
  {http://arxiv.org/abs/0908.0540} {arXiv:0908.0540 [astro-ph.CO]} \BibitemShut
  {NoStop}%
\bibitem [{\citenamefont {Li}\ \emph {et~al.}(2020)\citenamefont {Li},
  \citenamefont {Puglisi}, \citenamefont {Madavacheril},\ and\ \citenamefont
  {Alvarez}}]{websky-rs}%
  \BibitemOpen
  \bibfield  {author} {\bibinfo {author} {\bibfnamefont {Zack}\ \bibnamefont
  {Li}}, \bibinfo {author} {\bibfnamefont {Giuseppe}\ \bibnamefont {Puglisi}},
  \bibinfo {author} {\bibfnamefont {Mathew}\ \bibnamefont {Madavacheril}}, \
  and\ \bibinfo {author} {\bibfnamefont {Marcelo}\ \bibnamefont {Alvarez}},\
  }\bibfield  {title} {\enquote {\bibinfo {title} {{Websky: simulated catalogs
  and maps of radio galaxies}},}\ }\href@noop {} {\bibfield  {journal}
  {\bibinfo  {journal} {in prep.}\ } (\bibinfo {year} {2020})}\BibitemShut
  {NoStop}%
\bibitem [{\citenamefont {Abazajian}\ \emph
  {et~al.}(2019{\natexlab{b}})\citenamefont {Abazajian} \emph
  {et~al.}}]{s4dsr}%
  \BibitemOpen
  \bibfield  {author} {\bibinfo {author} {\bibfnamefont {Kevork}\ \bibnamefont
  {Abazajian}} \emph {et~al.},\ }\bibfield  {title} {\enquote {\bibinfo {title}
  {{CMB-S4 Science Case, Reference Design, and Project Plan}},}\ }\href@noop {}
  {\  (\bibinfo {year} {2019}{\natexlab{b}})},\ \Eprint
  {http://arxiv.org/abs/1907.04473} {arXiv:1907.04473 [astro-ph.IM]}
  \BibitemShut {NoStop}%
\bibitem [{\citenamefont {Naess}(2020)}]{naess-private}%
  \BibitemOpen
  \bibfield  {author} {\bibinfo {author} {\bibfnamefont {Sigurd}\ \bibnamefont
  {Naess}},\ }\bibfield  {title} {\enquote {\bibinfo {title} {Private
  communication},}\ }\href@noop {} {\  (\bibinfo {year} {2020})}\BibitemShut
  {NoStop}%
\bibitem [{\citenamefont {Reichardt}\ \emph {et~al.}(2020)\citenamefont
  {Reichardt} \emph {et~al.}}]{reichardt2020}%
  \BibitemOpen
  \bibfield  {author} {\bibinfo {author} {\bibfnamefont {C.L.}\ \bibnamefont
  {Reichardt}} \emph {et~al.} (\bibinfo {collaboration} {SPT}),\ }\bibfield
  {title} {\enquote {\bibinfo {title} {{An Improved Measurement of the
  Secondary Cosmic Microwave Background Anisotropies from the SPT-SZ + SPTpol
  Surveys}},}\ }\href@noop {} {\  (\bibinfo {year} {2020})},\ \Eprint
  {http://arxiv.org/abs/2002.06197} {arXiv:2002.06197 [astro-ph.CO]}
  \BibitemShut {NoStop}%
\bibitem [{\citenamefont {{Tegmark}}\ and\ \citenamefont {{de
  Oliveira-Costa}}(2001)}]{Tegmark2001}%
  \BibitemOpen
  \bibfield  {author} {\bibinfo {author} {\bibfnamefont {Max}\ \bibnamefont
  {{Tegmark}}}\ and\ \bibinfo {author} {\bibfnamefont {Angelica}\ \bibnamefont
  {{de Oliveira-Costa}}},\ }\bibfield  {title} {\enquote {\bibinfo {title}
  {{How to measure CMB polarization power spectra without losing
  information}},}\ }\href {\doibase 10.1103/PhysRevD.64.063001} {\bibfield
  {journal} {\bibinfo  {journal} {\prd}\ }\textbf {\bibinfo {volume} {64}},\
  \bibinfo {eid} {063001} (\bibinfo {year} {2001})},\ \Eprint
  {http://arxiv.org/abs/astro-ph/0012120} {arXiv:astro-ph/0012120 [astro-ph]}
  \BibitemShut {NoStop}%
\bibitem [{\citenamefont {Smith}(2006)}]{Smith:2005gi}%
  \BibitemOpen
  \bibfield  {author} {\bibinfo {author} {\bibfnamefont {Kendrick~M.}\
  \bibnamefont {Smith}},\ }\bibfield  {title} {\enquote {\bibinfo {title}
  {{Pseudo-$C_\ell$ estimators which do not mix E and B modes}},}\ }\href
  {\doibase 10.1103/PhysRevD.74.083002} {\bibfield  {journal} {\bibinfo
  {journal} {Phys. Rev.}\ }\textbf {\bibinfo {volume} {D74}},\ \bibinfo {pages}
  {083002} (\bibinfo {year} {2006})},\ \Eprint
  {http://arxiv.org/abs/astro-ph/0511629} {arXiv:astro-ph/0511629 [astro-ph]}
  \BibitemShut {NoStop}%
\bibitem [{\citenamefont {{Grain}}\ \emph {et~al.}(2009)\citenamefont
  {{Grain}}, \citenamefont {{Tristram}},\ and\ \citenamefont
  {{Stompor}}}]{grain2009}%
  \BibitemOpen
  \bibfield  {author} {\bibinfo {author} {\bibfnamefont {J.}~\bibnamefont
  {{Grain}}}, \bibinfo {author} {\bibfnamefont {M.}~\bibnamefont {{Tristram}}},
  \ and\ \bibinfo {author} {\bibfnamefont {R.}~\bibnamefont {{Stompor}}},\
  }\bibfield  {title} {\enquote {\bibinfo {title} {{Polarized CMB power
  spectrum estimation using the pure pseudo-cross-spectrum approach}},}\ }\href
  {\doibase 10.1103/PhysRevD.79.123515} {\bibfield  {journal} {\bibinfo
  {journal} {\prd}\ }\textbf {\bibinfo {volume} {79}},\ \bibinfo {eid} {123515}
  (\bibinfo {year} {2009})},\ \Eprint {http://arxiv.org/abs/0903.2350}
  {arXiv:0903.2350 [astro-ph.CO]} \BibitemShut {NoStop}%
\bibitem [{\citenamefont {{Fert{\'e}}}\ \emph {et~al.}(2013)\citenamefont
  {{Fert{\'e}}}, \citenamefont {{Grain}}, \citenamefont {{Tristram}},\ and\
  \citenamefont {{Stompor}}}]{ferte2013}%
  \BibitemOpen
  \bibfield  {author} {\bibinfo {author} {\bibfnamefont {A.}~\bibnamefont
  {{Fert{\'e}}}}, \bibinfo {author} {\bibfnamefont {J.}~\bibnamefont
  {{Grain}}}, \bibinfo {author} {\bibfnamefont {M.}~\bibnamefont {{Tristram}}},
  \ and\ \bibinfo {author} {\bibfnamefont {R.}~\bibnamefont {{Stompor}}},\
  }\bibfield  {title} {\enquote {\bibinfo {title} {{Efficiency of
  pseudospectrum methods for estimation of the cosmic microwave background
  B-mode power spectrum}},}\ }\href {\doibase 10.1103/PhysRevD.88.023524}
  {\bibfield  {journal} {\bibinfo  {journal} {\prd}\ }\textbf {\bibinfo
  {volume} {88}},\ \bibinfo {eid} {023524} (\bibinfo {year} {2013})},\ \Eprint
  {http://arxiv.org/abs/1305.7441} {arXiv:1305.7441 [astro-ph.CO]} \BibitemShut
  {NoStop}%
\bibitem [{\citenamefont {{Fert{\'e}}}\ \emph {et~al.}(2015)\citenamefont
  {{Fert{\'e}}}, \citenamefont {{Peloton}}, \citenamefont {{Grain}},\ and\
  \citenamefont {{Stompor}}}]{ferte2015}%
  \BibitemOpen
  \bibfield  {author} {\bibinfo {author} {\bibfnamefont {A.}~\bibnamefont
  {{Fert{\'e}}}}, \bibinfo {author} {\bibfnamefont {J.}~\bibnamefont
  {{Peloton}}}, \bibinfo {author} {\bibfnamefont {J.}~\bibnamefont {{Grain}}},
  \ and\ \bibinfo {author} {\bibfnamefont {R.}~\bibnamefont {{Stompor}}},\
  }\bibfield  {title} {\enquote {\bibinfo {title} {{Detecting the
  tensor-to-scalar ratio with the pure pseudospectrum reconstruction of B
  -mode}},}\ }\href {\doibase 10.1103/PhysRevD.92.083510} {\bibfield  {journal}
  {\bibinfo  {journal} {\prd}\ }\textbf {\bibinfo {volume} {92}},\ \bibinfo
  {eid} {083510} (\bibinfo {year} {2015})},\ \Eprint
  {http://arxiv.org/abs/1506.06409} {arXiv:1506.06409 [astro-ph.CO]}
  \BibitemShut {NoStop}%
\bibitem [{\citenamefont {Akrami}\ \emph {et~al.}(2020)\citenamefont {Akrami}
  \emph {et~al.}}]{Akrami:2018mcd}%
  \BibitemOpen
  \bibfield  {author} {\bibinfo {author} {\bibfnamefont {Y.}~\bibnamefont
  {Akrami}} \emph {et~al.} (\bibinfo {collaboration} {Planck}),\ }\bibfield
  {title} {\enquote {\bibinfo {title} {{Planck 2018 results. IV. Diffuse
  component separation}},}\ }\href {\doibase 10.1051/0004-6361/201833881}
  {\bibfield  {journal} {\bibinfo  {journal} {Astron. Astrophys.}\ }\textbf
  {\bibinfo {volume} {641}},\ \bibinfo {pages} {A4} (\bibinfo {year} {2020})},\
  \Eprint {http://arxiv.org/abs/1807.06208} {arXiv:1807.06208 [astro-ph.CO]}
  \BibitemShut {NoStop}%
\bibitem [{\citenamefont {Aghanim}\ \emph
  {et~al.}(2016{\natexlab{c}})\citenamefont {Aghanim} \emph
  {et~al.}}]{Aghanim:2016pcc}%
  \BibitemOpen
  \bibfield  {author} {\bibinfo {author} {\bibfnamefont {N.}~\bibnamefont
  {Aghanim}} \emph {et~al.} (\bibinfo {collaboration} {Planck}),\ }\bibfield
  {title} {\enquote {\bibinfo {title} {{Planck intermediate results. XLVIII.
  Disentangling Galactic dust emission and cosmic infrared background
  anisotropies}},}\ }\href {\doibase 10.1051/0004-6361/201629022} {\bibfield
  {journal} {\bibinfo  {journal} {\aap}\ }\textbf {\bibinfo {volume} {596}},\
  \bibinfo {pages} {A109} (\bibinfo {year} {2016}{\natexlab{c}})},\ \Eprint
  {http://arxiv.org/abs/1605.09387} {arXiv:1605.09387 [astro-ph.CO]}
  \BibitemShut {NoStop}%
\bibitem [{\citenamefont {Zonca}\ \emph {et~al.}(2019)\citenamefont {Zonca},
  \citenamefont {Singer}, \citenamefont {Lenz}, \citenamefont {Reinecke},
  \citenamefont {Rosset}, \citenamefont {Hivon},\ and\ \citenamefont
  {Gorski}}]{Zonca2019}%
  \BibitemOpen
  \bibfield  {author} {\bibinfo {author} {\bibfnamefont {Andrea}\ \bibnamefont
  {Zonca}}, \bibinfo {author} {\bibfnamefont {Leo}\ \bibnamefont {Singer}},
  \bibinfo {author} {\bibfnamefont {Daniel}\ \bibnamefont {Lenz}}, \bibinfo
  {author} {\bibfnamefont {Martin}\ \bibnamefont {Reinecke}}, \bibinfo {author}
  {\bibfnamefont {Cyrille}\ \bibnamefont {Rosset}}, \bibinfo {author}
  {\bibfnamefont {Eric}\ \bibnamefont {Hivon}}, \ and\ \bibinfo {author}
  {\bibfnamefont {Krzysztof}\ \bibnamefont {Gorski}},\ }\bibfield  {title}
  {\enquote {\bibinfo {title} {healpy: equal area pixelization and spherical
  harmonics transforms for data on the sphere in python},}\ }\href {\doibase
  10.21105/joss.01298} {\bibfield  {journal} {\bibinfo  {journal} {Journal of
  Open Source Software}\ }\textbf {\bibinfo {volume} {4}},\ \bibinfo {pages}
  {1298} (\bibinfo {year} {2019})}\BibitemShut {NoStop}%
\bibitem [{\citenamefont {{G{\'o}rski}}\ \emph {et~al.}(2005)\citenamefont
  {{G{\'o}rski}}, \citenamefont {{Hivon}}, \citenamefont {{Banday}},
  \citenamefont {{Wand elt}}, \citenamefont {{Hansen}}, \citenamefont
  {{Reinecke}},\ and\ \citenamefont {{Bartelmann}}}]{healpix}%
  \BibitemOpen
  \bibfield  {author} {\bibinfo {author} {\bibfnamefont {K.~M.}\ \bibnamefont
  {{G{\'o}rski}}}, \bibinfo {author} {\bibfnamefont {E.}~\bibnamefont
  {{Hivon}}}, \bibinfo {author} {\bibfnamefont {A.~J.}\ \bibnamefont
  {{Banday}}}, \bibinfo {author} {\bibfnamefont {B.~D.}\ \bibnamefont {{Wand
  elt}}}, \bibinfo {author} {\bibfnamefont {F.~K.}\ \bibnamefont {{Hansen}}},
  \bibinfo {author} {\bibfnamefont {M.}~\bibnamefont {{Reinecke}}}, \ and\
  \bibinfo {author} {\bibfnamefont {M.}~\bibnamefont {{Bartelmann}}},\
  }\bibfield  {title} {\enquote {\bibinfo {title} {{HEALPix: A Framework for
  High-Resolution Discretization and Fast Analysis of Data Distributed on the
  Sphere}},}\ }\href {\doibase 10.1086/427976} {\bibfield  {journal} {\bibinfo
  {journal} {\apj}\ }\textbf {\bibinfo {volume} {622}},\ \bibinfo {pages}
  {759--771} (\bibinfo {year} {2005})},\ \Eprint
  {http://arxiv.org/abs/astro-ph/0409513} {arXiv:astro-ph/0409513 [astro-ph]}
  \BibitemShut {NoStop}%
\bibitem [{\citenamefont {Harris}\ \emph {et~al.}(2020)\citenamefont {Harris}
  \emph {et~al.}}]{2020NumPy-Array}%
  \BibitemOpen
  \bibfield  {author} {\bibinfo {author} {\bibfnamefont {Charles~R.}\
  \bibnamefont {Harris}} \emph {et~al.},\ }\bibfield  {title} {\enquote
  {\bibinfo {title} {{Array programming with NumPy}},}\ }\href {\doibase
  10.1038/s41586-020-2649-2} {\bibfield  {journal} {\bibinfo  {journal}
  {Nature}\ }\textbf {\bibinfo {volume} {585}},\ \bibinfo {pages} {357--362}
  (\bibinfo {year} {2020})},\ \Eprint {http://arxiv.org/abs/2006.10256}
  {arXiv:2006.10256 [cs.MS]} \BibitemShut {NoStop}%
\bibitem [{\citenamefont {Virtanen}\ \emph {et~al.}(2020)\citenamefont
  {Virtanen} \emph {et~al.}}]{2020SciPy-NMeth}%
  \BibitemOpen
  \bibfield  {author} {\bibinfo {author} {\bibfnamefont {Pauli}\ \bibnamefont
  {Virtanen}} \emph {et~al.},\ }\bibfield  {title} {\enquote {\bibinfo {title}
  {{SciPy 1.0--Fundamental Algorithms for Scientific Computing in Python}},}\
  }\href {\doibase 10.1038/s41592-019-0686-2} {\bibfield  {journal} {\bibinfo
  {journal} {Nature Meth.}\ }\textbf {\bibinfo {volume} {17}},\ \bibinfo
  {pages} {261} (\bibinfo {year} {2020})},\ \Eprint
  {http://arxiv.org/abs/1907.10121} {arXiv:1907.10121 [cs.MS]} \BibitemShut
  {NoStop}%
\bibitem [{\citenamefont {Hunter}(2007)}]{Hunter:2007}%
  \BibitemOpen
  \bibfield  {author} {\bibinfo {author} {\bibfnamefont {J.~D.}\ \bibnamefont
  {Hunter}},\ }\bibfield  {title} {\enquote {\bibinfo {title} {Matplotlib: A 2d
  graphics environment},}\ }\href {\doibase 10.1109/MCSE.2007.55} {\bibfield
  {journal} {\bibinfo  {journal} {Computing in Science \& Engineering}\
  }\textbf {\bibinfo {volume} {9}},\ \bibinfo {pages} {90--95} (\bibinfo {year}
  {2007})}\BibitemShut {NoStop}%
\bibitem [{\citenamefont {Lewis}\ \emph {et~al.}(2011)\citenamefont {Lewis},
  \citenamefont {Challinor},\ and\ \citenamefont {Hanson}}]{Lewis:2011fk}%
  \BibitemOpen
  \bibfield  {author} {\bibinfo {author} {\bibfnamefont {Antony}\ \bibnamefont
  {Lewis}}, \bibinfo {author} {\bibfnamefont {Anthony}\ \bibnamefont
  {Challinor}}, \ and\ \bibinfo {author} {\bibfnamefont {Duncan}\ \bibnamefont
  {Hanson}},\ }\bibfield  {title} {\enquote {\bibinfo {title} {{The shape of
  the CMB lensing bispectrum}},}\ }\href {\doibase
  10.1088/1475-7516/2011/03/018} {\bibfield  {journal} {\bibinfo  {journal}
  {\jcap}\ }\textbf {\bibinfo {volume} {1103}},\ \bibinfo {pages} {018}
  (\bibinfo {year} {2011})},\ \Eprint {http://arxiv.org/abs/1101.2234}
  {arXiv:1101.2234 [astro-ph.CO]} \BibitemShut {NoStop}%
\bibitem [{\citenamefont {Challinor}\ and\ \citenamefont
  {Chon}(2002)}]{Challinor:2002cd}%
  \BibitemOpen
  \bibfield  {author} {\bibinfo {author} {\bibfnamefont {Anthony}\ \bibnamefont
  {Challinor}}\ and\ \bibinfo {author} {\bibfnamefont {Gayoung}\ \bibnamefont
  {Chon}},\ }\bibfield  {title} {\enquote {\bibinfo {title} {{Geometry of weak
  lensing of CMB polarization}},}\ }\href {\doibase 10.1103/PhysRevD.66.127301}
  {\bibfield  {journal} {\bibinfo  {journal} {\prd}\ }\textbf {\bibinfo
  {volume} {66}},\ \bibinfo {pages} {127301} (\bibinfo {year} {2002})},\
  \Eprint {http://arxiv.org/abs/astro-ph/0301064} {arXiv:astro-ph/0301064
  [astro-ph]} \BibitemShut {NoStop}%
\end{thebibliography}%

\end{document}